\documentclass[aps,prb,superscriptaddress,twocolumn,floatfix]{revtex4-1}

\usepackage{amsmath}
\usepackage{amssymb}
\usepackage{amsthm}
\usepackage{graphicx}
\usepackage{graphics}
\usepackage{subfigure}
\usepackage{color}
\usepackage{bm}
\usepackage{hyperref}
\usepackage{rotating}

\newcommand{\abs}[1]{\left\vert#1\right\vert}

\newcommand{\<}{\langle}
\renewcommand{\>}{\rangle}
\newcommand{\be}{\begin{equation} }
\newcommand{\ee}{\end{equation} }
\newcommand{\ba}{\begin{eqnarray} }
\newcommand{\ea}{\end{eqnarray} }
\newcommand{\n}{\nonumber \\ }

\newcommand{\mac}{\mathcal}
\newcommand{\mb}{\mathbf}

\newcommand{\bpm}{\begin{pmatrix}}
\newcommand{\epm}{\end{pmatrix}}
\newcommand{\bmm}{\begin{matrix}}
\newcommand{\emm}{\end{matrix}}

\renewcommand{\v}[1]{\boldsymbol{#1}}
\begin{document}

\title{Exactly soluble models for fractional topological insulators in 2 and 3 dimensions}
\author{Michael Levin}
\affiliation{Condensed Matter Theory Center, Department of Physics, University of Maryland, College Park, Maryland 20742, USA}

\author{F. J. Burnell}
\affiliation{Rudolf Peierls Centre for Theoretical Physics, 1 Keble Rd, Oxford OX1 3NP, UK}
\affiliation{All Souls College, Oxford, OX1 4AL, UK}

\author{Maciej Koch-Janusz}
\author{Ady Stern}
\affiliation{Department of Condensed Matter Physics, Weizmann Institute of Science, Rehovot 76100, Israel}


\begin{abstract}
We construct exactly soluble lattice models for fractionalized, time reversal invariant electronic insulators in 2 and 3
dimensions. The low energy physics of these models is exactly equivalent to a non-interacting topological 
insulator built out of fractionally charged fermionic quasiparticles. We show that some of our 
models have protected edge modes (in 2D) and surface modes (in 3D), and are thus fractionalized 
analogues of topological insulators. We also find that some of the 2D models do not have protected 
edge modes -- that is, the edge modes can be gapped out by appropriate time reversal invariant, charge
conserving perturbations. (A similar state of affairs may also exist in 3D). We show that all of our models 
are topologically ordered, exhibiting fractional statistics as well as ground state degeneracy on a torus. 
In the 3D case, we find that the models exhibit a fractional magnetoelectric effect.
\end{abstract}


\maketitle

\section{Introduction}
One of the more surprising discoveries of the past decade has been that time-reversal invariant band 
insulators come in two kinds: topological insulators and trivial insulators.
These two families of insulators exist in both two\cite{KaneMele,KaneMele2,BernevigZhang,HasanKaneRMP} and
three\cite{Roy,FuKaneMele,MooreBalents,HasanKaneRMP} dimensional systems. They are distinguished 
by the fact that the interface of a topological insulator
with the vacuum always carries a gapless edge mode (in two dimensions) or surface mode
(in three dimensions), while no such protected boundary modes exist for the trivial
insulator.

Though much of our current understanding of topological insulators has focused on non-interacting or weakly
interacting systems, it is natural to consider the fate of this physics in the presence of strong interactions. 
Strongly interacting insulators can be divided into two classes: systems that can be 
adiabatically connected to (non-interacting) band insulators without closing the bulk gap, and those that 
cannot. In the former case it has been shown (explicitly in 2D\cite{FuKanepump}, and implicitly in 
3D\cite{FuKaneHall}) that the gapless boundary modes of a topological insulator are stable to strong interactions 
as long as time reversal symmetry and charge conservation are not broken (explicitly or spontaneously). 
Therefore there is a well-defined notion of interacting topological insulators 
in systems that are adiabatically connected to band insulators.

Here we will consider the second possibility: strongly interacting, time reversal invariant electron 
systems whose ground state {\it cannot} be adiabatically connected to a band insulator.  
The same question can be posed: do some of these systems have protected gapless edge modes?
This question is particularly interesting in light of the fact that such phases can be {\it fractionalized}, leading to a 
great diversity of possibilities. That is, such phases need not have excitations that resemble electrons; in general, 
the quasiparticles may have fractional charge and statistics. 

Our understanding of these fractionalized insulators is, however, limited. 
Focusing on the two dimensional case, Ref. \onlinecite{LevinStern} analyzed
a class of strongly interacting toy models\cite{BernevigZhang} where spin-up and spin-down electrons
each form fractional quantum Hall states with opposite chiralities.\cite{Freedmanetal} Ref. \onlinecite{LevinStern} 
concluded that some of these strongly interacting, time reversal invariant insulators have 
protected edge modes, while some do not. The two kinds of insulators were dubbed ``fractional topological insulators'' and
``fractional trivial insulators'' since they are analogous to non-interacting
topological and trivial insulators, but they contain quasiparticle excitations with
fractional charge and fractional statistics. These models demonstrate that
fractionalized analogues of topological insulators are possible in principle,
but they do not exhaust all the possibilities for these phases.

In the three dimensional case, even less is known. Refs. \onlinecite{Swingle,Maciejko}
used a parton construction to build time reversal invariant insulators with a 
fractional magnetoelectric effect. However, this work did not prove that these states have
protected surface modes (though Ref. \onlinecite{Swingle} did conjecture that
this is the case). Also, Refs. \onlinecite{Swingle,Maciejko} did
not construct a microscopic Hamiltonian realizing these phases--a standard limitation
of parton or slave particle approaches.

In this paper, we address both of these issues. First, we construct   
a set of exactly soluble lattice electron models -- in both two and 
three dimensions -- that realize time reversal invariant insulators with fractionally charged 
excitations. Second, we prove that some of these fractionalized electronic 
insulators have protected edge or surface modes (that is, we show that 
perturbations cannot gap out the boundary modes without breaking time reversal 
symmetry or charge conservation symmetry, explicitly or spontaneously). In this 
sense, these models provide concrete examples of ``fractional topological insulators''
in both two and three dimensions. An added bonus of our analysis is that the 
argument we use to establish the robustness of the edge or surface modes is not specific 
to our exactly soluble models, and can be applied equally well to more general 
fractionalized (or unfractionalized) insulators. 

The low energy physics of our models is exactly equivalent to a non-interacting 
topological insulator built out of fractionally charged fermions. Surprisingly,
however, some of the models do \emph{not} have protected boundary modes. Specifically, 
we find that in some 2D models (namely those for which our protected-edge argument
breaks down) the edge modes can be gapped out by adding an appropriate 
time reversal invariant, charge conserving perturbation. In the 3D case, our understanding is more 
limited: we can prove that some of the models have protected surface modes, but we do 
not know the fate of the surface modes in the remaining models.

The models that we construct and solve are rather far from describing systems that are 
presently accessible to experiments. Nevertheless, they are of value for two reasons. 
First, they allow for a study of matters of principle, such as the existence of fractional 
topological insulators, their properties and their stability. And second, since the phases
we consider are robust to arbitrary deformations of the Hamiltonian that do not close the
bulk gap (or break time reversal or charge conservation symmetry), these phases may 
also be realized by models that are significantly different from the ones discussed here.

The paper is organized as follows. In section \ref{physpict} we describe the basic physical
picture and summarize our results. In section \ref{ModelSect}, we construct models in the 2D case. 
In section \ref{2dprop}, we analyze the physical properties of
these models, including the structure of the edge modes and the topological order in the bulk.
In section \ref{ModelSect3D}, we consider the same models in the 3D case.
We then analyze the physical properties of the 3D models in section \ref{PhysPropSect},
including the structure of the surface modes, the topological order in the bulk, and
the nature of the magnetoelectric effect. In the final part of the paper, section \ref{SurfaceSect},
we investigate whether the boundary modes in our models are robust to arbitrary time reversal
invariant, charge conserving perturbations. The Appendix contains some of the more technical calculations. 
Table \ref{Tab1} lists the various symbols that we use in the text.  

\begin{table}
\begin{tabular}{|p{1.7cm}|p{5.4cm}|p{1.0cm}|}
	\hline
Symbol &  Description & Section  \\
	\hline
$N_{\text{site}}$ & number of sites $s$ in the lattice & \ref{physpict} \\
$N_{\text{link}}$ & number of links $\<ss'\>$ in the lattice & \ref{physpict} \\
$N_{\text{plaq}}$ & number of plaquettes $P$ in the lattice & \ref{physpict} \\
\hline
$m$ & integer parameter in boson model & \ref{latbose} \\
$b_s^\dagger$ & boson creation operator on site $s$  & \ref{latbose} \\
$b_{ss'}^\dagger$ & boson creation operator on link $\<ss'\>$ & \ref{latbose} \\
$n_s$ & boson occupation number on site $s$ & \ref{latbose} \\
$n_{ss'}$ & boson occupation number on link $\<ss'\>$ & \ref{latbose} \\
$Q_s$   & cluster charge on site $s$ in boson model & \ref{latbose} \\
$B_P$   & ring exchange term on plaquette $P$ & \ref{latbose} \\
$\alpha_s$ & sublattice weighting factor for site $s$ & \ref{latbose} \\
$U_{ss'}$ & hopping term on link $\<ss'\>$ & \ref{latbose} \\
$q_s$   & eigenvalue of $Q_s$ & \ref{SolveSect} \\
$b_P$   & eigenvalue of $B_P$ & \ref{SolveSect} \\
$|q_s, b_P \>$ & simultaneous eigenstate of $Q_s, B_P$ & \ref{SolveSect} \\
$|q_s \>$ & simultaneous eigenstate $|q_s, b_P = 1\>$ & \ref{bosefrch} \\
$q_{\text{ch}}$ & electric charge of charge excitation & \ref{bosefrch} \\
\hline
$c_{s\sigma}^\dagger$ & electron creation operator & \ref{latelec} \\
$n_{s\sigma}$ & electron occupation number & \ref{latelec} \\
$n_{s,e}$ & total number of electrons on site $s$ & \ref{latelec} \\
$\tilde{Q}_s$ & cluster charge in electron model & \ref{latelec} \\
$k$ & integer parameter in electron model & \ref{latelec} \\
$\tilde{q}_s$ & eigenvalue of $\tilde{Q}_s$ & \ref{solveelec} \\
$|\tilde{q}_s, b_P, n_{s\sigma}\>$ & simultaneous eigenstate of $\tilde{Q}_s, B_P, n_{s\sigma}$ & \ref{solveelec} \\
$|n_{s\sigma},\text{elec}\>$ & electron state with occupation $\{n_{s\sigma}\}$ & \ref{elecfrch} \\
$|n_{s\sigma}\>$ & eigenstate $|\tilde{q}_s=0,b_P =1, n_{s\sigma}\>$ & \ref{elecfrch} \\
$q_{\text{f}}$ & electric charge of fermion excitation & \ref{elecfrch} \\
$H_{\text{hop}}$ & hopping term for fermion excitations & \ref{buildfrtop} \\
$t_{ss',\sigma\sigma'}$ & hopping amplitudes for fermions & \ref{buildfrtop} \\ 
$d_{s\sigma}^\dagger$ & creation operator for fermions & \ref{buildfrtop} \\
\hline
$\theta_{\text{ch,fl}}$ & Mutual statistics of charges and fluxes & \ref{2dtopord} \\
$\theta_{\text{f,fl}}$ & Statistics of fermions and fluxes & \ref{2dtopord} \\
\hline
$e^*$ & smallest charged excitation & \ref{monopolech} \\	
\hline
\end{tabular}
\caption{ \label{Tab1}List of symbols, their description, and the section where they are defined.}
\end{table}

\section{Summary of results} \label{physpict}
This section is aimed at introducing the reader to our exactly
soluble models, and to the main results we find by analyzing these models.
We will emphasize the physical picture, leaving the detailed calculations
to the following sections.

\subsection{Constructing exactly soluble models for fractional topological insulators}
To obtain candidate fractional topological insulators, we build models with two important properties: (1) fractionally charged 
fermionic quasiparticles and (2) a topological insulator band structure for these excitations. Our construction has three steps. 
In the first step we construct lattice boson models with fractionally charged bosonic excitations. In the second step 
we add electrons to the lattice, and define an electron-boson interaction that binds each electron to fractionally charged 
excitations of the bosonic model, thus creating a fractionally charged fermion. In the third step we construct a hopping term 
on the lattice that allows the fractionally charged fermion to hop between lattice sites without exciting other degrees of 
freedom. We then choose the hopping terms so that these fermions have a topological insulator band structure.

The boson models we construct are similar in spirit to the ``toric code'' model and its $Z_m$
generalizations\cite{KitaevToric}, but are built out of bosonic charged particles whose total charge is conserved.
In this sense, these models are a hybrid between the toric code model (which is exactly soluble but not charge
conserving) and the fractionalized bosonic insulators of Refs. \onlinecite{MotrunichSenthil,SenthilMotrunich, 
ReadChakraborty,MoessnerSondhiFradkin, NayakShtengel} (which are charge conserving, but not exactly soluble).

The construction of the models is based on the following recipe.
We consider a system of bosons that live on the sites $s$ and links $\<ss'\>$ of a bipartite square (or in 3D, cubic) lattice. 
We construct a bosonic Hamiltonian composed of two parts: $H_B=H_1+H_2$. Each term has an associated energy scale whose 
magnitude is of minor significance to our discussion. Both, however, depend on an integer parameter $m$ which plays a crucial role,
as it determines the fractional charge carried by the quasiparticle excitations.

The first term $H_1$ is the ``charging'' Hamiltonian. This term depends only on the number of bosons on each site $n_s$
and on each link $n_{ss'}$. Each boson is made of two electrons, of a charge $e$ each. The charging Hamiltonian assigns
different energies to different charge configurations $\{n_s,n_{ss'}\}$, by coupling the charge on a site to the charges 
on the four links neighboring the site. The spectrum of $H_1$ is discrete, as expected
from a charging Hamiltonian. The spectrum is also highly degenerate, since many charge configurations have the same energy
cost. In fact, the number of degenerate eigenstates of the lowest eigenvalue of $H_1$ is
$m^{N_{\text{link}}-N_{\text{site}}+1}$ with $N_{\text{link}}$ being the number of links in the lattice and
$N_{\text{site}}$ being the number of sites.

The second term $H_2$ is the ``hopping'' Hamiltonian. This term makes bosons hop between neighboring lattice sites and
links. A crucial aspect of our model is that the two parts are mutually commuting: $[H_1,H_2]=0$. Thus, the hopping
Hamiltonian $H_2$ only has matrix elements between degenerate states of the charging Hamiltonian $H_1$, and splits the
degeneracy for the ground state.

As we want to build an insulator, we need the ground state of $H_B$ to be separated from the excited
states by a finite energy gap. Furthermore, because we want fractionally charged excitations, $H_B$
must be topologically ordered\cite{WenReview,WenBook} (in gapped systems, fractional charge implies the existence of topological 
order). The presence of topological order means that the degeneracy of the ground state must depend on 
the topology of the system.\cite{WenReview,WenBook,Einarsson} More specifically, we need the degeneracy of the ground state to be 
independent of the system size, and to be different for a system with open and periodic boundary 
conditions.

The first condition -- existence of an energy gap -- is guaranteed in our model by having the spectra of the charging
Hamiltonian $H_1$ and the hopping Hamiltonian $H_2$ discrete. Note that this is not a common feature to hopping
Hamiltonians. The continuous spectrum of the Josephson Hamiltonian is a representative example to the contrary. To
make the spectrum discrete, we need to choose a carefully tailored hopping operator. While a conventional hopping 
Hamiltonian allows a single particle to hop between two neighboring sites, the hopping term we introduce allows only 
for a simultaneous correlated hopping of several particles around a single plaquette.

The second condition -- a ground state degeneracy that depends on the topology of the system -- is a consequence
of the way that the hopping Hamiltonian splits the degeneracy of the ground state of the charging Hamiltonian. For
example, consider the case of the 2D system defined on a torus. Each of the terms in $H_2$ describes hopping around
one of the $N_{\text{plaq}}$ plaquettes of the lattice and only one out of $m^{N_{\text{plaq}}-1}$ ground states of $H_1$
is also a ground state of $H_2$. Thus, the ground state degeneracy of the Hamiltonian $H_B$ on a torus is
$m^{N_{\text{link}}-N_{\text{site}}-N_{\text{plaq}}+2}$. By Euler's theorem, this number is exactly $m^2$. A similar
calculation in a 2D open geometry yields a ground state degeneracy of $1$. In the 3D case, the analysis is similar. One
finds that the ground state degeneracy in a 3D open geometry is again $1$, while on a 3D torus it is
$m^3$.

This counting agrees with the generalized ``toric code'' model with gauge group $G = Z_m$.\cite{KitaevToric}
The quasiparticle excitations of the boson model are also similar to the $Z_m$ toric code: there are
two types of quasiparticle excitations -- ``charge'' particles and
``flux'' particles -- which are individually bosons but have fractional mutual statistics.
Also, like the toric code model, the boson model does not have gapless edge modes. The main difference from
the $Z_m$ toric code model is that the ``charge'' quasiparticles carry a fractional electric
charge, $2e/m$.

After constructing the bosonic models, we next introduce single electron degrees of freedom that live on
the lattice sites. The electrons couple to the bosons through the charging energy,
and the electron-boson coupling is characterized by a second integer parameter $k$. We design this coupling so that
it energetically binds an electron to a composite of $k$ fractionally charged bosonic quasiparticles, each carrying
charge $2e/m$. The resulting composite particle then has a fractional charge of $q_{\text{f}}= e(1+2k/m)$, and follows fermionic
statistics. We denote the Hamiltonian of this modified lattice model by $H_e$.

In order for the composite particle to be a stable degree of freedom, it must be able to hop between lattice sites
``in one piece'', i.e. without affecting the other types of excitations. In the final step of the construction,
we find a hopping term $H_{\text{hop}}$ that does just that. We then add $H_{\text{hop}}$ to the Hamiltonian $H_e$,
choosing the hopping amplitudes so that the composite particles have a band structure of a topological insulator.
The energy gap between the bands is a parameter of $H_{\text{hop}}$, and we assume it to be much
smaller than the energy gap of the bosonic excitations.

This construction results in a system of non-interacting fermions of spin-$1/2$ and charge $q_{\text{f}} = e(1+2k/m)$ in a
topological insulator band structure in either two or three dimensions. The smallest charged excitation in the system
carries a charge $e^*=2e/m$ when $m$ is even, and a charge of $e^*=e/m$ when $m$ is odd. In the former case, this is 
a bosonic excitation. In the latter, it is a composite of fermionic and bosonic excitations.

\subsection{Properties of the models}

\subsubsection{The two-dimensional case}
In two dimensions our models realize quantized spin Hall states, with a pair of gapless edge modes and a spin-Hall
conductivity of $\frac{e}{2\pi}\left(1+\frac{2k}{m}\right )$. The topological order characterizing the states
originates from the bosonic models underlying them. The ground state degeneracy on a torus is $m^2$. In the bulk there are three
types of excitations: the bosonic charge excitation with electric charge $2e/m$, the bosonic flux excitation which
is neutral, and the fermion excitation with charge $q_{\text{f}} = e(1+2k/m)$. We find the flux excitation to have a
non-trivial mutual statistics with the other two types of excitations. When a bosonic charge excitation of charge
$2e/m$ winds around a flux excitation, it accumulates a phase of $2\pi/m$. Consequently, when a fermion, which is a
composite of an electron and $k$ bosonic charge excitations, winds around a flux particle, it accumulates a phase of
$2\pi k/m$.

In certain limits the only active degrees of freedom are those of the fermions at the edge, where a gapless mode
exists. The system can then be described as a topological insulator built out of non-interacting fermions of fractional
charge $q_{\text{f}} = e(1+2k/m)$. In particular, this description holds when the system is driven at low frequencies and
long wave lengths by a weak electromagnetic field or when thermodynamical properties are probed at low temperatures.
Under these conditions, the system would show a two-terminal conductance of $2 q_{\text{f}}^2/h$, the shot noise
associated with tunneling between edges would correspond to a charge of $q_{\text{f}}$, and the heat capacity would be
linear in temperature and proportional to the system's circumference, as expected from a 2D topological insulator
of non-interacting charge $q_{\text{f}}$ fermions.

When deviating from these conditions, the bosonic degrees of freedom can become active. Examples    
include the application of a magnetic flux of the order of a flux quantum, $\Phi_0 = hc/e$, 
per plaquette, the application of bias charges of order of $e/2m$ to particular sites and the application of an electromagnetic 
field at frequencies that correspond to the gap to bosonic excitations.

\subsubsection{The three dimensional case}
In three dimensions our models are strong topological insulators built out of charge $q_{\text{f}} = e(1+2k/m)$
fermions, with a gapless Dirac cone on each surface. When time reversal symmetry is broken on
the surface, the models exhibit a surface Hall effect with a fractional Hall conductivity of $q_{\text{f}}^2/2h$.
As in the two dimensional case, the topological order in the 3D model originates from the topological
order of the underlying bosonic system. The charged excitations carry electric charges of $2e/m$ and
$e (1+2k/m)$ and are identical to those in the 2D case, but the flux excitation becomes a flux loop
rather than the point particle it is in 2D. The ground state degeneracy on a 3D torus is $m^3$.
Again, in certain limits the bosonic degrees of freedom may be neglected and the only active degrees
of freedom are the fermionic ones. The conditions for these limits to hold are similar to those of
the two dimensional case.

The bosonic degrees of freedom are active in several cases, one of which is of particular interest.
In a 3D topological insulator of non-interacting electrons, a magnetic monopole in the bulk of the
insulator binds a half integer electric charge. \cite{WittenEffect,Wilczek,RosenbergFranz} Hence, a 
monopole/anti-monopole pair -- which may
be created by a finite-length solenoid carrying a flux quantum $\Phi_0$ and positioned within the bulk --
creates an electric dipole with a half-integer electric charge at its ends. In our model we find
that such a solenoid leads to a dipole with a charge which is a half-integer multiple of $q_{\text{f}}^2/e$.
Unlike the non-interacting case, however, the energy involved in creating the dipole is proportional 
to its length -- indicating that the two ends of the dipole cannot be effectively separated from
one another. The two ends of the dipole can be separated only when the flux carried by the solenoid is $e/e^*$
flux quanta. Furthermore, because we could presumably trap any number of additional charge $e^*$
quasiparticles near the ends of the solenoid by adding an appropriate local potential, the only quantity
which is independent of microscopic details is the monopole charge \emph{modulo} $e^*$. Calculating this quantity,
we find that the charge at the end of the $(e/e^*)\Phi_0$ solenoid is a half-integer multiple of $e^*$
for the models where $q_{\text{f}}/e^*$ is odd, and an integer multiple of $e^*$ for the models where $q_{\text{f}}/e^*$ 
is even.

\subsection{The stability of the edge or surface modes}
In conventional topological insulators, the edge or surface modes are protected as long as time
reversal symmetry and charge conservation are not broken. \cite{HasanKaneRMP} If either of these symmetries is broken,
e.g. by a Zeeman magnetic field that couples to the electron spin or by a proximity-coupling to
a superconductor that allows for Cooper pairs to tunnel into and out of the edge or surface modes,
these modes may be gapped. The breaking of time reversal symmetry may be spontaneous rather than
explicit, induced for example by the Fock term of electron-electron interaction. The stability of
the edge or surface modes to perturbations that do not break these symmetries is the distinguishing
feature of topological insulators in 2D and strong topological insulators in 3D.

An important question is whether the phases we study here have protected edge or surface modes similar to
conventional topological insulators. We find that some of the models do indeed have edge or surface modes protected
by time reversal symmetry and charge conservation, while some do not. (Independent of this difference, all the
models are topologically ordered, as demonstrated by their topological ground state degeneracy).

\subsubsection{The two dimensional case}
In the 2D case, we find that our models conform to the general rule derived in Ref. \onlinecite{LevinStern}: that
is, the edge modes are protected if and only if the ratio $\sigma_{sH}/e^*$ is odd, where $\sigma_{sH}$
is the spin-Hall conductivity in units of $e/2\pi$ and $e^*$ is the elementary charge in units of $e$.
In our models, this criterion is equivalent to the condition that the ratio $q_{\text{f}}/e^*$
is odd. 

We establish the stability of the edge modes for the models with odd $q_{\text{f}}/e^*$ by a general flux insertion
argument similar to the used in Ref. \onlinecite{LevinStern}, and establish the instability in the
case of even $q_{\text{f}}/e^*$ by explicitly constructing the perturbations whose combination gaps the edge.
This combination is rather interesting. As defined, the models have two fermionic edge modes of opposite chiralities 
-- the bosonic excitations are gapped at the edge. In order to gap the fermionic edge modes, 
we introduce one perturbation whose role is to close the gap of the bosonic excitations
at the edge, and then two additional perturbations that couple the bosonic and fermionic modes, gapping them both.

For the closure of the bosonic gap at the edge we apply a perturbation aimed at turning the edge of 
the bosonic system from an insulator into a superfluid. The natural way of doing that is by introducing 
a hopping Hamiltonian that allows fractionally charged bosonic excitations at the edge to hop from one 
site to another. When the hopping term is strong enough it can overcome the charging term described by $H_1$,
thereby closing the gap at the edge. As for the perturbations that couple the bosons and the fermions at the edge, 
the first such perturbation breaks a spinless boson of charge $2e$ into two electrons of opposite spin directions 
on the same lattice site. The second of these perturbations flips the direction of an integer number of electrons' spins,
while simultaneously operating on the flux degrees of freedom on the edge. Both of these perturbations make use
of the bosonic degrees of freedom and therefore do not have analogues in non-interacting electron systems. 

\subsubsection{The three dimensional case}
Just as in the 2D case, we find that the 3D models with odd $q_{\text{f}}/e^*$ have protected surface modes. 
We establish this result using a 3D generalization of the flux insertion
argument of Ref. \onlinecite{LevinStern}. We note that this argument is of interest beyond the particular
models discussed here, and can be applied to more general fractionalized and conventional insulators.
Unlike the 2D case, we are not able to determine the stability of the surface modes for the models with even 
$q_{\text{f}}/e^*$. Addressing this question requires either the construction of specific perturbations that gap out
the surface, or an argument proving that the surface modes are protected. 

\section{Lattice models for 2D fractional topological insulators} \label{ModelSect}

\subsection{Step 1: 2D lattice boson models with fractional charge} \label{latbose}
In this section we describe a collection of exactly soluble lattice boson models with fractionally
charged excitations---one for each integer $m \geq 2$. The models can be defined on 
any bipartite lattice in $2$ or higher dimensions. Here, for simplicity, we will focus on the case 
of the square lattice. Later, when we construct 3D models, we will consider the cubic lattice case.

\begin{figure}[t]
\centerline{
\includegraphics[width=0.7\columnwidth]{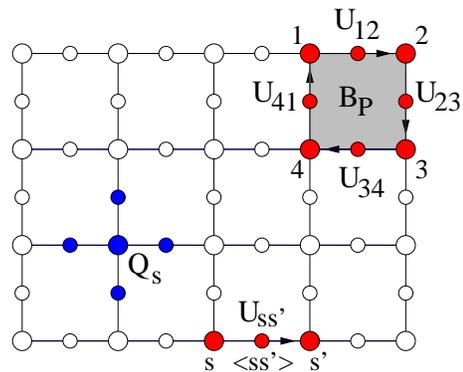}
}
\caption{
In the lattice boson model, bosons live on both the sites $s$ and links $\<ss'\>$ of the square lattice.
The Hamiltonian $H_B$ (\ref{HBose}) is a sum of a $Q_s$ term (\ref{Qsop}), which acts on four links
$\<ss'\>$ and one site $s$, and a $B_P$ term (\ref{Bpop}), which acts on the four sites and links 
adjacent to a plaquette $P$. The $B_P$ term is a product of four link operators $U_{ss'}$ (\ref{Uop}) 
which each act on the sites $s,s'$ and the link $\<ss'\>$.
}
\label{bosefig} 
\end{figure}

The basic degrees of freedom in these models are charge $2e$ spinless bosons which live on the sites
$s$ and links $\<ss'\>$ of the square lattice. We denote the boson creation operators on the
sites and links by $b_s^\dagger$ and $b_{ss'}^\dagger$ and the corresponding boson occupation numbers 
by $n_s$ and $n_{ss'}$. The Hamiltonian $H_B$ can be written as a sum of two terms, one associated with 
sites $s$, and the other associated with 
plaquettes $P$ of the square lattice 
(Fig. \ref{bosefig}):
\begin{eqnarray} \label{HBose}
H_B &=& H_1 + H_2 \nonumber \\
&=& V \sum_s Q_s^2 - \frac{u}{2} \sum_P (B_P + B_P^\dagger)
\end{eqnarray}
We will take $u, V > 0$ but otherwise arbitrary. The $Q_s$ term is a ``cluster charge'' term
which measures the total charge on the site $s$ and the four neighboring links $\<ss'\>$ with
appropriate weighting factors. It is defined as the sum
\begin{equation}  
Q_s = \alpha_s \sum_{s'} n_{ss'} + m \cdot n_s
\label{Qsop}
\end{equation}
where 
\be
\alpha_s = \begin{cases} 1 & \mbox{if }s \in A \\
m-1 & \mbox{if } s\in B
\end{cases}
\ee
and $A$ and $B$ are the two sublattices of the square lattice.
Since $V$ is positive, $V Q_s^2$ describes a short range repulsive
interaction between the bosons. This interaction breaks the sublattice symmetry between the $A$ and $B$
sublattices, except in the case $m = 2$. The $B_P$ term can be thought of as a ring exchange term. It is 
defined as the product
\begin{equation}   
B_P = U_{12} U_{23} U_{34} U_{41}
\label{Bpop}
\end{equation}
where $U_{ss'}$ is a boson hopping term on the link $\<ss'\>$:
\begin{equation}
U_{ss'} = \left(b_s^\dagger \right)^{\alpha_s-1}b_{s'}^\dagger b_{ss'}^{\alpha_{s}}
+ b_{s'}^{\alpha_{s'}-1}b_s \left(b_{ss'}^\dagger \right)^{\alpha_{s'}}
\label{Uop}
\end{equation}    
The hopping term $U_{ss'}$ describes processes where bosons hop from the sites $s,s'$
to the link $\<ss'\>$ and vice versa. It is designed so that it has two special properties.
First, $U_{ss'}$ changes the number of bosons on the site at the center of the link $\<ss'\>$ by $\pm 1 \text{ (mod m)}$ with 
the $+$ sign when $s \in B$ and the $-$ sign when $s \in A$. This change is compensated by a 
corresponding increase or decrease in the number of bosons in the two neighboring sites $s, s'$ 
so that the total number of bosons is conserved. Second, $U_{ss'}$ decreases the cluster charge 
$Q_s$ by $1$ and increases the cluster charge $Q_{s'}$ by $1$ and doesn't affect the charge on any other site:
\begin{equation} \label{QOCom}
[Q_r, U_{ss'}] =  ( \delta_{rs'}-\delta_{rs}) U_{ss'}   
\end{equation}
An important consequence of this relation is that $Q_s$ commutes with the product of $U_{ss'}$ around any set of 
closed loops, and in particular,
\begin{equation}
[Q_s, B_P] = 0 \ \ \ . \label{QBCom}
\end{equation}
Equation (\ref{QBCom}) is at the root of why our system is an insulator: the $B_P$ operator has no effect on 
the cluster charges $Q_s$ and hence does not provide for the long-distance transport of electric charge. 

The final component of the model is our definition of the boson creation operators
$b_s^\dagger, b_{ss'}^\dagger$. For the site bosons $b_s^\dagger$, we use a rotor 
representation, letting $b^\dagger_s = e^{i\theta_s}$ with $[\theta_s, n_s] = i$. The
boson occupation number on the sites can therefore be any integer, $n_s \in (-\infty, \infty)$. On the other hand, 
we take the link bosons $b_{ss'}^\dagger$ to be a kind of generalized hard-core boson, restricting the 
boson occupation number to $n_{ss'} \in 
\{0,1,...,m-1\}$, and defining $b_{ss'}^\dagger$ to be the 
$m \times m$ matrix
\begin{align}
b_{ss'}^\dagger = \bpm 0 & 1 & 0 & \cdots & 0 \\
                   0 & 0 & 1 & \cdots & 0 \\
                   \vdots & \vdots & \vdots & \vdots & \vdots \\
                   0 & 0 & 0 & \cdots & 1 \\
                   0 & 0 & 0 & \cdots & 0 \epm \ 
\end{align}
when written in the normalized number basis $\{|m-1\>,...,|0\>\}$ on the link $\<ss'\>$. We note that
while these generalized hard-core bosons are unconventional, they can arise as an effective description of a 
conventional boson system in an appropriate limit. For example, if 
the number of bosons on the link $\<ss'\>$ is very large but is restricted to a set of $m$ contiguous values 
$\{\mathcal{N}, \mathcal{N}+1,...,\mathcal{N} + m-1\}$ by appropriate energetics, then the above definition of 
$b_{ss'}^\dagger$ becomes a good approximation to conventional bosons (up to an overall normalization factor).
To summarize, the full Hilbert space for our model is spanned by the occupation number states $|n_s,n_{ss'}\>$ with 
$n_s \in (-\infty, \infty)$ and $0 \leq n_{ss'} \leq m-1$. 

\subsubsection{Solving the boson model} \label{SolveSect}
We will now show that the Hamiltonian (\ref{HBose}) is exactly soluble,
and compute its exact energy spectrum. We are already part way there, having 
established that $Q_s, B_P$ commute with each other (\ref{QBCom}). Next, we note that
\begin{align} \label{EqOp}
\left[ U_{ss'}, U_{r r'} \right] = 0 \ , \ U_{ss'}^\dagger = U_{ss'}^{-1} = U_{s's}
\end{align}
from which it follows that
\begin{align}
[B_P, B_{P'}] = [B_P, B_{P'}^\dagger] = 0 \ \ \  .
\end{align} 
Combining these results with the obvious relation
\begin{equation}
[Q_s, Q_{s'}] = 0,
\end{equation}
we conclude that $\{Q_s, B_P, B_P^\dagger\}$ all commute, and therefore can be diagonalized simultaneously. 

The simultaneous eigenstates of these operators can be labeled as $|q_s, b_P\>$, where
\begin{eqnarray}
Q_s |q_s, b_P\> &=& q_s |q_s, b_P\> \nonumber \\
B_P |q_s, b_P\> &=& b_P |q_s, b_P\> \nonumber \\
B_P^\dagger |q_s, b_P\> &=& b_P^* |q_s, b_P\> 
\end{eqnarray}
The corresponding energies are          
\begin{equation}
E = V \sum_s q_s^2 - \frac{u}{2} \sum_P (b_P + b_P^*) \ \ \ .
\end{equation}
It is clear from the definition (\ref{Qsop}) that $Q_s$ has integer eigenvalues so
$q_s$ is an integer. Similarly, using the fact that $U_{ss'}$ changes the occupation number $n_{ss'}$ by
$\pm 1 \text{ (mod m)}$, we can show that
\be
B_P^m = 1
\ee
so $b_P$ must be a $m$th root of unity. This relation guarantees what we promised in section \ref{physpict}: 
the spectrum of $H_2$ is \emph{discrete}.

\begin{figure}[t]
\centerline{
\includegraphics[width=0.8\columnwidth]{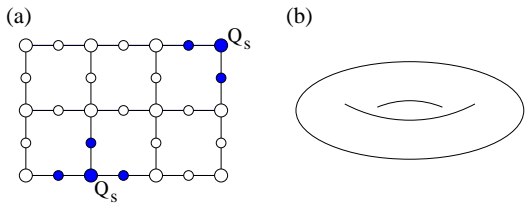}
}
\caption{
We consider the lattice boson model $H_B$ (\ref{HBose}) in two geometries: 
(a) a rectangular geometry with open boundary conditions, and (b) a periodic 
(torus) geometry. In the rectangular geometry (a), the $Q_s$ operators at the
corners act on two links $\<ss'\>$, while those at the edge act on three links.
}
\label{openbc}
\end{figure}

The only remaining question is to determine the degeneracy of each of the
$q_s, b_P$ eigenspaces. This degeneracy depends on the geometry we consider. We first consider the case of
a rectangular piece of square lattice with open boundary conditions (Fig. \ref{openbc}(a)). We show in 
Appendix \ref{degopen} that in this geometry, there is a unique eigenstate for each collection of 
$\{q_s, b_P\}$ satisfying the global constraint
\begin{equation}
\sum_s q_s \equiv 0 \ \ \text{(mod m)} 
\label{globconstr}
\end{equation}
In other words, $\{q_s, b_P\}$ are independent and complete quantum numbers, except for this
single global constraint. A similar result holds for a (periodic) torus geometry (Fig. \ref{openbc}(b)).
In this case, we find that there are $m^2$ states for each collection of $\{q_s, b_P\}$
satisfying (\ref{globconstr}) as well as the additional constraint $\prod_P b_P = 1$ (see Appendix \ref{degtorus}).
This degeneracy is a consequence of the fact that the system is topologically ordered (see Section \ref{2dtopord}).
Below we will focus exclusively on the open boundary condition geometry, unless otherwise indicated.

Putting this all together, we conclude that the ground state of (\ref{HBose}) is the unique 
state with $q_s = 0, b_P = +1$ everywhere. There 
are two types of elementary excitations: ``charge'' excitations where $q_s =1$ for some site $s$, 
and ``flux'' excitations where $b_P = e^{2\pi i/m}$ for some plaquette $P$. The total number of
charge excitations, in the bulk and edge together, must sum up to $0$ modulo $m$. The charge excitations cost 
an energy of $V$, while the flux loop excitations cost an energy of $u (1-\cos(2\pi/m))$. 
In particular, as long as $u, V > 0$, then the ground state is gapped.

\subsubsection{Fractional charge in the boson model} \label{bosefrch}
An important property of the boson model (\ref{HBose}) is that the charge excitations carry fractional
electric charge $q_{\text{ch}} = 2e/m$. We can derive this result by explicitly calculating the electric 
charge distribution in these states. Consider an eigenstate $|q_s\> \equiv |q_s, b_P = 1\>$
with some arbitrary configuration of charges $\{q_s\}$, and with no fluxes.
It is straightforward to show that the (un-normalized) microscopic wave function for this state in the 
occupation number basis is given by 
\begin{equation}
\<n_{ss'},n_s|q_s \> = 
\begin{cases} 
1 & \mbox{if }\alpha_s \sum_{s'} n_{ss'} + m n_s = q_s \mbox{ for all }s\\
0 & \mbox{otherwise }
\end{cases}
\label{qswf}
\end{equation}
(Again, to be precise, this wave function applies to a system with open boundary conditions).
This wave function has a nice property: if we were to measure the occupation number $n_{ss'}$ on any link $\<ss'\>$, then
we would find each of the $m$ possible values, $n_{ss'} = 0,1,... m-1$, with equal probability, $1/m$. This is true 
on every link $ss'$, \emph{independent} of the configuration of charges $\{q_s\}$. It follows that the expectation
value of $n_{ss'}$ in the state $|q_s\>$ is
\begin{equation}
\<n_{ss'}\>_{q_s} = \frac{0+1+...+m-1}{m} = \frac{m-1}{2}
\label{chglink}
\end{equation}
Using this result, together with the constraint $\alpha_s \sum_{s'} n_{ss'} + m n_s = q_s$, we deduce
\begin{eqnarray}
\<n_s\>_{q_s} 
&=& \frac{q_s - \alpha_s \sum_{s'} \<n_{ss'}\>}{m} \nonumber \\
&=& \frac{q_s - z\alpha_s(m-1)/2}{m}
\label{chgsite} 
\end{eqnarray}
where $z$ is the coordination number of lattice ($z = 4$ for the square lattice).

We are now in a position to compute the fractional charge $q_{\text{ch}}$. Consider the case of an 
isolated charge excitation---that is, a state where $q_s = 1$ at some site $s_0$ and $q_s = 0$ at all nearby sites. 
Denote this state by $|q_{s_0}=1\>$. Similarly, consider an eigenstate $|q_{s_0} = 0\>$  
where $q_s = 0$ both at $s_0$ and at all nearby sites. Then
the electric charge $q_{\text{ch}}$ carried by the excitation is given by the
difference in expectation values
\begin{equation}
q_{\text{ch}} = 2e \left[\<N_\mac{S}\>_{1} - \<N_\mac{S}\>_{0} \right]
\label{defchg}
\end{equation}
where $N_\mac{S}$ is an operator which measures the total number of bosons in some 
large area $\mac{S}$ containing $s_0$:
\begin{equation}
N_\mac{S} = \sum_{s \in \mac{S}} n_s + \sum_{s,s' \in \mac{S}} n_{ss'}
\end{equation}
Using the above results (\ref{chglink}-\ref{chgsite}) we derive
\begin{eqnarray}
\<n_{ss'}\>_{1} - \<n_{ss'}\>_{0} &=& 0 \nonumber \\
\<n_{s}\>_{1} - \<n_{s}\>_{0} &=& \frac{\delta_{s_0 s}}{m}
\end{eqnarray}
implying that $q_{\text{ch}} = \frac{2e}{m}$. Furthermore, we can see from these expressions
that this charge is perfectly localized to the site $s_0$. This perfect localization
is specific to the exactly soluble model: in a generic gapped system we expect 
excitations to have a finite size of order the correlation length.

Alternatively, we can derive the fractional charge using a simple identity: 
for any set of sites $\mac{S}$ in the square lattice, we have the relation 
\begin{eqnarray} \label{Eq_Ns2}
\sum_{s \in \mac{S} } Q_s &=& m \left(\sum_{s \in \mac{S}} n_s + \sum_{s, s' \in \mac{S}} n_{ss'} \right)
 + \sum_{s \in \mac{S}, s' \in \mac{S}^c } \alpha_s n_{ss'} \nonumber \\
&=& m N_\mac{S} + \sum_{s \in \mac{S}, s' \in \mac{S}^c } \alpha_s n_{ss'}
\end{eqnarray}
Taking expectation values of both sides in the two states $|q_s=1\>, |q_s=0\>$, and subtracting gives
\begin{equation}
\<N_\mac{S}\>_{1} - \<N_\mac{S}\>_{0} = \frac{1}{m}  
- \sum_{s \in \mac{S}, s' \in \mac{S}^c } \alpha_s [\<n_{ss'}\>_{1} - \<n_{ss'}\>_{0}]
\end{equation}
To complete the calculation, we note that the second term on the right hand side vanishes in the 
limit that $\mac{S}$ becomes large, since in that case, the sum only involves links $ss'$ that are far
from $s_0$, and the excess charge $(\<n_{ss'}\>_{1} - \<n_{ss'}\>_{0})$ must vanish at large
distances from $s_0$. It follows that $q_{\text{ch}} = 2e/m$, as claimed.

\subsection{Step 2: 2D lattice electron models with fractional charge} \label{latelec}
We are now ready to construct a model with fractionally charged spin-$1/2$ fermionic quasiparticle
excitations. We accomplish this task by modifying the lattice boson model (\ref{HBose}) and coupling the bosons to 
additional (unpaired) electron degrees of freedom. 

The Hilbert space for this modified model is very similar to the boson model, except that we
introduce an additional spin-$1/2$ electron degree of freedom at each lattice site $s$. 
In addition, we now think of the lattice bosons as being charge $2e$, spin-singlet pairs of electrons. 
This microscopic picture for the bosons is important conceptually because
ultimately we want a model for a fractional topological insulator which is constructed out of 
electron degrees of freedom.

We will denote the creation and annihilation operator for the (unpaired)
electrons by $c_{s\sigma}^\dagger, c_{s \sigma}$,
and their occupation number by $n_{s\sigma}$. We will use $n_{s,e}$ to 
denote the total number of the (unpaired) electrons on site $s$:
\begin{equation}
n_{s,e} = \sum_\sigma n_{s\sigma} = \sum_\sigma c_{s\sigma}^\dagger c_{s\sigma}
\end{equation}
Later, we will be interested in models with multiple orbitals on each lattice site $s$. In that
case, we will let $\sigma$ include both the spin and orbital degrees of freedom.

In this notation, the Hamiltonian $H_e$ for the electron model is a sum of three terms:
\begin{equation} \label{HFermi}
H_e = V \sum_s \tilde{Q}_s^2 - \frac{u}{2} \sum_P (B_P + B_P^\dagger) -\mu \sum_{s \sigma} n_{s\sigma}
\end{equation}
where
\begin{equation}
\tilde{Q}_s = Q_s - k \cdot n_{s,e}
\end{equation}
and $B_P, Q_s$ are defined as in Eq. (\ref{Qsop}-\ref{Uop}). We will take $V, u > 0$, but will 
consider both positive and negative $\mu$, and arbitrary integer $k$.

\subsubsection{Solving the electron model} \label{solveelec}
The electron model (\ref{HFermi}) can be solved in the same way as the original boson model. Just as before,
\begin{equation}
[\tilde{Q}_s, \tilde{Q}_{s'}] = [B_P, B_{P'}] = [B_P, B_{P'}^\dagger] = [\tilde{Q}_s, B_P] = 0
\end{equation}
Also, it is clear that
\begin{equation}
[\tilde{Q}_{s'}, n_{s\sigma}] = [B_P, n_{s\sigma}] = 0
\end{equation}
Hence, we can simultaneously diagonalize $\{\tilde{Q}_s, B_P, B_P^\dagger, n_{s\sigma}\}$. Let
$|\tilde{q}_s, b_P, n_{s\sigma}\>$
denote the simultaneous eigenstates, where $q_s$ is an integer, $b_P$ is a $m$th root of unity, and
$n_{s\sigma} = 0,1$.
The corresponding energies are: 
\begin{equation}
E = V \sum_s \tilde{q}_s^2 - \frac{u}{2} \sum_P (b_P + b_P^*) -\mu \sum_{s\sigma} n_{s\sigma}
\end{equation}
By our analysis of the lattice boson model, we know that there is a unique state
for each choice of $\tilde{q}_s, b_P, n_{s\sigma}$ satisfying the global constraint
\begin{equation}
\sum_s \tilde{q}_s + k\sum_{s\sigma} n_{s\sigma} \equiv 0 \ \ \text{(mod m)}
\end{equation}
(Again, we are assuming a geometry with open boundary conditions).

Putting this all together, and taking $\mu < 0$ for simplicity, we conclude that the ground state 
is the unique state with $\tilde{q}_s = 0, b_P = 1, n_{s\sigma} = 0$. The system has 
three types of elementary excitations: ``charge'' excitations with $\tilde{q}_s = 1$ for some site $s$, 
``flux'' excitations where $b_p = e^{2\pi i/m}$ on some plaquette $p$, and 
spin-$1/2$ ``fermion'' excitations, where $n_{s\sigma} = 1$ for some site $s$ and spin 
$\sigma = \uparrow, \downarrow$. The charge excitations cost an energy of $V$, the flux 
excitations cost an energy of $u (1-\cos(2\pi/m))$, and the fermion excitations cost energy 
$-\mu$. In particular, as long as $u, V > 0$ and $\mu < 0$, the ground state is gapped. 

\subsubsection{Fractional charge in the electron model} \label{elecfrch}
Our next task is to show that the fermion excitations carry fractional charge.
To do this, we first need to introduce some notation. Let $|n_{s\sigma},elec\>$ denote
the electron occupation basis state
\begin{equation} 
|n_{s\sigma}, elec\> = \prod_s (c_{s \sigma}^\dagger)^{n_{s\sigma}} |0\>
\label{basis}
\end{equation}
where $|0\>$ is the empty state. Also, let $|n_{ss'},n_s\>$ denote the boson occupation
state defined in section \ref{latbose}. A complete basis for the Hilbert space of the
electron model is given by tensor product states $|n_{s\sigma},elec\> \otimes |n_s, n_{ss'}\>$.

In addition to these general basis states, we will also find it useful to think about the set 
of eigenstates $|n_{s\sigma}\> \equiv |\tilde{q}_s = 0, b_P = 1, n_{s\sigma}\>$ 
made up of some arbitrary configuration of fermions $\{n_{s\sigma}\}$ with no charge or flux excitations.
(To construct these states we must impose the global constraint 
$\sum_{s\sigma} n_{s\sigma} \equiv 0 \text{ (mod m)}$). The microscopic wave function for these states is given by 
\begin{equation} \label{PsiLow}
|n_{s \sigma}\> = |n_{s \sigma},elec\> \otimes |q_s =  k n_{s,e}\>
\end{equation}
where $|q_s\>$ are the boson eigenstates defined in (\ref{qswf}). 

In order to compute the fractional charge carried by the fermion, it suffices to consider a state 
with an \emph{isolated} fermion excitation--- that is, suppose $n_{s\sigma} = 1$ at some site $s_0$ 
and $n_{s\sigma}$ vanishes at all nearby sites. Denote this state by $|n_{s_0\sigma} = 1\>$. According to the definition
(\ref{PsiLow}), this state is a tensor product of an electron state and a boson
state:
\begin{equation}
|n_{s_0\sigma}=1\> = |n_{s_0\sigma} = 1,elec\> \otimes |q_{s_0} = k\>
\label{composite}
\end{equation}
We can see that $|n_{s_0\sigma} = 1,elec\>$ consists of a single spin-$\sigma$ electron at site $s_0$, while
$|q_{s_0} = k\>$ corresponds to $k$ ``charge'' excitations at site
$s_0$. Thus, the fermion excitation is a composite particle made of an electron and
$k$ charge excitations. To compute the total charge of the fermion, we need
to add together the contributions coming from these two pieces. By our analysis 
of the fractional charge in the bosonic model, we know that each charge excitation
carries charge $2e/m$. On the other hand, the electron clearly 
has charge $e$. Adding together these two contributions, 
we conclude that the fermion excitation has charge 
\begin{equation}
q_{\text{f}}= e(1 + 2k/m)
\label{frcharge}
\end{equation}

\subsubsection{Time reversal symmetry and the electron model}
To construct candidate fractional topological insulators, it will be important
to understand how the fermionic excitations in our model transform under time reversal. 
We use the usual convention for $\mathcal{T}$, where the electron 
creation operators transform according to:
\begin{align}
\mathcal{T}: \ c^\dagger_{s \uparrow} \rightarrow c^\dagger_{s\downarrow} \ , 
c^\dagger_{s \downarrow} \rightarrow -c^\dagger_{s\uparrow}
\end{align}
In this convention, the bosons in (\ref{HFermi}) transform trivially, since they are spin 
singlet pairs of electrons:
\begin{align}
b^\dag_{ss'} \Rightarrow b^\dag_{ss'} \ , \  b^\dagger_s  \rightarrow b^\dagger_s \
\end{align}
Applying these transformation laws, we see that the fermion excitations transform like 
spin-$1/2$ electrons.

\subsection{Step 3: Building candidate 2D fractional topological insulators}
\label{buildfrtop}
In the last two sections, we have shown that the fermionic excitations of the electron model 
(\ref{HFermi}) carry spin-$1/2$ and transform under time reversal just like 
electrons. In fact, they are virtually indistinguishable from electrons except for the fact 
that they carry fractional charge $q_{\text{f}}$ (\ref{frcharge}).
Given these properties, it is easy to build a candidate fractional topological insulator: 
we simply put the fractionally charged fermions into 
a non-interacting topological insulator band structure. We accomplish this by
adding a new term to the electron Hamiltonian $H_e$ (\ref{HFermi}): 
\begin{eqnarray}
H &=& H_e + H_{\text{hop}} \nonumber \\
&=& \left( V \sum_s \tilde{Q}_s^2 - \frac{u}{2} \sum_P (B_P + B_P^\dagger) - \mu \sum_{s \sigma} n_{s\sigma} \right) \nonumber \\
&+& H_{\text{hop}}
\label{Hfrtop}
\end{eqnarray}
where
\begin{equation}
H_{\text{hop}} = -\sum_{\<ss'\>} (t_{ss' \sigma \sigma'} c_{s' \sigma'}^\dagger c_{s \sigma} U^{k}_{ss'} + h.c.)
\label{Hhop}
\end{equation}
and $\tilde{Q}_s, B_P, U_{ss'}$ are defined as before. The new term $H_{\text{hop}}$ gives
an amplitude for the fermion excitations to hop from site to site without affecting any of
the other degrees of freedom, as we now show. We will assume that $t_{ss' \sigma \sigma'} \ll u,V$
so that the bandwidth of the fermion excitations is much smaller than the gap to the bosonic excitations.

We can understand the effect of $H_{\text{hop}}$ by computing the matrix elements of this operator
between different eigenstates of $H_e$,
\begin{equation}
\<\tilde{q}'_s, b'_P, n'_{s\sigma}| H_{\text{hop}}|\tilde{q}_s, b_P, n_{s\sigma}\>
\end{equation}
This computation is considerably simplified by the fact that
\begin{equation}
[\tilde{Q}_s, H_{\text{hop}}] = [B_P, H_{\text{hop}}] = 0
\end{equation}
implying that the matrix elements are only nonzero when $\tilde{q}'_s = \tilde{q}_s$, and $b'_P = b_P$.
In what follows, we specialize to the $\tilde{q}_s = \tilde{q}'_s = 0, b_P = b'_P = 1$ case,
since these are the lowest energy states and this is all we will need to understand the low energy physics. 
These states contain only fermions and no other excitations. As 
in (\ref{PsiLow}), we will denote a state with some arbitrary configuration of fermions 
$\{n_{r\tau}\}$ using the abbreviated notation $|n_{r\tau}\> \equiv |\tilde{q}_s = 0, b_P = 1, n_{r\tau}\>$ (Here 
$r$ labels the sites of the lattice, while $\tau = \uparrow, \downarrow$ labels the two possible
spin states). To find the matrix elements $\<n'_{r\tau}| H_{\text{hop}} | n_{r\tau}\>$, we write
\begin{equation}
c_{s'\sigma}^\dagger c_{s\sigma} U^{k}_{ss'} |n_{r\tau}\> =
c_{s'\sigma'}^\dagger c_{s\sigma} |n_{r \tau},elec\> \otimes U^{k}_{ss'} |q_r =   k n_{r,e}\>
\label{umat}
\end{equation}
and then analyze each of these two pieces in turn. Using the explicit form of 
$|q_r\>$ (\ref{qswf}), we find
\begin{equation}
U^{k}_{ss'} |q_r \> = |q_r'\>
\end{equation}
where
\begin{equation}
q_r' = q_r - k \delta_{rs} + k \delta_{rs'}
\end{equation}
Similarly, we have
\begin{equation}
c_{s'\sigma'}^\dagger c_{s\sigma} |n_{r \tau},elec\> = \pm |n'_{r \tau}, elec\>
\end{equation}
where
\begin{equation}
n'_{r\tau} = n_{r\tau} - \delta_{rs} \delta_{\sigma\tau} + \delta_{rs'}\delta_{\sigma' \tau}
\label{ndef}
\end{equation}
and the $\pm$ sign depends on the ordering of the electron creation and annihilation operators in 
Eq. (\ref{basis}). Combining these two results with (\ref{umat}), we derive
\begin{equation}
c_{s'\sigma}^\dagger c_{s\sigma} U^{k}_{ss'} |n_{r\tau}\> = \pm |n'_{r\tau},elec\> \otimes |q'_{r}\> 
\equiv \pm |n'_{r\tau}\>
\end{equation}
where $n'_{r\tau}$ is defined as in (\ref{ndef}). This relation establishes what we promised earlier: $H_{\text{hop}}$ 
gives an amplitude for the fermion excitations to hop from site to site, but does not affect the other
types of excitations. (We should not take these properties for granted: for example, if we had not included the 
operator $U^k_{ss'}$ in the definition of $H_{\text{hop}}$ (\ref{Hhop}) then $H_{\text{hop}}$ would have affected the bosonic 
charge excitations).

Denoting the creation operators for the fermion excitations by $d_{s\sigma}^\dagger$, we conclude
that the matrix elements of $H$ within the low energy $\tilde{q}_s = 0, b_P = 1$ subspace are
given by the free fermion Hamiltonian
\begin{equation} \label{HNI}
H_{\text{eff}} = -\sum_{\< ss'\> } (t_{ss' \sigma \sigma'} d_{s' \sigma'}^\dagger d_{s \sigma} + h.c.) 
- \mu \sum_{s\sigma} d_{s\sigma}^\dagger d_{s\sigma} 
\end{equation}
To complete the construction, we choose the hopping amplitudes $t_{ss' \sigma \sigma'}$
and the chemical potential $\mu$ so that $H_{\text{eff}}$ is a non-interacting topological
insulator. The low energy physics is then described by a topological insulator built out of 
\emph{fractionally} charged fermions. 

There are many possible choices for $t_{ss' \sigma \sigma'}, \mu$, but to be concrete 
we will focus our discussion on the following tight binding model on the square 
lattice\cite{BernevigHughesZhang}. We consider a model with two orbitals, whose hopping 
matrix elements are related by time reversal symmetry: letting $(\sigma,\sigma')$ denote the 
fermion spin, and $(1,2)$ denote the orbital index, we have:
\be
t_{ss' \sigma \sigma',1} = t_{ss' \sigma \sigma',2}^*
\ee
and there is no hopping matrix element connecting the two orbitals.  
For each spin, the hopping matrix elements for orbital $1$ are: 
\begin{eqnarray}
t_{s s' \sigma \sigma',1} &=& \sum_{\hat{e}_i = \hat{x}, \hat{y} } \delta_{s-s', \hat{e}_i} 
\left( t  \sigma_z+  i\lambda  \sigma_i \right) + \delta_{s-s', 0} t (\kappa -2 )  \sigma_z \n
 \mu &=& 0 
\label{2dtiband}
\end{eqnarray}
where $\sigma_{x,y,z}$ are Pauli matrices acting on the spin indices.  Here $t$ 
parametrizes the spin-independent hopping terms, and $\lambda$ is a spin-orbit coupling.
The constant $\kappa$ sets the band gap at the momenta $(0,0), (\pm \pi,0), ( 0, \pm \pi )$; in the regime 
$ 0 < \kappa < 2 $, the model is in a topologically insulating phase.

In addition to being a topological insulator, the model (\ref{2dtiband}) conserves the total
$z$ component of the spin. The ground state 
consists of a filled spin-up band with Chern number $\nu = +1$, and a filled spin-down band 
with Chern number $\nu = -1$. As a result, the model exhibits a spin-Hall conductivity of
\begin{equation}
\sigma_{sH} = \frac{q_{\text{f}}}{2\pi}
\label{spinHall}
\end{equation}

\subsection{Effect of an electromagnetic field} \label{emeffect}
In this section, we investigate the response of the fractionalized insulator (\ref{Hfrtop}) to an applied
electromagnetic field. We show that for weak, slowly varying fields, the model behaves like 
a non-interacting system of fractionally charged fermions. On the other hand, for stronger fields, we find 
that other, non-fermionic, degrees of freedom contribute to the response.

\begin{figure}[t] 
\centerline{
\includegraphics[width=0.5\columnwidth]{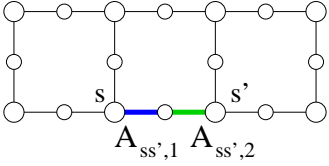}
}
\caption{
In order to include an electromagnetic vector potential in $H$ (\ref{Hfrtop}), we need to
define a vector potential for each of the two halves of each link $\<ss'\>$.
We denote these vector potentials by $A_{ss',1}$, $A_{ss',2}$.
}
\label{gaugefield}
\end{figure}

The first step is to understand how to incorporate a vector potential $A$ into the
Hamiltonian. As the model (\ref{Hfrtop}) contains charged bosons that live on links of the lattice in addition to those that live 
on the sites, we define a lattice vector potential $A_{ss',1}, A_{ss',2}$ for each of the two halves of each link 
$\<ss'\>$ (see Fig. \ref{gaugefield}). The hopping operator $U_{ss'}$ is then\cite{Kogut}
\ba
U_{ss'} &=& \left(b_s^\dagger\right)^{\alpha_s-1} b_{s'}^\dagger b_{ss'}^{\alpha_{s}} 
e^{ 2ie(1-\alpha_s) A_1} e^{2ieA_2 }
\\ &&
+ b_{s'}^{\alpha_{s'}-1}b_s\left( b_{ss'}^\dagger \right)^{\alpha_{s'}} e^{2ie(1-\alpha_{s'}) A_2 } e^{2ieA_1} 
\nonumber 
\ea
We can simplify this expression with the help of the unitary transformation
\begin{align*}
W_A = \exp \left[-\frac{2ie}{m} \cdot \sum_{\<ss'\>} n_{ss'} \cdot (\alpha_s A_{ss',1} - \alpha_{s'} A_{ss',2}) \right]
\end{align*}
A little algebra shows that
\begin{eqnarray}
W_A U_{ss'} W_A^{-1} &=& \left [ \left(b_s^\dagger\right)^{\alpha_s-1} b_{s'}^\dagger b_{ss'}^{\alpha_{s}} \right. \nonumber \\
&+& \left. b_{s'}^{\alpha_{s'}-1}b_s\left( b_{ss'}^\dagger \right)^{\alpha_{s'}} \right] e^{i \frac{2e}{m} A_{ss'} }
\end{eqnarray}
where $A_{ss'} = A_{ss',1} + A_{ss',2}$ is the total vector potential on the link $\<ss'\>$. In retrospect,
this expression is to be expected as $U_{ss'}$ hops a charge $2e/m$ from $s$ to $s'$, and as such, should
be multiplied by a phase factor $e^{2ieA_{ss'}/m}$ in the presence of an electromagnetic vector potential.
Substituting this expression into $H$ (\ref{Hfrtop}), we find that the Hamiltonian can be written as
\begin{eqnarray} \label{HfrtopA}
W_A H W_A^{-1} &=& V \sum_s \tilde{Q}_s^2 - \frac{u}{2} \sum_P (B_P e^{2i e \phi_P/m} + h.c.) \nonumber \\
&-&\mu \sum_{s \sigma} n_{s\sigma} + H_{\text{hop}}
\end{eqnarray}  
where $\phi_P = A_{12}+ A_{23}+ A_{34}+ A_{41}$,
\be
H_{\text{hop}} = -\sum_{\langle ss'\rangle} (t_{ss' \sigma \sigma'} e^{i q_{\text{f}} A_{ss'}} 
c_{s' \sigma'}^\dagger c_{s \sigma} U^{k}_{ss'} + h.c.) 
\ee
and $\tilde{Q}_s, B_P, U_{ss'}$ are defined as in the $A=0$ case. 

We now analyze this Hamiltonian in several cases. First, we consider the case where $A$ is time independent
and the flux $\phi_P$ through each plaquette is small compared with a unit flux quantum.
We proceed by simultaneously diagonalizing 
$\{\tilde{Q}_s, B_P, B_P^\dagger, n_{s\sigma}\}$, denoting the eigenstates by $|\tilde{q}_s, b_P, n_{s\sigma}\>$.
We then define rotated states
\begin{equation}
|\tilde{q}_s,b_P,n_{s\sigma},A_{ss'}\> \equiv W_A|\tilde{q}_s, b_P, n_{s\sigma}\>
\label{eigenabbrev}
\end{equation}
We note that in the absence of the fermion hopping term $H_{\text{hop}}$, these states are exact 
eigenstates of $H$ with energies
\begin{eqnarray} \label{EmEs}
E &=& V \sum_{s} \tilde{q}_s^2 
- \frac{u}{2} \sum_P ( b_P e^{2i e \phi_P/m} + b_P^* e^{-2i e \phi_P/m}) \nonumber \\
&-& \mu \sum_{s\sigma} n_{s\sigma}
\end{eqnarray}
When the flux $\phi_P$ is small (specifically, $|\phi_P| < \Phi_0/4$), the lowest energy states 
are those with $\tilde{q}_s = 0, b_P = 1$. 
We will denote these (purely fermionic) states using the abbreviated notation
\begin{equation}
|n_{r\tau},A_{ss'}\> \equiv W_A|\tilde{q}_s = 0, b_P = 1, n_{r\tau}\>
\label{fermAst}
\end{equation} 
To obtain the low energy effective Hamiltonian $H_{\text{eff}}$, we project
$H$ (\ref{HfrtopA}) to the subspace spanned by these states (\ref{fermAst}). 
By the calculation in the previous section, the matrix elements of (\ref{HfrtopA}) 
within this subspace are given by
\begin{eqnarray} \label{EqFermiGauge}
H_{\text{eff}} &=& -\sum_{\< ss'\> } 
(t_{ss' \sigma \sigma'} e^{i q_{\text{f}} A_{ss'}} d_{s' \sigma'}^\dagger d_{s \sigma} + h.c.) \nonumber \\ 
&-& \mu \sum_{s\sigma} d_{s\sigma}^\dagger d_{s\sigma} 
\end{eqnarray}
This result proves that the model behaves like a non-interacting system of fractionally charged fermions, 
even in the presence of a weak, time independent vector potential $A$.

In fact, the low energy effective Hamiltonian (\ref{EqFermiGauge}) is also valid for
a weak, \emph{time dependent} $A$, as long as $A$ varies slowly compared with
the energy gap $u,V$ of the bosonic excitations. One way to derive this result is to note that
when $A$ is slowly varying, we can make an adiabatic approximation and can assume 
that the system always remains in the instantaneous low energy subspace spanned
by the fermion states $|n_{r\tau},A_{ss'}\>$. The time evolution is therefore 
described by projecting the Hamiltonian (\ref{HfrtopA}) to the instantaneous low energy subspace,
and we again obtain the low energy effective Hamiltonian $H_{\text{eff}}$ (\ref{EqFermiGauge}).
As for Berry phase effects, one can check that non-abelian Berry connection
$\<n_{r\tau},A_{ss'} | i \partial_A | n_{r'\tau'},A_{ss'}\>$ reduces to a overall $c$-number 
phase factor and hence can be neglected for our purposes. (In other words, Berry phase effects
only contribute a global phase factor to the time evolution of the wave function).

We emphasize, however, that the above low energy effective theory (\ref{EqFermiGauge}) is only valid when 
the flux through each plaquette is small compared to $\Phi_0$. Examining (\ref{EmEs}), we see that when
the flux becomes comparable to $\Phi_0$, the states with $b_P \neq 1$ become energetically favorable. Hence, 
in this regime, we cannot describe the low energy physics in terms of the fermions alone: we also need to keep 
track of the other types of excitations. 

One case where these additional degrees of freedom are particularly relevant -- and which will play an important
role in our analysis in section \ref{SurfaceSect} -- is in flux insertion thought experiments. Imagine 
we take a geometry with open boundary conditions, and we adiabatically increase the flux through a plaquette $P_0$ 
from $\phi_{P_0} = 0$ to $\phi_{P_0} = \Phi_0$, while keeping the flux through all other plaquettes constant at 
$\phi_P = 0$. For simplicity, let us assume that the system is initialized in the state 
$|\tilde{q}_s=0, b_P = 1, n_{s\sigma}=0,A_{ss'}\>$ when $\phi_{P_0} = 0$, and let us neglect the fermion hopping 
term $H_{\text{hop}}$. From (\ref{EmEs}), we can see that there are two level crossings during the flux 
insertion process -- one at $\phi_{P_0} = \Phi_0/4$ and one at $\phi_{P_0} = 3\Phi_{0}/4$. At the first 
level crossing, the state with $b_{P_0} = e^{-2\pi i/m}$ becomes lower in energy than the state with 
$b_{P_0} = 1$, while at the second level crossing, the state with $b_{P_0} = e^{-4\pi i/m}$ becomes lower 
in energy than $b_{P_0} = e^{-2\pi i/m}$. For a finite system size, we expect that these crossings will 
be avoided crossings, at least if we add a generic local perturbation to the Hamiltonian. Therefore, 
if we insert flux sufficiently slowly, the system will follow the lowest energy state at all times, 
finally evolving into a state with $b_{P_0} = e^{-4\pi i/m}$ and $\phi_{P_0} = \Phi_0$. 
One can check using the definition (\ref{eigenabbrev}) that this final state is the same as the initial state. Hence, 
the insertion of a unit flux quantum returns us to our original starting point -- as expected from general
considerations (e.g. the Byers-Yang theorem). 

An important point, however, is that the gap at these avoided level crossings vanishes in the limit that 
the plaquette $P_0$ is far from the edge of the system, since the states 
involved in the level crossings have different values of $b_{P_0}$ and therefore only couple to one 
another via a process where a flux quasiparticle tunnels from the edge to $P_0$. As a result, 
the above picture is only valid at extremely long time scales. 
If we insert the flux at a rate which is fast compared with the gap at the level crossing
(but still slow compared with the bulk gap) the outcome is different. In this
case, the system passes through the level crossings unaffected, leading to a final state with 
$b_{P_0} = 1$ at $\Phi_{P_0} = \Phi_0$. One can check using (\ref{eigenabbrev}) that this final state corresponds to having $2$ 
flux quasiparticles on the plaquette $P_0$. This result shows
that other degrees of freedom beyond the fermions come into play in flux insertion experiments.
In addition, it implies that we need to insert $m$ flux quanta (if $m$ is odd) or $m/2$ flux 
quanta (if $m$ is even) for the system to return to its original configuration. In this sense,
our models have a reduced flux periodicity -- like fractional quantum Hall states. \cite{GefenThouless}

\section{Physical properties of the 2D models} \label{2dprop}
In the previous section, we introduced a Hamiltonian $H$ (\ref{Hfrtop}) whose
low energy physics is exactly described by a non-interacting 2D topological insulator of fractionally
charged fermions. We now describe the physical properties of this fractionalized insulator, and derive
an effective Chern-Simons field theory which summarizes them. 

\subsection{Edge states}
One of the distinguishing features of non-interacting 2D topological insulators is that they have gapless
edge states which are protected by time reversal symmetry and charge conservation. \cite{KaneMele,KaneMele2,HasanKaneRMP} 
Here we discuss the analogous edge states for the 2D fractionalized insulator $H$ (\ref{Hfrtop}).

Fortunately, the exact mapping between the low energy physics of $H$ and the non-interacting 
topological insulator $H_{\text{eff}}$ (\ref{HNI}) holds for any lattice, including geometries with 
an edge. Hence, our problem reduces to understanding the edge structure of the insulator $H_{\text{eff}}$. 
We focus on the specific band structure (\ref{2dtiband}), for simplicity. This insulator has two filled bands: 
a spin-up band with Chern number $\nu = 1$, and a spin-down band with Chern number $\nu = -1$. Therefore, 
by the usual bulk-boundary correspondence, $H_{\text{eff}}$ must have one spin-up edge mode and one spin-down 
edge mode with opposite chiralities. \cite{HasanKaneRMP} Translating this result over to the fractionalized insulator 
$H$, we conclude that $H$ also has a pair of counterpropagating free fermion edge modes. 
The only difference from the non-interacting case is that the low energy fermions carry charge $q_{\text{f}}$ instead of 
charge $e$. The low energy edge Hamiltonian for $H$ is thus of the form
\begin{equation}
H_{\text{edge}} = v d^\dagger_\uparrow (i\partial_x + q_{\text{f}} A_x) d_\uparrow - v d^\dagger_\downarrow (i\partial_x + q_{\text{f}} A_x) 
d_\downarrow
\label{2dedge}
\end{equation}
where $v$ is the velocity of the two edge modes. We leave the discussion of the stability of these edge modes to
section \ref{SurfaceSect}.

\subsection{Topological order} \label{2dtopord}
To fully characterize our model, we must also analyze the topological order of the fractionalized insulator $H$ 
(\ref{Hfrtop}). We first recall the concept of topological order in 2D systems.\cite{WenReview,WenBook} Topologically 
ordered systems in two dimensions have three important physical properties. First, these systems have a finite energy
gap separating the ground state(s) from excited states. Second, when a topologically ordered system
is defined in a torus geometry (periodic boundary conditions in both directions), the
ground state is typically multiply degenerate. This degeneracy is not a consequence of any symmetry and is
robust to arbitrary perturbations.\cite{WenReview,WenBook,Einarsson} Third, and perhaps most importantly, these systems have
quasiparticle excitations with fractional statistics: when one such quasiparticle is braided
around another, it acquires a nonvanishing Berry phase. 

Before analyzing the fractionalized insulator $H$ (\ref{Hfrtop}), it is useful to first understand the topological
order in the lattice boson model $H_B$ (\ref{HBose}). As we discussed earlier, this model contains
two types of elementary excitations: ``charge'' excitations where $q_s = 1$ at some site $s$,
and ``flux'' excitations where $B_P = e^{2\pi i/m}$ on some plaquette $P$.
We will now show that these excitations have nontrivial \emph{mutual} statistics:
when a charge moves around a flux, it acquires a Berry phase of $\theta_{\text{ch,fl}} = 2\pi/m$.

\begin{figure}[t]
\centerline{
\includegraphics[width=0.6\columnwidth]{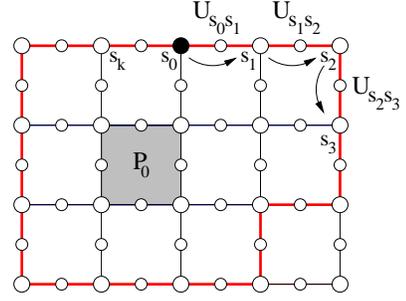}
}
\caption{
We can move a charge at position $s_0$ around a flux at position $P_0$ 
by applying a string of $U_{ss'}$ operators along a closed path $C = s_0s_1...s_ks_0$ 
encircling $P_0$.
}
\label{topordfig}
\end{figure}

Imagine we have a flux on plaquette $P_0$ and a charge on some site $s_0$. We can denote
this state by 
\be
|s_0,P_0\>\equiv |n_{s_0 \sigma} =1, b_{P_0} =1 \> \ \ \ .
\ee
 We need to adiabatically move the charge around the flux and 
compute the resulting Berry phase. To this end, we note that the operator $U_{s_0s_1}$ 
moves the charge from site $s_0$ to a neighboring site $s_1$:
\begin{equation}
U_{s_1s_0} |s_0,P_0\> = |s_1,P_0\>
\end{equation}
Indeed, this follows from the commutation relations between $Q_t, B_P$ and $U_{ss'}$.
Therefore, we can move the charge around the flux by applying a string of $U_{ss'}$ 
operators along some closed path $C = s_0s_1...s_ks_0$ encircling $P_0$ (see Fig. \ref{topordfig}):
\begin{equation}
U_{s_k s_0}...U_{s_1s_2} U_{s_0s_1} |s_0, P_0\>
\end{equation}
To find the accumulated Berry phase, we use an operator identity:
\begin{eqnarray}
U_{s_k s_0}...U_{s_1s_2} U_{s_0s_1} |s_0, P_0\> &=& \prod_{P \in C} B_P |s_0, P_0\> \nonumber \\
&=& e^{2\pi i/m} |s_0, P_0\>
\label{bphase1}
\end{eqnarray}
Here, the second equality follows from the fact that $b_P = 1$ everywhere except for $P = P_0$,
where $b_{P_0} = e^{2\pi i/m}$. 

To isolate the statistical Berry phase from other geometric phases,
we need to compare this phase with the phase accumulated when the flux is \emph{not} enclosed by the 
path $C$. In that case, the same operator identity gives
\begin{equation}
U_{s_k s_0}...U_{s_1s_2} U_{s_0s_1} |s_0, P_0\> = |s_0, P_0\>
\label{bphase2}
\end{equation}
Comparing (\ref{bphase1}),(\ref{bphase2}), we conclude that the statistical Berry phase is precisely
$\theta_{\text{ch,fl}} = 2\pi/m$, as we claimed. 

In a similar way, we can check that there is no Berry phase associated with exchanging 
a pair of charge or flux excitations. \cite{LevinWenHop} In other words, the charge and flux excitations are bosons. 
To complete our analysis of the topological order, we need to find the ground state degeneracy on a torus. 
We describe this calculation in Appendix \ref{degtorus}. There, we show that the lattice boson model has 
exactly $D = m^2$ degenerate ground states in a torus geometry.

The above ground state degeneracy and quasiparticle statistics are identical to
the statistics and ground state degeneracy of the generalized ``toric code'' model with 
gauge group $G = Z_m$. \cite{KitaevToric} Thus, the fractionalized bosonic insulator $H_B$ 
(\ref{HBose}) has the same topological order as the $Z_m$ toric code, or equivalently 2D 
$Z_m$ gauge theory coupled to bosonic matter.

Given these results, we can now easily analyze the topological order of the fractionalized insulator
$H$ (\ref{Hfrtop}). This model contains three types of elementary excitations -- charges, fluxes, and fermions.
The charge and flux excitations are identical to the excitations in the lattice boson model 
and obey the same statistics. On the other hand the fermion excitation is a composite of an electron and 
$k$ ``charge'' excitations, as we argued in equation (\ref{composite}). This decomposition implies that 
the fermion excitation has mutual statistics $\theta_{\text{f,fl}} = 2\pi k/m$ with respect to the flux excitations, 
but no mutual statistics with respect to the charges. As for the ground state degeneracy
on a torus, it is easy to check that $D =m^2$ just as in the lattice boson model.

\subsection{Chern-Simons field theory description} \label{csfield}
In this section, we derive a field theoretical description of the fractionalized insulator
$H$ (\ref{Hfrtop}). This field theory is useful as it captures all of the (universal) properties of the 
fermionic phase--including both the topological order in the bulk, and the edge modes at the boundary 
(see section \ref{2dcsedge}).

We begin by writing down a field theory description of the lattice boson 
model $H_B$ (\ref{HBose}). As we discussed in section \ref{2dtopord}, the topological order in this 
model is the same as $Z_m$ gauge theory coupled to bosonic matter. Previous work\cite{KouLevinWen} has shown that this 
kind of topological order is described by the Chern-Simons theory,
\begin{equation}
L_{B} = \frac{m}{4\pi} \epsilon^{\lambda \mu \nu} (\alpha_\lambda \partial_\mu \beta_\nu + \beta_\lambda \partial_\mu \alpha_\nu)
- \frac{2e}{2\pi} \epsilon^{\lambda \mu \nu} A_\lambda \partial_\mu \beta_\nu
\label{csZp}
\end{equation}
where $A_\lambda$ is the electromagnetic gauge field. (This particular form of $U(1) \times U(1)$ Chern-Simons 
theory is sometimes referred to as 2D BF theory\cite{MooreBF,SondhiSC}). In this description, the boson 
number current is given by
\begin{equation}
j^\lambda_{boson} = \frac{1}{2\pi} \epsilon^{\lambda \mu \nu} \partial_\mu \beta_\nu
\end{equation}
Also, the two types of quasiparticle excitations, ``charges'' $(q_s = 1)$ and ``fluxes'' $(b_p = e^{2\pi i/m})$, are 
described by coupling (\ref{csZp}) to bosonic particles carrying unit $\alpha_\mu$ and $\beta_\mu$ charge respectively.
 
This field theory correctly describes the fractional charges and statistics of the quasiparticle excitations 
in the lattice boson model, as well as the ground state degeneracy on a torus. We can verify this 
using the ``$K$-matrix'' formalism\cite{WenBook,WenReview,WenKmatrix} for abelian Chern-Simons theory and abelian fractional 
quantum Hall states. First we write $L_B$ in $K$-matrix notation:
\begin{equation}
L_{B} = \frac{K_{IJ}}{4\pi} \epsilon^{\lambda \mu \nu} a_{I \lambda} \partial_\mu a_{J \nu} 
- \frac{1}{2\pi} t_I \epsilon^{\lambda \mu \nu} A_{\lambda} \partial_\mu a_{I \nu} 
\label{kmat}
\end{equation}
where
\begin{align}
K_{IJ} = \bpm 0 & m  \\
	      m & 0  \epm \ , \ t_I = \bpm 0 \\ 2e \epm \ , \ a_I = \bpm \alpha \\ \beta \epm 
\end{align}
In this notation, the gauge charge carried by the two types of particles (charges and fluxes) can be 
represented in terms of the vectors
\begin{align}
l_{\text{ch}} = \bpm 1 \\ 0 \epm \ , \ l_{\text{fl}} = \bpm 0 \\ 1 \epm \ 
\end{align}
According to the $K$-matrix formalism, the physical electric charge of each excitation is given by
\begin{equation}
q_l = l^T K^{-1} t
\label{cschg}
\end{equation}
while the mutual statistics associated with braiding one particle around another is given by
\begin{equation}
\theta_{l l'} = 2\pi l^T K^{-1} l'
\label{csstat}
\end{equation}
(The statistical phase associated with exchanging two identical particles is $\theta_l = \theta_{ll}/2$).
Applying this to $l_{\text{ch}}$ and $l_{\text{fl}}$, we see that
\begin{align}
q_{\text{ch}} = \frac{2e}{m} \ , \ q_{\text{fl}} = 0 \ , \ \theta_{\text{ch,fl}} = \frac{2\pi}{m} \ , \ \theta_{\text{ch}} = \theta_{\text{fl}} = 0
\end{align}
in agreement with the properties of the lattice boson model. Furthermore, according to the $K$-matrix
formalism, the ground state degeneracy on a torus is given by
\begin{equation}
D = \abs{det(K)} = m^2
\label{gsd}
\end{equation}
Again, this is in agreement with the properties of the lattice boson model. We conclude that
the Chern-Simons theory (\ref{csZp}) does indeed describe the topological order in the lattice boson model.

To obtain a field theory description of the fractionalized insulator $H$ (\ref{Hfrtop}), we think of this system
as a non-interacting topological insulator built out of fractionally charged fermions. These fermions
are composite particles composed of $k$ ``charge'' excitations of the lattice boson model together with 
one electron. In terms of the field theory (\ref{csZp}) we can describe these composite particles by coupling 
(\ref{csZp}) to fermionic particles, each of which carries $k$ units of $\alpha_\mu$ charge and $e$ units of electric 
charge. We conclude that the Lagrangian for $H$ is of the form
\begin{equation}
L = L_B + L_{\text{TI}}[k \alpha_\mu + e A_\mu]
\label{csfti1}
\end{equation}
where $L_{\text{TI}}[k \alpha_\mu + e A_\mu]$ is the Lagrangian of a non-interacting topological insulator 
coupled to an external gauge field $k \alpha_\mu + e A_\mu$. 

To complete our derivation, we need to construct a field theory $L_{TI}$ for the non-interacting topological
insulator coupled to an external gauge field. We can write down such a field theory explicitly in the case of 
the band structure (\ref{2dtiband}). In this case, the spin-up fermions form a band insulator
with Chern number $\nu = +1$, while the spin-down fermions form an insulator with Chern number $\nu = -1$, so the
appropriate field theory\cite{WenBook} is a sum of two decoupled Chern-Simons theories:
\begin{eqnarray}
L_{\text{TI}}[k \alpha_\mu + e A_\mu] &=& \frac{1}{4\pi} \epsilon^{\lambda \mu \nu} (\gamma_{\uparrow \lambda} 
\partial_\mu \gamma_{\uparrow \nu} 
- \gamma_{\downarrow \lambda} \partial_\mu \gamma_{\downarrow \nu}) \nonumber \\
- \frac{1}{2\pi} \epsilon^{\lambda \mu \nu} (k \alpha_\lambda 
&+& e A_\lambda) \partial_\mu (\gamma_{\uparrow \nu} + \gamma_{\downarrow \nu})
\label{csti}
\end{eqnarray}
In this description, the spin-up and spin-down fermion currents are given by
\begin{eqnarray}
j^\lambda_\uparrow &=& \frac{1}{2\pi} \epsilon^{\lambda \mu \nu} \partial_\mu \gamma_{\uparrow \nu} \\
j^\lambda_\downarrow &=& \frac{1}{2\pi} \epsilon^{\lambda \mu \nu} \partial_\mu \gamma_{\downarrow \nu}
\end{eqnarray} 
while the spin-up and spin-down fermion excitations can be described by coupling (\ref{csti})
to bosonic sources carrying unit $\gamma_{\uparrow \mu}$ and 
$\gamma_{\downarrow \mu}$ charge respectively.

Combining (\ref{csZp}), (\ref{csfti1}), (\ref{csti}), we see that the fractionalized electronic insulator $H$ is described
by the $4$ component Chern-Simons theory
\begin{equation}
L = \frac{K_{IJ}}{4\pi} \epsilon^{\lambda \mu \nu} a_{I \lambda} \partial_\mu a_{J \nu} 
- \frac{1}{2\pi} t_I \epsilon^{\lambda \mu \nu} A_{\lambda} \partial_\mu a_{I \nu} 
\label{cs4}
\end{equation}
where
\begin{align} \label{Kmat2d}
K_{IJ} = \bpm 0 & m & -k & -k \\
	      m & 0 & 0 & 0 \\
	      -k & 0 & 1 & 0 \\
              -k & 0 & 0 & -1 \epm, \ t_I = \bpm 0 \\ 2e \\ e \\ e \epm, \ a_I = \bpm \alpha \\ \beta \\ 
\gamma_\uparrow \\ \gamma_\downarrow \epm 
\end{align}
As in the bosonic lattice model case (\ref{kmat}), this Chern-Simons theory contains all the information about the 
fractional charge and fractional statistics of the three types of excitations, as well as the ground state degeneracy
on the torus. The ``charge'' excitations correspond to particles with unit $a_1$ charge,
the ``flux'' excitations correspond to particles with unit $a_2$ charge, and the spin-up/spin-down fermions
correspond to particles with unit $a_3$ or $a_4$ charge. We represent these particles with the vectors
\begin{align}
l_{\text{ch}} = \bpm 1 \\ 0 \\ 0 \\ 0 \epm, \ l_{\text{fl}} = \bpm 0 \\ 1 \\ 0 \\ 0 \epm, \ l_{\uparrow} = 
\bpm 0 \\ 0 \\ 1 \\ 0 \epm, \
l_{\downarrow} = \bpm 0 \\ 0 \\ 0 \\ -1 \epm
\label{csqp}
\end{align}
Applying the formulas (\ref{cschg}-\ref{csstat}), we can check that the quasiparticle charges and statistics 
are
\begin{align}
q_{\text{ch}} &= \frac{2e}{m} \ , \ q_{\text{fl}} = 0 \ , \ q_{\text{f}} = \frac{e(2k+m)}{m} \\
\theta_{\text{ch,fl}} &= \frac{2\pi}{m} \ , \ \theta_{\text{f,fl}} = \frac{2\pi k}{m} \ , \ \theta_{\text{f}} = \pi \ , 
\ \theta_{\text{ch}} = \theta_{\text{fl}} = 0 \nonumber
\end{align}
in agreement with our previous calculations. Also, the ground state degeneracy on a torus is given by
\begin{equation}
D = \abs{det(K)} = m^2
\end{equation}
as expected.

\section{Lattice models for 3D fractional topological insulators} \label{ModelSect3D}
In this section we construct exactly soluble lattice models for 
3D fractionalized insulators. The 3D construction is a simple generalization of the 2D case, and much of
our analysis carries over without any change. As before, we proceed in three steps,
beginning by constructing lattice boson models with fractional charge. 

\subsection{Step 1: 3D lattice boson models with fractional charge}
\begin{figure}[t]
\centerline{
\includegraphics[width=0.7\columnwidth]{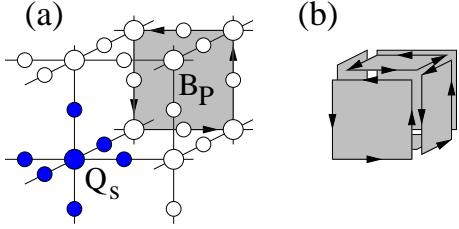}
}
\caption{
(a) The lattice boson model defined on the cubic lattice. As in the square lattice case,
the Hamiltonian $H_B$ (\ref{HBose3d}) is a sum of $Q_s$ operators and 
$B_P$ operators, but now the $Q_s$ operators act on six links $\<ss'\>$ and one
site $s$, and the $B_P$ operators act on the plaquettes of the cubic lattice.
(b) The $B_P$ operators obey the identity $\prod_{P \in C} B_P = 1$ where
the product runs over the six plaquettes $P$ adjacent to a ``cube'' $C$, and
where we choose appropriate orientations on these plaquettes. 
}
\label{cublatfig}
\end{figure}

The 3D lattice boson models are identical to the 2D models (\ref{HBose}),
except that we consider these systems on the cubic lattice instead of the square lattice. 
In other words, the Hamiltonian is still
\begin{equation} 
H_B = V \sum_s Q_s^2 - \frac{u}{2} \sum_P (B_P + B_P^\dagger)
\label{HBose3d}
\end{equation}
but now the sums over $s$ and $P$ run over the sites $s$ and plaquettes $P$ of the cubic lattice
(Fig. \ref{cublatfig}(a)).

This model is exactly soluble just as in the 2D case, since $Q_s, B_P, B_P^\dagger$
all commute with one another. Again, we can choose simultaneous eigenstates $|q_s, b_P\>$ where
$q_s$ is an integer and $b_P$ is a $m$th root of unity, and these states have
energies
\begin{equation}
E = V \sum_s q_s^2 - \frac{u}{2} \sum_P (b_P + b_P^*)
\end{equation}
Furthermore, as before, we can find the degeneracy of the $q_s, b_P$ eigenspaces for particular system geometries 
-- say a rectangular slab of cubic lattice with open boundary conditions.
The only difference from the 2D case is that the $B_P$ operators satisfy the identity
\begin{equation}
\prod_{P \in C} B_P = 1
\end{equation}
where the product runs over the six plaquettes $P$ adjacent to a ``cube'' $C$ in the cubic
lattice, and where we choose appropriate orientations on these plaquettes (Fig. \ref{cublatfig}(b)). Therefore, the
$b_P$ quantum numbers obey the local constraint
\begin{equation}
\prod_{P \in C} b_P = 1
\label{bpconstr}
\end{equation}
for every ``cube'' $C$ in the cubic lattice, in addition to the global constraint
\begin{equation}
\sum_s q_s \equiv 0 \text{ (mod m)}
\end{equation}
We can check that for each set of $\{q_s, b_P\}$ satisfying these constraints, there is a unique 
eigenstate $|q_s, b_P\>$ (see Appendix \ref{degopen3}).

As in the square lattice case, the ground state of the 3D cubic lattice model (with open boundary conditions) is the 
unique state with $q_s = 0, b_P = 1$,
and there are two types of elementary excitations. The first kind of excitation are ``charges''
where $q_s = 1$ at some site $s$. These excitations are very similar to the charge excitations in the 2D model.
In particular, they carry the same fractional charge $2e/m$, as can be verified using the arguments in section 
\ref{bosefrch}. On the other hand, the second kind of excitation is slightly different from the 2D case: instead
of particle-like ``flux'' excitations, the 3D model has ``flux loop'' excitations where $b_P = e^{2\pi i/m}$
along some closed loop in the dual cubic lattice. The reason that these excitations are loop-like in this case
is the constraint (\ref{bpconstr}) which ensures that the flux is divergence free and hence must form closed loops.

\subsection{Step 2: 3D lattice electron models with fractional charge}
The next step, as in the 2D case, is to modify the 3D boson model (\ref{HBose3d}) by including additional spin-$1/2$
electron degrees of freedom at each site $s$. 
The resulting 3D electron model is identical to the square lattice case (\ref{HFermi}).
The Hamiltonian is still
\begin{equation} \label{HFermi3d}
H_e = V \sum_s \tilde{Q}_s^2 - \frac{u}{2} \sum_P (B_P + B_P^\dagger) -\mu \sum_{s \sigma} n_{s\sigma}
\end{equation}
except now the sums over $s$, $P$ run over sites and plaquettes of the cubic lattice.

As in the 2D case, we can choose simultaneous eigenstates $|\tilde{q}_s, b_P, n_{s\sigma}\>$ with 
corresponding energies
\begin{equation}
E = V \sum_s \tilde{q}_s^2 - \frac{u}{2} \sum_P (b_P + b_P^*) - \mu \sum_{s \sigma} n_{s \sigma}
\end{equation}
There is a unique state for each choice of $\tilde{q}_s, b_P, n_{s\sigma}$ satisfying the two constraints
\begin{align}
\prod_{P \in C} b_P = 1 \ , \ \sum_s \tilde{q}_s + k\sum_{s\sigma} n_{s\sigma} \equiv 0 \ \ \text{(mod m)}
\end{align}
(assuming a geometry with open boundary conditions).

Assuming $\mu < 0$ for simplicity, the ground state is the unique state with $\tilde{q}_s = 0, b_P = 1, n_{s\sigma} = 0$.
There are three types of elementary excitations, just as in the 2D case: ``charge'' excitations with $\tilde{q}_s = 1$,
``flux loop'' excitations with $b_P = e^{2\pi i/m}$ along some loop in the dual lattice, and spin-$1/2$ ``fermion'' 
excitations where $n_{s\sigma} = 1$. The spin-$1/2$ fermion excitations are similar to the fermion excitations
in the 2D model, and in particular they carry the same fractional charge $q_{\text{f}} = e(1 + 2k/m)$ (\ref{frcharge}), 
and transform under time reversal in the same way.

\subsection{Step 3: Building candidate 3D fractional topological insulators}
The final step is to modify the lattice electron model (\ref{HFermi3d}) so that the fractionally charged fermions
form a non-interacting topological insulator. As in the 2D case, we accomplish this
by adding a new term, $H_{\text{hop}}$ to the Hamiltonian:
\begin{equation}
H = H_e + H_{\text{hop}}
\label{Hfrtop3d}
\end{equation}
where $H_{\text{hop}}$ is defined just as in the 2D case (\ref{Hhop}):
\begin{equation}
H_{\text{hop}} = -\sum_{\<ss'\>} (t_{ss' \sigma \sigma'} c_{s' \sigma'}^\dagger c_{s \sigma} U^{k}_{ss'} + h.c.)
\label{HNI3d}
\end{equation}
Again, we assume that $t_{ss' \sigma \sigma'} \ll u,V$ so that the bandwidth of the fermion excitations is
small compared with the gap to the bosonic excitations. Just as in the 2D case, we can show that the low energy physics 
of $H$ is described by the non-interacting fermion model
\begin{equation} 
H_{\text{eff}} = -\sum_{\< ss'\> } (t_{ss' \sigma \sigma'} d_{s' \sigma'}^\dagger d_{s \sigma} + h.c.) 
- \mu \sum_{s\sigma} d_{s\sigma}^\dagger d_{s\sigma} 
\label{HNIA3d}
\end{equation}

We can then complete the construction, just as in the 2D case: we simply choose the hopping amplitudes 
$t_{ss' \sigma \sigma'}$ and chemical potential $\mu$ so that $H_{\text{eff}}$ is a non-interacting topological 
insulator. The low energy physics is then described by a topological insulator built out of \emph{fractionally} charged
fermions. There are many possible choices for $t_{ss' \sigma \sigma'}$, 
but to be concrete, we consider a $2$-orbital model on the cubic lattice, introduced by 
Ref. \onlinecite{RosenbergFranz}. The $4$ fermions on each site can be expressed as a single 
vector $(d_{s1\uparrow}, \ d_{s1\downarrow}, \ d_{s2\uparrow}, \ d_{s2\downarrow})$. 
In this basis the hopping matrix 
elements can be expressed as a tensor product of matrices $\tau_i$ acting on the orbital indices $(1,2)$ 
and $\sigma_i$ acting on the spin indices $(\uparrow, \downarrow)$.  We take 
\ba \label{Hkin}
t_{ss' \sigma \sigma'} &=& \sum_{\hat{e}_i=  \hat{x}, \hat{y}, \hat{z} } \delta_{s-s', \hat{e}_i}   
\left[  i \lambda  \left( \tau_z \otimes \sigma_i \right )_{\sigma \sigma'} 
+ t \left(\tau_x \otimes \mb{1} \right )_{\sigma \sigma'} \right .   \n
&&  \left. + h.c.  \right] + m  \delta_{s-s', 0} \left(\tau_x \otimes \mb{1} \right )_{\sigma \sigma'}  \ \ \ 
\ea
where $\sigma, \sigma' \in \{ (1,\uparrow), (2,\uparrow), (1,\downarrow), (2,\downarrow) \}$ specify
both the spin and orbital indices.  The first term is a spin-orbit coupling whose sign differs for the 
two orbitals; the second term is a nearest-neighbor hopping from one orbital to the other, and the third 
term is an on-site hopping between the two orbitals.  Taking $\mu=0$, together with $t>0$ and $\lambda>0$, this yields a 
topologically insulating band structure, provided that
$2 t < | m| < 6t$\cite{RosenbergFranz}.  

\section{Physical properties of the 3D models} \label{PhysPropSect}
In the previous section, we introduced an electron Hamiltonian $H$ (\ref{Hfrtop3d}) whose 
low energy physics is exactly described by a non-interacting topological insulator of fractionally
charged fermions. We now describe the physical properties of the resulting fractionalized 
insulator.  

\subsection{Surface states}
One of the most important properties of 3D non-interacting topological insulators is that they have gapless
surface states which are protected by time reversal symmetry and charge conservation. \cite{HasanKaneRMP}
In this section, we discuss the analogue of these surface states for the fractionalized insulator realized by $H$ 
(\ref{Hfrtop3d}).

To begin, we review the structure of the surface states of the non-interacting topological insulator with
band structure (\ref{Hkin}). To be concrete, we consider this model in an $L_x \times L_y \times L_z$ slab 
of cubic lattice which is periodic in the $x,y$ directions and has open boundary conditions at $z=0$ and 
$z = L_z$. (This geometry is sometimes referred to as a ``Corbino donut''). This system is 
translationally invariant in the $x,y$ directions, so we can still define
a 2D band structure as a function of $k_x, k_y$. It has been shown that the surface band structure for the 
tight binding model (\ref{Hkin}) has two Dirac points---one for each surface. The surface states have a 
particularly simple structure if the chemical potential is tuned near these Dirac points. In that case, 
the low energy surface states on each surface are described by the Dirac Hamiltonian
\begin{equation}
H^{\text{TI}}_{\text{surf}} = v c^\dagger_{\alpha}  \left [  \v{\sigma}_{\alpha\beta} \cdot ( i  \v{\nabla} + e \v{A})  \right ] 
c_{ \beta} - \mu c^\dagger_{ \alpha} c_{ \alpha}  
\label{nonintsurf}
\end{equation}
where $\alpha, \beta = \uparrow, \downarrow$, and $\v{\sigma} = (\sigma_x, \sigma_y, \sigma_z)$. 

The surface states for the fractionalized insulator $H$ (\ref{Hfrtop3d}) are almost identical 
to the non-interacting case, since the low energy physics of $H$ is exactly described by 
the free fermion model $H_{\text{eff}}$. The only difference is that the 
fermions excitations carry fractional charge $q_{\text{f}}$. Therefore, the low energy surface Hamiltonian for $H$
near the Dirac points is given by exactly (\ref{nonintsurf}), except with $e$ replaced by $q_{\text{f}}$ (and
$c$ replaced by $d$):
\begin{equation}
H_{\text{surf}} = v d^\dagger_{\alpha}  \left [  \v{\sigma}_{\alpha\beta} \cdot (i \v{\nabla} + q_{\text{f}} \v{A}) \right ] d_{\beta} - 
\mu d^\dagger_{\alpha} d_{\alpha}  
\label{frtopsurf}
\end{equation}
We leave the discussion of the stability of these surface modes to section \ref{SurfaceSect}.

\subsection{Surface quantum Hall effect}
\label{surfqhe}
The presence of a single Dirac cone in the surface mode spectrum has an interesting consequence if 
we break time reversal on this surface, for example by attaching to it a thin magnetic film.  Let us first
review what this does to a non-interacting topological insulator with band structure (\ref{Hkin}). Assuming that 
the magnetic film is only weakly coupled to topological insulator, we can model its effect by including a 
Zeeman term $H_Z = B d^\dagger_\alpha \sigma^z_{\alpha \beta} d_\beta$ in the surface Dirac Hamiltonian 
(\ref{nonintsurf}). This term opens up a gap in the Dirac spectrum of size $2|B|$. If the chemical potential 
lies within this gap, then the surface states are completely gapped and the system exhibits a half-integer 
surface Hall conductivity\cite{FuKaneHall,HasanKaneRMP}
\begin{equation}
\sigma^{\text{TI}}_{xy} = \frac{e^2}{2h}
\end{equation}
The total Hall conductivity on the two surfaces is $e^2/h$ or $0$ depending on whether the magnetic field $B$
has the same or opposite signs at $z = 0$ and $z = L_z$.

The behavior of the fractionalized insulator $H$ (\ref{Hfrtop3d}) is similar. As in the 
non-interacting case, the magnetic film introduces a Zeeman term, thereby gapping the spectrum of the surface Dirac
Hamiltonian (\ref{frtopsurf}). If the chemical potential lies within this gap, then the system
exhibits a quantized surface Hall conductivity. The only difference is that the fermionic excitations carry
charge $q_{\text{f}}$ so the Hall conductivity is a half-integer in units of $q_{\text{f}}^2/h$:
\begin{equation}\label{sigxy}
\sigma_{xy} = \frac{q_{\text{f}}^2}{2h} 
\end{equation}

The total Hall conductivity on both surfaces is $q_{\text{f}}^2/h$ or $0$ depending 
on whether the magnetic field $B$ has the same or opposite signs at $z=0$ and $z=L_z$.
The first case is particularly interesting, since in this case the total Hall conductivity is 
fractional. Hence, if we view our geometry as a quasi-2D system, taking the thermodynamic limit 
$L_x , L_y \rightarrow \infty$ while keeping $L_z$ finite, then $H$ realizes a fractional 
quantum Hall state. The properties of this state can be derived using the same approach as
section \ref{csfield}.

\subsection{Charge acquired by a magnetic monopole} \label{monopolech}
Another important property of non-interacting topological insulators is their response 
in the presence of a magnetic monopole. Imagine threading a microscopic solenoid into
a topological insulator which is so small that it fits within a single plaquette of
the lattice. If we insert a full flux quantum $\Phi_0 =hc/e$, we can effectively
construct a monopole/anti-monopole pair at the two ends of the solenoid. An interesting property of
non-interacting topological insulators is that this process causes a half-integer charge
\begin{equation}
q_{\text{TI}} = (n+1/2) e
\end{equation}
to be localized at each end of the solenoid\cite{WittenEffect,Wilczek,RosenbergFranz}. In other words, a magnetic 
monopole (or anti-monopole) in a topological insulator carries half-integer charge. (Here the 
integer $n$ depends on the microscopic form of the Hamiltonian near the ends of the solenoid. As such, it can
be varied without encountering a phase transition in the bulk).

We consider the analogous question for the fractionalized insulators (\ref{Hfrtop3d}). In this case, we have to
be careful because the Dirac quantization condition for monopoles is modified due to the presence
of fractionally charged excitations. Rather than requiring monopoles to carry flux which is an integer multiple of
$\Phi_0$, the Dirac argument implies that monopoles carry flux which is a multiple of
$(e/e^*)\Phi_0$ where $e^*$ is the smallest charged quasiparticle excitation in the system. In our case,
the minimal charge $e^*$ is given by
\begin{equation}
e^* = \begin{cases} \frac{e}{m} & \mbox{if m is odd}\\
\frac{e}{m/2} & \mbox{if m is even}
\end{cases}
\label{mincharge}
\end{equation}
When $m$ is even, the bosonic charge excitations carry the minimal charge $e^*$; when 
$m$ is odd, an excitation with minimal charge $e^*$ can be constructed by forming a composite of an electron and
$(1-m)/2$ charge excitations.  

If one violates the Dirac condition and inserts a monopole/anti-monopole pair whose flux 
is not a multiple of $(e/e^*)\Phi_0$, the result is that the monopole is joined by a physically 
observable Dirac string to the anti-monopole partner. One way to see this is to calculate the ground state
energy in the presence of the monopole/anti-monopole pair using the 3D analogue of the eigenvalue
spectrum (\ref{EmEs}). Examining the $b_P$ term, we can see that a Dirac string with flux $j\Phi_0$ will have an energy
cost $\sim u L (1-\cos(4j\pi/m))$ proportional to its length $L$ unless $2j/m$ is an integer -- that is, unless $j$ is a multiple
of $e/e^*$ (\ref{mincharge}). (We cannot eliminate this energy cost by changing the value of 
$b_P$ along these plaquettes due to the constraint $\prod_{P \in C} b_P$.)  Intuitively, the point is that, 
unless this condition is satisfied, a charge $e^*$ quasiparticle excitation will acquire a nonvanishing Berry phase 
when it travels around the Dirac string. 

Hence, the natural quantity in the fractional case is the electric charge carried by a flux $(e/e^*)\phi_0$ monopole. 
To compute this charge, we again use the fact that the low energy physics of $H$ (\ref{Hfrtop3d}) is described by the 
free fermion model (\ref{HNIA3d}). Because the fermions carry charge $q_{\text{f}}$, this system behaves exactly like a 
non-interacting topological insulator with a flux $(q_{\text{f}}/e^*)\phi_0$ monopole. In this case, we know that 
$(n + 1/2)q_{\text{f}}/e^*$ fermions are localized near the monopole, as we discussed above. We conclude that the flux 
$(e/e^*)\phi_0$ monopole in the fractionalized insulator $H$ carries electric charge 
\begin{equation}
q = (n +1/2)q_{\text{f}}^2/e^*
\label{frmon}
\end{equation}

On the other hand, we could presumably trap any number of additional charge $e^*$ quasiparticles near the monopole, 
by adding an appropriate local potential. Therefore, the only quantity which is independent of microscopic details 
is the charge of the monopole \emph{modulo} $e^*$. This quantity is given by
\begin{equation}
q = \begin{cases} \frac{e^*}{2} \text{ (mod $e^*$)} & \mbox{if $q_{\text{f}}/e^*$ is odd}\\
0 \text{ (mod $e^*$)} & \mbox{if $q_{\text{f}}/e^*$ is even} 
\end{cases}
\label{frmonfinal}
\end{equation}
More explicitly, if we use the expressions for $q_{\text{f}}$ (\ref{frcharge}) and $e^*$ (\ref{mincharge}), we find that
$q_{\text{f}}/e^*$ is even if and only if $2k+m$ is divisible by $4$.

\subsection{Topological order} \label{3dtopord}
In the previous two sections, we showed that the 3D model (\ref{Hfrtop3d}) exhibits
a fractional surface Hall conductivity (\ref{sigxy}) and a fractional charge bound to a magnetic
monopole (\ref{frmonfinal}). These phenomena are manifestations of a \emph{fractional} magnetoelectric 
effect\cite{QiHughesZhang,MooreEssin}. On the other hand, it was argued in Ref. \onlinecite{Swingle}
that the only way a fractional magnetoelectric effect can be consistent with time reversal symmetry is if the ground state
on a three dimensional torus is multiply degenerate. In this section, we show that
our model does indeed have multiply degenerate ground states on a 3D torus. This ground state degeneracy originates
from the fact that the model is topologically ordered, as we now discuss.

We begin by analyzing the topological order of the 3D lattice boson model (\ref{HBose3d}).
To derive the quasiparticle statistics in this model, we recall that (\ref{HBose3d}) contains 
two types of elementary excitations: point-like ``charge'' excitations where $q_s = 1$ at some site $s$, and ``flux-loop'' excitations where 
$B_p = e^{2\pi i/m}$ along some loop in the dual lattice. Using the same approach as in the 2D
case (section \ref{2dtopord}), one can show that when one braids a charge around a flux-loop, it acquires a Berry phase of 
$e^{\pm 2\pi i/m}$ where the sign depends on the orientation of the braiding trajectory. We can 
think of this as the 3D analogue of mutual statistics. It is also straightforward 
to show that there is no Berry phase associated with exchanging two charges, so the ``charge'' excitations 
are bosons. 

In addition to nontrivial statistics, the lattice boson model (\ref{HBose3d}) also displays the 
second signature of topological order, namely ground state degeneracy. In particular, if we define 
this model in a 3D torus geometry, we find $m^3$ degenerate ground states (see Appendix 
\ref{degtorus3}). 

The quasiparticle statistics and ground state degeneracy of (\ref{HBose3d}) is identical to the
the statistics and ground state degeneracy of the 3D toric code \cite{KitaevToric,HammaZanardiWen} 
with gauge group $G = Z_m$. Thus, the lattice boson model (\ref{HBose3d}) has the same topological order as the $Z_m$
toric code in three dimensions, or equivalently 3D $Z_m$ gauge theory coupled
to bosonic matter.

Given these results, it is easy to analyze the topological order in the fractionalized insulator 
$H$ (\ref{Hfrtop3d}). This model contains three types of elementary excitations -- charges, flux loops, and fermions.
The charge and flux loop excitations are identical to the excitations in the lattice boson model 
and obey the same statistics. On the other hand the fermion excitation is a composite of an electron and 
$k$ ``charge'' excitations, just as in the 2D case (\ref{composite}). This decomposition implies that 
the fermion excitation has mutual statistics $e^{2\pi ik/m}$ with respect to the flux loop excitations. 
As for the ground state degeneracy on a 3D torus, we can show that this is $m^3$, just as for the
lattice boson model.

\section{Gapless boundary modes: protected or unprotected?} \label{SurfaceSect}
In the previous sections we have constructed fractionalized insulators (\ref{Hfrtop},\ref{Hfrtop3d}) whose low energy physics
is equivalent to a non-interacting topological insulator built out of fractionally charged fermions.
We have also seen that these models have gapless boundary modes. However these two properties are
not by themselves sufficient to declare these models ``fractional topological insulators'': we
must also show that the boundary modes are \emph{protected}. That is, we must show that the modes
are robust to arbitrary time reversal invariant, charge conserving perturbations.

In this section we investigate this robustness, and find two key results.
First, we show that in both $2$ and $3$ dimensions, the models for which the ratio $q_{\text{f}}/e^*$ is odd
have protected boundary modes. Thus, these models are indeed authentic ``fractional topological insulators.'' 
Second -- and perhaps more unexpectedly -- we show that in the 2D case, the models with even $q_{\text{f}}/e^*$ do 
{\it not} have protected edge modes, and are thus ``fractional trivial insulators.'' We do not 
have an analogous result in the 3D case. That is, we do not establish either the stability or instability of the surface modes 
when $q_{\text{f}}/e^*$ is even.

In the 2D case, our analysis of edge modes closely follows the arguments of Ref. \onlinecite{LevinStern}. 
Our conclusion is also very similar to Ref. \onlinecite{LevinStern}: in that paper, the authors found that 
a certain class of $s^z$ conserving models have protected edge modes if and only if $\sigma_{sH}/e^*$ is odd, 
where $\sigma_{sH}$ is the spin-Hall conductivity in units of $e/2\pi$, and $e^*$ is the elementary charge 
in units of $e$. Here, we find that our exactly soluble models have protected edge modes if and only if $q_{\text{f}}/e^*$ is
odd. The two criteria agree exactly, since as one can easily check, $q_{\text{f}}/e^* = \sigma_{sH}/e^*$ (\ref{spinHall}) 
for the models discussed here.  

\subsection{2D flux insertion argument} \label{2dfluxarg}
In this section we present a general argument that shows that when $q_{\text{f}}/e^*$ is odd, the 
2D fractionalized insulator $H$ (\ref{Hfrtop}) has protected gapless edge modes.
The argument is very similar to the one given in Ref. \onlinecite{LevinStern}, which is in
turn a generalization of the flux insertion argument of Ref. \onlinecite{FuKanepump}.

To be precise, we prove the following statement: we consider the exactly soluble model $H$ in 
a cylindrical geometry with an even number of electrons and zero flux through
the cylinder. We assume that the ground state is time reversal invariant and 
does not have a Kramers degeneracy on either of the two edges. (We discuss the meaning of this
Kramers degeneracy assumption in section \ref{niflux} below). 
Given these assumptions, we show that if $q_{\text{f}}/e^*$ is odd, then there is always at least one
excited state whose energy gap vanishes in the thermodynamic limit. Furthermore,
this excited state is robust if we add an arbitrary time reversal invariant, 
charge conserving, local perturbation to the Hamiltonian, as long as we do not close the bulk gap. 
Finally, this excited state has the important property that it is in the same ``topological 
sector'' as the ground state, as we explain below. We interpret this low lying excited state 
as evidence for a protected gapless edge mode -- that is, a mode which cannot be gapped out 
without breaking time reversal symmetry or charge conservation, explicitly or spontaneously.

\subsubsection{Non-interacting case} \label{niflux}
\begin{figure}[t]
\centerline{
\includegraphics[width=0.7\columnwidth]{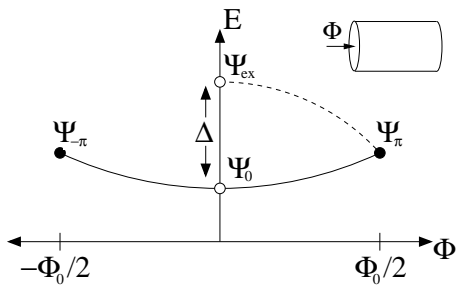}
}
\caption{
The flux insertion argument in the 2D non-interacting case: we start in the ground state $\Psi_0$ and adiabatically 
insert $\pm \Phi_0/2$ flux through the cylinder, 
obtaining two states $\Psi_{\pm \pi}$. The state $\Psi_{\pi}$ has a Kramers degeneracy
at the two ends of the cylinder, and is thus degenerate in energy with three other states, one of which is
$\Psi_{-\pi}$. If we start in one of these three degenerate states and then adiabatically reduce the flux to $0$, 
we obtain an excited state $\Psi_{ex}$ whose energy gap vanishes in the thermodynamic limit. 
}
\label{adiabatic}
\end{figure}

We first review the argument for the case of non-interacting topological
insulators. \cite{FuKanepump} The proof that the system always has a low lying state begins as follows: 
we consider the system on a cylinder, and start in the many-body ground state $\Psi_0$ at zero flux. We 
imagine adiabatically inserting $\Phi_0/2 = hc/2e$ flux through the cylinder. We call the resulting state $\Psi_\pi$. Similarly,
we let $\Psi_{-\pi}$ be the state obtained by adiabatically inserting $-\Phi_0/2$ flux
(See Fig. \ref{adiabatic}). We note that $\Psi_\pi$ and $\Psi_{-\pi}$ are time reversed partners. 

To proceed further, we recall that 2D topological
insulators are characterized by a $Z_2$ invariant $\nu = -1$. This value of the 
invariant means that $\Psi_\pi$ has a Kramers degeneracy at the two ends of
the cylinder if and only if $\Psi_0$ does not. \cite{FuKanepump} 
(At an intuitive level, when we say that a many-body state has a Kramers degeneracy at the
two ends of the cylinder, we mean that it can be brought to a state which is time reversal invariant by removing an
odd number of electrons from each of the two ends. Assuming the ends are far apart, we can treat them as separate systems, each of
which must have a two-fold degeneracy according to Kramers theorem. A precise definition of this
notion of ``local Kramers degeneracy'' is given in Ref. \onlinecite{LevinSternprep}). By our assumption above,
$\Psi_0$ has no Kramers degeneracy so $\Psi_\pi$ must have a Kramers degeneracy at the two 
ends of the cylinder. Hence, as long as the two ends are well separated, Kramers theorem
guarantees that $\Psi_\pi$ is part of a multiplet of four states (two associated with each end) which are nearly 
degenerate in energy. More precisely, these four states are separated by an energy splitting which vanishes exponentially as
the distance between the two ends grows. We note that $\Psi_{-\pi}$ is one of these degenerate 
states, being the time reversed partner of $\Psi_\pi$.

Next, we imagine starting with the system at $\Phi_0/2$ flux
and then adiabatically decreasing the flux to $0$. 
If we start with the state $\Psi_\pi$, then adiabatic flux removal takes us to the ground state $\Psi_0$. 
However, if we start with $\Psi_{-\pi}$ (or one of the other two degenerate states), the result 
is a eigenstate $\Psi_{ex}$ of the zero flux Hamiltonian, which is \emph{distinct} from $\Psi_0$ 
(see Fig. \ref{adiabatic}). At the same time, it necessarily has low energy since the energy change $\Delta E$ 
associated with an adiabatic insertion of flux through a cylinder must vanish in the 
thermodynamic limit as long as charge conservation is not broken (in a gapless system 
with linear dispersion, we expect a scaling like $\Delta E \sim 1/L$, while in a gapped 
system, $\Delta E \sim e^{-const. \cdot L}$). 

To complete the argument, we now imagine adding an arbitrary time reversal invariant,
charge conserving, local perturbation to the system (for example, we could add short-ranged interactions between 
electrons). As long as the perturbation does not close the bulk gap, the above
picture must stay the same: the Kramers degeneracy between $\Psi_\pi$ and $\Psi_{-\pi}$ 
must remain intact, and hence $\Psi_{ex}$ must continue to be low in energy. We conclude
that the system always contains at least one low-lying excited state $\Psi_{ex}$ whose energy gap vanishes
in the thermodynamic limit. 

\subsubsection{Fractionalized case}
We now explain the analogous argument for our fractionalized insulator $H$ (\ref{Hfrtop}). The crucial difference
from the non-interacting case is that this model has excitations with fractional charge $e^*$. As a result,
we need to modify the flux insertion argument, inserting $\pm \frac{e}{e^*} \cdot \frac{\Phi_0}{2}$ flux 
through the cylinder instead of $\pm \frac{\Phi_0}{2}$ flux. \cite{LevinStern}

To understand why this is so, imagine we insert $\pm \Phi_0/2$ flux instead. Then the resulting states 
$\Psi_\pi, \Psi_{-\pi}$ differ by the insertion of a single flux quantum. But the insertion of a single flux quantum
changes the Berry phase $\theta$ associated with moving a charge $e^*$ particle around the cylinder by
$\Delta \theta = 2\pi \frac{e^*}{e}$. As $\Delta \theta$ is not a multiple of $2\pi$, $\Psi_\pi$ and $\Psi_{-\pi}$ 
are in different ``topological sectors.'' That is, to get from one state to the other, we need to
transfer a fractionalized excitation (in this case, two flux quasiparticles -- see section \ref{emeffect}) 
from one end of the cylinder to the other. \cite{LevinStern,GefenThouless} If we then try to construct a low lying excited 
state at zero flux in the usual way (i.e. by starting from $\Psi_{-\pi}$ and adiabatically reducing the flux 
to zero), the resulting state $\Psi_{ex}$ is in a different topological sector from $\Psi_0$. Unfortunately, 
this property means that we cannot conclude much from the existence of $\Psi_{ex}$. The reason is that 
topologically ordered systems like $H$ \emph{always} have a finite number of low lying states that lie in different
topological sectors. \cite{WenReview,WenBook,Einarsson} These states have nothing to do with time reversal symmetry and continue 
to exist even if time reversal symmetry is broken and the edge is gapped. Instead, these states are ``topologically protected'' 
and are analogous to the degenerate ground states in a torus geometry. Therefore, it is important that we construct a 
low-lying state \emph{in the same topological sector} as the ground state. This will establish the existence of 
an ``unexpected'' low lying state, which can plausibly be taken as evidence for a time reversal protected gapless 
edge mode. 

For this reason, we insert $\pm \frac{N \Phi_0}{2}$ flux through the cylinder 
where $N = e/e^*$: when we insert this amount of flux, the states $\Psi_{N\pi}, \Psi_{-N\pi}$
lie in the same topological sector, so we do not run into the issue discussed above. Other 
than this simple modification, the argument proceeds as before: we note
that the low energy physics of $H$ is described by a topological insulator
built out of charge $q_{\text{f}}$ fermions, so our insertion of $\pm \frac{e}{e^*} \cdot \Phi_0/2$ flux is equivalent to
inserting $\pm \frac{q_{\text{f}}}{e^*} \cdot \frac{\Phi_0}{2}$ flux for a non-interacting topological
insulator. If $q_{\text{f}}/e^*$ is odd, then this process changes the time reversal
polarization at the two ends of the cylinder, implying that $\Psi_{N\pi}, \Psi_{-N\pi}$ are Kramers degenerate.
As before, we can then construct a protected low lying state $\Psi_{ex}$ at zero flux by starting 
with $\Psi_{-N\pi}$ and adiabatically inserting $-\frac{N \Phi_0}{2}$ flux quanta. Furthermore, $\Psi_{ex}$
is in the same topological sector as the ground state $\Psi_0$ since $\Psi_{N\pi},\Psi_{-N\pi}$ are in
the same topological sector. We conclude that when $q_{\text{f}}/e^*$ is odd, the fractionalized
insulator $H$ has a protected low lying excited state in the same topological sector as the ground state.
This is what we wanted to show.

\subsection{3D flux insertion argument}
Next, we show that a similar flux insertion argument can be used to establish the 
existence of protected low-lying surface states in the 3D models (\ref{Hfrtop3d}) with $q_{\text{f}}/e^*$ odd. The 
precise statement we will prove is this: we consider the 3D models in a ``Corbino donut'' geometry, periodic
in the $x$ and $y$ directions, and with open boundary conditions in the $z$ direction (see Fig. \ref{3Dflux}a). We
assume that the number of electrons is even and that the ground state is time reversal invariant
when there is zero flux through the two holes of the Corbino donut. In particular, we assume that
the ground state does not have a Kramers degeneracy on the $z = 0$, $z =L_z$ surfaces. 
We then show that if $q_{\text{f}}/e^*$ is odd, then there is always at least one excited state whose
energy gap vanishes in the thermodynamic limit. Furthermore, this excited state is robust
if we add an arbitrary time reversal invariant, charge conserving, local perturbation to the Hamiltonian,
as long as we do not close the bulk gap. Finally, this excited state is in the same topological 
sector as the ground state. Just as in the 2D case, we interpret this low lying excited state as evidence for
a protected gapless surface mode--that is, a mode which cannot be gapped out without breaking 
time reversal symmetry or charge conservation, explicitly or spontaneously.

\subsubsection{Non-interacting case}
It is useful to first give the argument in the case of the non-interacting 3D topological
insulator. (This argument was alluded to in Ref. \onlinecite{FuKaneHall}, though not explicitly discussed). 
The proof that the system has a protected low lying state begins
similarly to the 2D case: we start in the many-body ground state at zero flux, which we call $\Psi_{(0,0)}$, and then 
imagine adiabatically inserting $\Phi_0/2$ flux through each of the two holes of the Corbino donut. In this way, we obtain
three new states $\Psi_{(\pi,0}), \Psi_{(0,\pi)}, \Psi_{(\pi,\pi)}$. 

\begin{figure}[t]
\centerline{
\includegraphics[width=0.7\columnwidth]{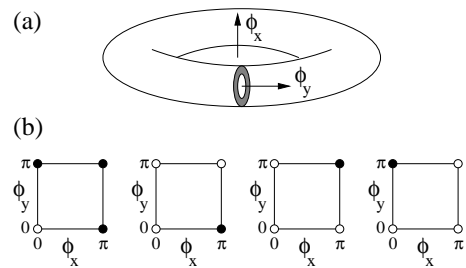}
}
\caption{(a) The 3D flux insertion argument makes use of a ``Corbino donut'' geometry, with flux $\phi_x, \phi_y$ 
through the two holes of the Corbino donut. (b) In the non-interacting case, we start in the ground state 
$\Psi_{(0,0)}$ and adiabatically insert $\Phi_0/2$ flux through each of the two holes, obtaining $3$ other 
states $\Psi_{(\pi,0)}, \Psi_{(0,\pi)}, \Psi_{(\pi,\pi)}$. An odd number of these states must have Kramers 
degeneracies at the two surfaces of the Corbino donut, leading to four possible degeneracy patterns (filled 
circles represent states with Kramers degeneracies). In particular, at least one state has a Kramers degeneracy, 
and hence can be used to construct a protected low lying excited state $\Psi_{ex}$, as in the 2D case.
}
\label{3Dflux}
\end{figure}

We next recall that 3D topological insulators are characterized by a $Z_2$ invariant
$\nu = -1$. Similar to the situation in 2D, this value of the invariant  
implies that an \emph{odd} number of the $4$ states $\Psi_{(0,0)}, \Psi_{(\pi,0)}, \Psi_{(0,\pi)}, \Psi_{(\pi,\pi)}$
have a Kramers degeneracy on the $z = 0$ and $z = L_z$ surfaces of the Corbino donut. \cite{FuKaneHall} 
(We give a simple derivation of this result at the end of this section). By our assumption above, $\Psi_{(0,0)}$ has
no Kramers degeneracy, so either all three of the other states have a degeneracy, or exactly
one of them does (see Fig. \ref{3Dflux}b). In particular, at least one of the three other
states has a Kramers degeneracy. Let us denote this state by $\Psi_{(\alpha,\beta)}$. Then, as long as $L_z$ is large 
so that the two surfaces of the Corbino donut are well separated, Kramers theorem 
guarantees that $\Psi_{(\alpha,\beta)}$ is part of a multiplet of four states (two associated with each surface) 
which are nearly degenerate in energy. More precisely, these four states are separated by an energy splitting that
vanishes exponentially as the distance between the two surfaces grows.

Just as in the 2D case, we can use this Kramers degenerate state to construct a low lying excited state
at zero flux: as in that case, we start in one of the three states which are degenerate with $\Psi_{(\alpha,\beta)}$, 
and then adiabatically reduce the flux to $(0,0)$. The result is an eigenstate $\Psi_{ex}$ of the zero flux 
Hamiltonian. This state is distinct from the ground state $\Psi_{(0,0)}$ and its energy gap vanishes in the 
thermodynamic limit. Furthermore, time reversal invariant perturbations cannot change this picture, since they
cannot split the degeneracy between $\Psi_{(\alpha,\beta)}$ and its Kramers partners. 
Hence, even in the presence of arbitrary time reversal invariant, charge conserving perturbations, the system always 
has at least one low lying excited state $\Psi_{ex}$.  

\begin{figure}[t]
\centerline{ \includegraphics[width=0.9\columnwidth]{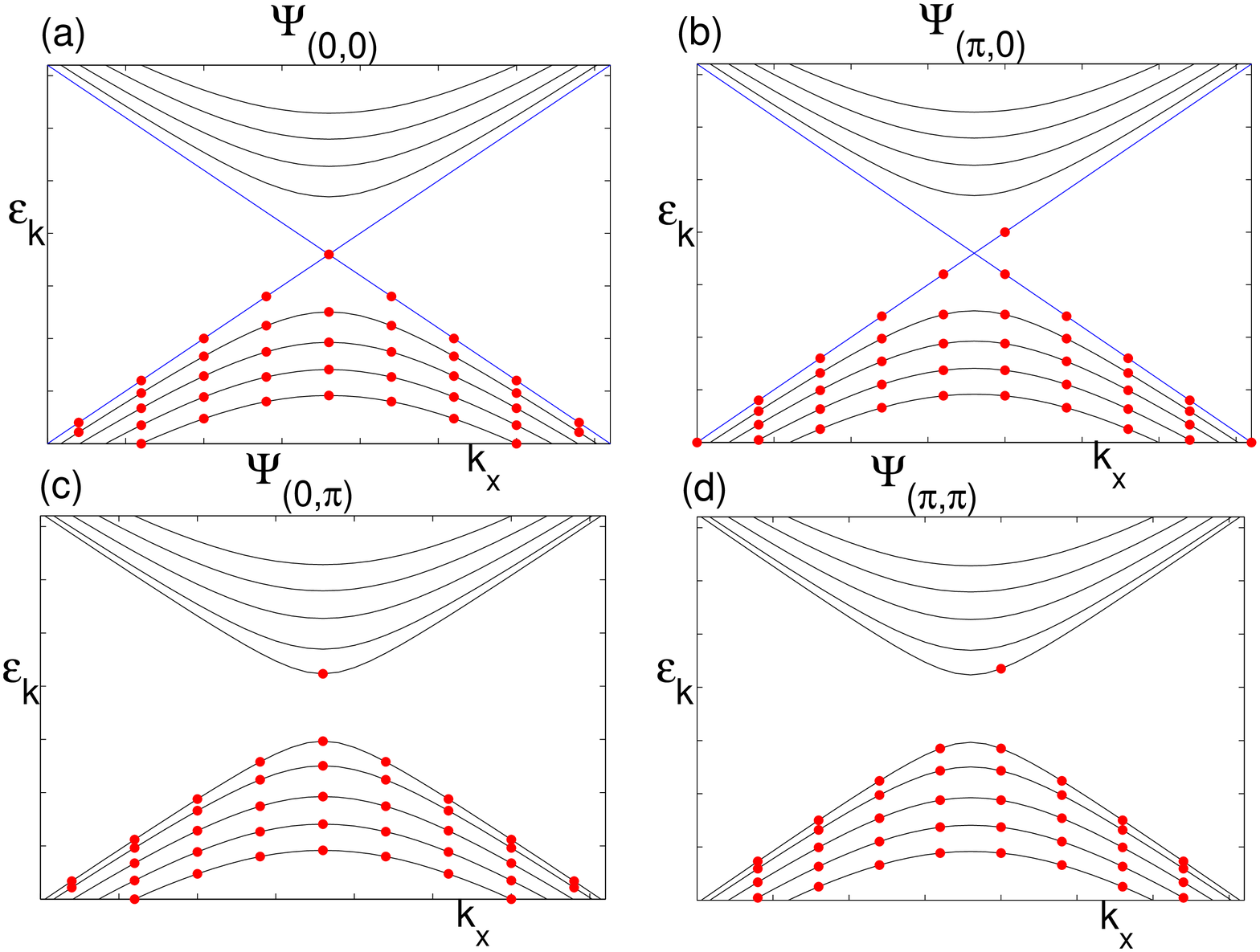}  }
\caption{Surface state configurations of $\Psi_{(0,0)}, \Psi_{(\pi,0)},\Psi_{(0,\pi)},\Psi_{(\pi,\pi)}$
for a system with a single Dirac cone on each surface, and a chemical potential just above the Dirac point. Here 
we plot the surface states as a function of $k_x$; the bands indicate different discrete values of $k_y$. The 
band that crosses the tip of the Dirac cone, corresponding to $k_y = 0$, is singly degenerate; all other bands, 
for which $k_y \neq 0$, are doubly degenerate. 
The $4$ panels show the arrangement of the discrete surface momenta relative to the tip of the Dirac cone in each 
of the $4$ flux sectors. The filled circles indicate occupied states. 
}
\label{DiracFig}
\end{figure}

For completeness we now explain why non-interacting topological insulators have the property that an odd number of 
the $4$ states $\Psi_{(0,0)}, \Psi_{(\pi,0)}, \Psi_{(0,\pi)}, \Psi_{(\pi,\pi)}$ have a Kramers degeneracy on the 
$z = 0$ and $z = L_z$ surfaces of the Corbino donut. This result was first established in Ref. \onlinecite{FuKaneHall}; 
here we give an alternative derivation using the fact that these systems have an odd number of Dirac cones on each 
surface. 

For simplicity we assume a band structure with a \emph{single} Dirac cone on each surface. 
We also assume that the chemical potential is tuned to lie just above the Dirac point. 
We begin with the system at zero flux: $(\phi_x, \phi_y) = (0,0)$. Letting $L_x$ and $L_y$ be the dimensions in the 
two periodic directions, the momenta $k_x, k_y$ are quantized as: 
\be
k_x  = \frac{ 2 \pi m_x}{L_x} \ \ \ \ \ \ k_y = \frac{2 \pi m_y}{L_y} 
\ee
where $m_x, m_y$ are arbitrary integers. 

Let us focus on one of the two surfaces, say the $z = 0$ surface. Suppose that the Dirac cone on that surface 
is located at $(k_x, k_y) = (0,0)$. Then the momentum states near the Dirac point will be arranged as in Fig. \ref{DiracFig}a. 
The wave function $\Psi_{(0,0)}$ will be a Slater determinant consisting of all the momentum states with $\epsilon_k \leq \mu$ --
including \emph{both} momentum states at $(k_x,k_y) = (0,0)$. If we now thread half a flux quantum through the 
$x$ or $y$ direction, the momenta will shift by half a unit in $k_x$ or $k_y$, leading to the states $\Psi_{(\pi,0)}, 
\Psi_{(0,\pi)}, \Psi_{(\pi,\pi)}$ shown in Fig. \ref{DiracFig}b-d. Examining these configurations, we can see that 
all three have a single filled state in the upper band, implying that all three of them have a Kramers degeneracy 
on the $z =0$ surface (these states must also have a Kramers degeneracy on the $z = L_z$
surface since the total number of electrons is even).  

On the other hand, if the Dirac cone is located at a different point in the Brillouin zone,
say $(k_x,k_y) = (\pi,\pi)$, then depending on the parity of $L_x, L_y$, the momentum states 
near the Dirac point at flux $(\phi_x, \phi_y) = (0,0)$, can be arranged in any of the three possibilities shown in   
Fig. \ref{DiracFig}b-d. Again, $\Psi_{(0,0)}$ will be a Slater determinant involving all states in the lower band. 
In these three cases, one can check that when one inserts flux through the two holes of the Corbino
donut, exactly one of the resulting states $\Psi_{(\pi,0)},\Psi_{(0,\pi)},\Psi_{(\pi,\pi)}$ has an unpaired momentum
state, and therefore a Kramers degeneracy on the $L_z = 0$ surface. Combining all of these cases, we conclude that the 
system always has a Kramers degeneracy in either $1$ or $3$ of the four possible states. This is what we wanted to show.

\subsubsection{Strong vs. weak topological insulators}
The reader may have noticed that, in the flux insertion argument, we only used the fact that \emph{at least one} of 
the states $\Psi_{(\pi,0)},\Psi_{(0,\pi)}, \Psi_{(\pi,\pi)}$ had Kramers degeneracies: we did not make use of the 
additional information that the number of such states was \emph{odd}. In other words, the argument would have 
worked equally well if $2$ of these states had degeneracies. 

We can understand the physical meaning of this observation as follows. Recall that, in addition to the usual 
3D topological insulator, there is another interesting kind of 3D time reversal invariant band insulator known as a 
``weak topological insulator.'' \cite{HasanKaneRMP} (In this terminology, the usual 3D topological insulator
is known as a ``strong topological insulator.'') Weak topological insulators can be thought of as layered systems, 
where each layer is a 2D topological insulator. The reason that weak topological insulators are relevant here is that 
these systems give examples where $2$ of the three states $\Psi_{(\pi,0)},\Psi_{(0,\pi)}, \Psi_{(\pi,\pi)}$ have 
Kramers degeneracies (assuming, as before, that $\Psi_{(0,0)}$ is not Kramers degenerate). 
More precisely, this pattern of degeneracies must arise in a weak topological insulator with an 
\emph{odd} number of layers, with the layers oriented perpendicular to the $z = 0, L_z$ surfaces. 

This example clarifies why the flux insertion argument extends to systems where $2$ of the
states $\Psi_{(\pi,0)},\Psi_{(0,\pi)}, \Psi_{(\pi,\pi)}$ have Kramers degeneracies: just like strong topological insulators, 
these odd-layer weak topological insulators have protected surface modes. One way to understand 
these surface modes is to note that each 2D topological insulator layer contributes a one dimensional edge mode 
at the boundary. These edge modes can be gapped out in pairs by appropriate perturbations, but if there 
are an odd number of layers, we will always be left with at least one gapless mode.  

At the same time, this example reveals a shortcoming of the flux insertion argument: the argument does not 
distinguish strong topological insulators from weak topological insulators with an odd number of layers, 
as both systems have protected surface modes. Yet it is clear intuitively that the surfaces of these two systems 
are topologically distinct: the surface modes of a strong topological insulator are in some sense 
``more two dimensional'' than those of a weak topological insulator with an odd number of layers. 

In the following, we refine the flux insertion argument so that it addresses this issue.
We use the additional information that an \emph{odd} number of $\Psi_{(0,0)},\Psi_{(\pi,0)}, \Psi_{(0,\pi)}, 
\Psi_{(\pi,\pi)}$ have Kramers degeneracies to show that our system not
only has a low lying surface mode, but this mode is delocalized in both the $x$ and $y$ 
directions. To be precise, we prove that the low lying surface modes in our system can never 
be localized to one dimensional strips. This property distinguishes our system from a weak 
topological insulator with an odd number of layers since in that case a carefully designed perturbation 
can gap out all low lying excitations except a single one dimensional edge mode. (As an aside, we
note that the difference between the surface modes of weak and strong topological insulators largely 
disappears if one considers the surfaces in the presence of a \emph{random} potential\cite{RingelKrausStern}). 

We give a proof by contradiction. Suppose that an appropriate perturbation could localize all 
low-lying surface modes to a single one dimensional strip. Let $W_x, W_y$ denote the number of 
times this strip winds around the (toroidal) surface in the $x$ and $y$ directions respectively. 
Then, when we insert flux $\phi_x, \phi_y = 0,\pi$ through the two holes of the Corbino donut, the low 
lying excitations will only couple to $\phi_x, \phi_y$ via the linear combination 
$\phi_{\text{strip}} = W_x \phi_x + W_y \phi_y$. In particular, it follows that the presence or absence 
of a Kramers degeneracy at $\Psi_{(\phi_x, \phi_y)}$ only depends on $\phi_x, \phi_y$ via 
$\phi_{\text{strip}} = W_x \phi_x + W_y \phi_y$. But this kind of $\phi_x, \phi_y$ dependence always 
leads to an \emph{even} number of $\Psi_{(0,0)},\Psi_{(\pi,0)}, \Psi_{(0,\pi)}, \Psi_{(\pi,\pi)}$ 
having Kramers degeneracies, as we now show.

It is useful to consider the cases where $W_x, W_y$ are even and odd separately.
First suppose that $W_x$ is even and $W_y$ is arbitrary. In this case, the effective flux 
$\phi_{\text{strip}}$ takes the same value at $(\phi_x, \phi_y) = (0,0), (\pi,0)$ and similarly at 
$(\phi_x,\phi_y) = (0,\pi), (\pi,\pi)$. Therefore, since $\phi_{\text{strip}}$ completely determines 
whether the system has a Kramers degeneracy, it follows that either $\Psi_{(0,0)}, \Psi_{(\pi,0)}$ 
both have Kramers degeneracies or they both have no degeneracy, and similarly for 
$\Psi_{(0,\pi)}, \Psi_{(\pi,\pi)}$. But then the Kramers degenerate states come in pairs, implying that an 
\emph{even} number of the four states have Kramers degeneracies. The same is true if $W_y$ is even 
and $W_x$ is arbitrary. The only remaining possibility is if $W_x, W_y$ are both odd. In this 
case, $\phi_{\text{strip}}$ takes the same value at $(\phi_x, \phi_y) = (0,0), (\pi,\pi)$ and similarly 
at $(\phi_x, \phi_y) = (0,\pi),(\pi,0)$. Again, the Kramers degeneracies come in pairs and we are 
guaranteed to get an even number of Kramers degenerate states. We conclude that a system where an 
odd number of $\Psi_{(0,0)},\Psi_{(\pi,0)}, \Psi_{(0,\pi)}, \Psi_{(\pi,\pi)}$ have Kramers degeneracies 
must have excitations which are truly two dimensional and are not localized to a one dimensional strip. 

\subsubsection{Fractionalized case}
The argument for the 3D fractionalized insulator $H$ (\ref{Hfrtop3d}) is completely analogous. For the same reason as in 2D, we
insert $N\Phi_0/2$ flux rather than $\Phi_0/2$ flux, where $N = e/e^*$. Inserting flux through each of the two holes
in the Corbino donut, we construct four states, $\Psi_{(0,0)}, \Psi_{(N\pi,0)}, \Psi_{(0,N\pi)}, \Psi_{(N\pi,N\pi)}$. Since 
the low energy physics is described by a topological insulator built out of charge $q_{\text{f}}$ fermions,
this flux insertion is equivalent to inserting $\frac{q_{\text{f}}}{e^*} \cdot\frac{\Phi_0}{2}$ flux for a non-interacting 
topological insulator. If $q_{\text{f}}/e^*$ is odd, then an odd number of 
$\Psi_{(0,0)}, \Psi_{(N\pi,0)}, \Psi_{(0,N\pi)}, \Psi_{(N\pi,N\pi)}$ have
Kramers degeneracies. We can then construct a protected low lying state $\Psi_{ex}$ at zero flux, by starting
in one of the Kramers degenerate states and adiabatically reducing the flux to $0$. Moreover, this state is 
in the same topological sector as the ground state, just as in the 2D case. This is
what we wanted to show. 

Finally, we note that we can refine the flux insertion argument just as in the non-interacting case. 
To be specific, by repeating the analysis in the previous section, we can prove that if $q_{\text{f}}/e^*$ is odd, 
the fractionalized insulators not only have low lying surface modes, but these modes are truly two dimensional 
and cannot be localized to a one dimensional strip. Just as in the non-interacting case, this allows us to 
distinguish these systems from ``weak fractional topological insulators'' with an odd number of layers.
 
\subsection{Microscopic theory of 2D edge} \label{2dmicroedge}
While the flux insertion argument shows that the edge modes are protected for the models (\ref{Hfrtop}) 
with odd $q_{\text{f}}/e^*$, it does not tell us anything about the converse statement---that is, whether the modes can be
gapped out in the models with even $q_{\text{f}}/e^*$. In order to establish this fact we now analyze the microscopic 
theory of the edge.

\subsubsection{Constructing the 2D edge theory} \label{2dcsedge}
The first step is to construct the edge theory. The free fermion theory (\ref{2dedge}) is unfortunately
not a good starting point since it is not truly a complete edge theory. The reason it is not complete
is that it doesn't include all the quasiparticle excitations with nontrivial statistics, such as
flux quasiparticles. As a result, this theory cannot be used to analyze the most general edge 
perturbations---for example, those that are large compared to the flux quasiparticle gap. 

A simple way to construct a complete edge theory is to use the general bulk-edge correspondence for 
abelian Chern-Simons theory. \cite{WenReview,WenBook} According to this correspondence, the edge of 
$H$ (\ref{Hfrtop}) is described by a four component chiral boson theory
\begin{eqnarray}
L &=& \frac{1}{4\pi} \partial_x \Phi_I \left( K_{IJ} \partial_t \Phi_J - V_{IJ} \partial_x\Phi_J \right) \nonumber \\
&+& \frac{1}{2\pi} t_I\epsilon^{\mu \nu} \partial_\mu \Phi_I A_\nu
\label{2dedgeboson}
\end{eqnarray}
Here $\Phi$ is a four component vector of fields, $V$ is the velocity matrix and 
$K$ and $t$ are defined as in (\ref{Kmat2d}).
In this language, the most general quasiparticle creation operators can be written as $e^{i l^T \Phi}$, 
where $l$ an integer valued four component vector. The charge carried by $e^{i l^T \Phi}$
is $l^T K^{-1} t$. The subset of these operators that are ``local'' 
(i.e. products of electron creation and annihilation operators) can be written as
$e^{i\Theta(\Lambda)}$ where $\Theta(\Lambda) \equiv \Lambda^T K \Phi$ and 
$\Lambda$ is an integer valued four component vector. 

The final component of the edge theory is the physical interpretation of the $e^{i\Theta(\Lambda)}$ operators in terms
of the microscopic model $H$. Using our understanding of the corresponding bulk Chern-Simons theory, we identify 
$\Lambda^T = (0,0,1,0), (0,0,0,1)$ with creation operators for spin-up and spin-down electrons, 
respectively, and $\Lambda^T = (0,1,0,0)$ with a creation operator for the charge $2e$ spinless 
boson. As for $(1,0,0,0)$, this creates a neutral spinless particle which can be thought 
of as composite of $m$ flux quasiparticles. The most general $e^{i\Theta(\Lambda)}$ is a composite of
these elementary operators.

These identifications have the added benefit of fixing the transformation properties of $\Phi$
under time reversal. If we require that the electron creation operators transform as
\begin{align}
\mathcal{T}: \ c^\dagger_{\uparrow} \rightarrow c^\dagger_{\downarrow} \ ,
c^\dagger_{\downarrow} \rightarrow -c^\dagger_{\uparrow}
\end{align}
and that the spinless charge $2e$ boson is invariant under time reversal, while the flux changes
sign, we deduce that 
$\Phi$ transforms as
\begin{equation}
\Phi \rightarrow T \Phi + \pi K^{-1} \chi 
\label{Tinv}
\end{equation}
where
\begin{align}
T = \bpm -1 & 0 & 0 & 0 \\ 
	 0 & 1 & 0 & 0 \\
	 0 & 0 & 0 & 1 \\
	 0 & 0 & 1 & 0 \epm \ ; \ \chi = \bpm 0 \\ 0 \\ 1 \\ 0 \epm
\end{align}
Equivalently, $\Theta(\Lambda)$ transforms as
\begin{equation}
\Theta(\Lambda) \rightarrow \Theta(-T\Lambda) - Q(\Lambda) \cdot \pi
\label{tinvth}
\end{equation}
where $Q(\Lambda) \equiv \Lambda^T \chi$.
To complete the story, we need to explain the relationship between the above edge theory (\ref{2dedgeboson}) and the 
edge (\ref{2dedge}) of the exactly soluble model $H$ (\ref{Hfrtop}). An important clue is that (\ref{2dedgeboson})
has four gapless modes while the free fermion edge (\ref{2dedge}) has only two modes. This mode counting suggests that 
(\ref{2dedge}) can be obtained from (\ref{2dedgeboson}) by adding a perturbation that gaps out two of the modes.
In appendix \ref{edgerelation}, we confirm this guess. We show that a time reversal invariant perturbation of the form
\begin{eqnarray}
U(\Lambda) &=& U(x)[\cos(\Theta(\Lambda) - \alpha(x)) \nonumber \\
&+& (-1)^{Q(\Lambda)}\cos(\Theta(T\Lambda) -\alpha(x))]
\label{scatter}
\end{eqnarray}
with $\Lambda_0^T = (1,0,0,0)$ does the job. That is, the perturbation $U(\Lambda_0)$ gaps out two of the 
edge modes of (\ref{2dedgeboson}) leaving behind exactly the free fermion edge (\ref{2dedge}). 

Turning this statement around, if we start with the exactly soluble model (\ref{Hfrtop}), we must be able to find
a perturbation that closes the gap to the bosonic excitations at the edge, and results in the four component
edge theory (\ref{2dedgeboson}). At a microscopic level, the following perturbation accomplishes this task:
\begin{equation}
\Delta H_{\text{edge}} = -J \sum_{\<ss'\> \in \partial X} (U_{ss'} + h.c)
\end{equation}
Here, the sum runs over links on the boundary $\partial X$ of the exactly soluble system. 
Physically, this term gives an amplitude for the bosonic charge excitations to hop. When $J \ll V$,
the bosonic charge excitations are gapped in both the bulk and the edge, and the only low energy edge 
excitations are the two chiral fermion modes in (\ref{2dedge}). However, when $J$ is sufficiently large, 
the gap to the bosonic charge excitations closes at the edge, resulting in a 1D superfluid. The resulting 
edge has four gapless modes and is described by (\ref{2dedgeboson}).

\subsubsection{Analysis of edge mode stability}
In the previous section we derived an edge theory (\ref{2dedgeboson}) for $H$ (\ref{Hfrtop}), and
showed it could be obtained from the exactly soluble edge by adding an appropriate perturbation. 
We now investigate whether this edge theory (\ref{2dedgeboson}) can be gapped out \emph{completely} by charge conserving, 
time reversal symmetric perturbations. We find that the edge can be gapped out if and only if $q_{\text{f}}/e^*$ is even, in 
agreement with the flux insertion argument.

Our analysis closely follows that of Ref. \onlinecite{LevinStern}. We focus on scattering terms of the form (\ref{scatter}). 
These perturbations can be divided into two different classes: perturbations where $\Lambda, T\Lambda$ are linearly independent,
and perturbations where $T\Lambda = \pm \Lambda$. Perturbations of the first type can gap out four edge modes, while
perturbations of the second type can gap out two edge modes. Therefore, in order to fully gap out the edge (\ref{2dedgeboson}), 
we either need one perturbation of the first type or two perturbations of the second type. Here, we will
focus on the second possibility (though we would obtain the same results if we considered the first kind of perturbation 
instead). That is, we will look for perturbations of the form 
\begin{equation}
U(\Lambda_1) + U(\Lambda_2) 
\label{pert2}
\end{equation}
where $T \Lambda_1 = \pm \Lambda_1$ and similarly for $\Lambda_2$.

To begin, we note that the most general charge conserving solution to $T \Lambda = \Lambda$ is $\Lambda^T = (0,x,-x,-x)$, while 
the most general solution to $T \Lambda = -\Lambda$ is $\Lambda^T = (y,0,z,-z)$. Therefore, in order to get two linearly 
independent $\Lambda$'s, we can either take
\begin{align}
\Lambda_1^T = (y_1,0,z_1,-z_1) \ , \  \Lambda_2^T = (y_2, 0, z_2, -z_2)
\label{type1}
\end{align}
or we can take 
\begin{align}
\Lambda_1^T = (0,x,-x,-x) \ , \ \Lambda_2^T = (y,0,z,-z)
\label{type2}
\end{align}
In the first case (\ref{type1}), the corresponding perturbation 
(\ref{pert2}) can indeed gap out the edge, but hand in hand with that it spontaneously breaks time reversal symmetry, as 
we now explain. We note that when (\ref{pert2}) gaps out the edge, it freezes the value of $\Theta(\Lambda_1), 
\Theta(\Lambda_2)$. It therefore also freezes the value of $\Theta(y_1 \Lambda_2 - y_2 \Lambda_1)$, which is a 
multiple of $\Theta(0,0,1,-1)$. But $\cos(\Theta(0,0,1,-1) - \alpha)$ is odd under time reversal
(this follows from (\ref{tinvth}) or alternatively from the fact that $\cos(\Theta(0,0,1,-1) - \alpha)$ describes 
a Zeeman field oriented in the $xy$ plane) so we conclude that this perturbation spontaneously breaks time reversal symmetry. 

Therefore, if we want to gap out the edge without breaking time reversal symmetry, we are led to perturbations of the 
form (\ref{type2}). According to Haldane's null vector criterion\cite{Haldane} such a 
perturbation can gap out the edge if $\Lambda_1, \Lambda_2$ satisfy
\begin{equation}
\Lambda_1^T K \Lambda_1 = \Lambda_2^T K \Lambda_2 = \Lambda_1^T K \Lambda_2 = 0
\label{hald}
\end{equation}
(The origin of the criterion (\ref{hald}) is that this condition guarantees that we can make a linear change of variables 
from $\Phi$ to $\Phi'$ such that (a) the action for $\Phi'$ consists of two decoupled non-chiral Luttinger liquids, and 
(b) the two perturbations correspond to backscattering terms for the two liquids. See appendix \ref{edgerelation} for a related 
example). For the above $\Lambda_1, \Lambda_2$, this leads to the condition
\begin{equation}
(2k+m)y - 2z = 0
\end{equation}
It is convenient to parameterize $y,z$ by $y = us$, $z = vs$ where $u,v$ have no common factors.
Then the above condition becomes
\begin{equation}
(2k+m)u - 2v = 0
\end{equation}
We now consider two cases: $2k+m$ divisible by $4$, and $2k+m$ not divisible by $4$. If $2k+m$ is not divisible by 
$4$ then we must have $v$ odd (since even $v$ implies even $u$, which contradicts the fact that $u,v$ have no common
factors). It then follows that the perturbation $U(\Lambda_2)$ spontaneously breaks time reversal symmetry:
the perturbation $U(\Lambda_2)$ freezes the values of $\Theta(\Lambda_2)$ which
is a multiple of $\Theta(u,0,v,-v)$. But $\cos(\Theta(u,0,v,-v)-\alpha)$ is odd under time reversal symmetry
(since $v$ is odd), so this perturbation must spontaneously break time reversal symmetry.
We conclude that when $2k+m$ is not divisible by $4$, we cannot gap out the edge using these perturbations
without breaking time reversal symmetry. We note that this agrees with the flux insertion argument since $q_{\text{f}}/e^*$ is 
odd precisely when $2k+m$ is not divisible by $4$.

On the other hand, when $2k+m$ is a multiple of $4$ (or equivalently, $q_{\text{f}}/e^*$ is even) the above condition suggests 
a natural solution for $(\Lambda_1, \Lambda_2)$: we can take $u = 1$, $v = k+m/2$, $s=1$, $x=1$, so that
\begin{eqnarray}
\Lambda_1^T &=& (0,1,-1,-1) \nonumber \\
\Lambda_2^T &=& (1,0,k+m/2,-k-m/2)
\end{eqnarray}
The physical meaning of the corresponding perturbations is clear: $U(\Lambda_1)$ describes a process where a charge $2e$ boson breaks 
into two electrons with opposite spin directions, while $U(\Lambda_2)$ describes a process which flips the spins of 
$k+m/2$ electrons while simultaneously creating a composite of $m$ flux quasiparticles. 

To see that the perturbation $U(\Lambda_1)+U(\Lambda_2)$
can gap out the edge, we note that the above $\Lambda_1, \Lambda_2$ obey Haldane's criterion (\ref{hald}). Moreover, this 
perturbation is time reversal invariant as long as $\alpha_1 = \pi/2$ or $3\pi/2$. Finally, this perturbation does not 
break time reversal symmetry spontaneously, as we now explain. The basic point is that, the only way that time reversal symmetry 
(or any other symmetry) can be spontaneously broken is if, for some $a,b$ with no common factors, the linear combination 
$a \Lambda_1 + b \Lambda_2$, is non-primitive -- that is, $a \Lambda_1 + b \Lambda_2 = n \Lambda$, where $\Lambda$ is an integer
vector, and $n$ is an integer larger than $1$. (To understand where this condition comes from, see the example of spontaneous symmetry breaking given
in the discussion after (\ref{type2})). But it is clear from the form of $\Lambda_1, \Lambda_2$ that all such linear combinations 
$a \Lambda_1 + b \Lambda_2$ are primitive. We conclude that $U(\Lambda_1)+U(\Lambda_2)$ gaps out the edge without breaking time 
reversal symmetry explicitly or spontaneously.

\section{Conclusion}
In this work, we have constructed a family of exactly soluble models for fractionalized, time reversal invariant insulators in
$2$ and $3$ dimensions. At low energies, these models behave like topological insulators made of fractionally charged 
fermions of charge $q_{\text{f}}$. As a result, the 2D models have two edge modes of opposite chiralities, while the 3D models
have a gapless Dirac cone on each surface. At high energies, the insulators possess other types of excitations, including 
quasiparticles with minimal charge $e^*$. As expected for systems with fractional charge, all of the 2D and 3D insulators are 
topologically ordered, exhibiting fractional statistics as well as ground state degeneracy in geometries with periodic boundary 
conditions.

An important characteristic of the 3D models is that they exhibit a fractional magnetoelectric effect. More specifically, if time 
reversal symmetry is broken at the surface, the 3D models exhibit a surface Hall effect with fractional Hall conductivity
$\sigma_{xy} = q_{\text{f}}^2/2h$. In addition, if a magnetic monopole is inserted into the bulk, it binds a fractional 
electric charge. Somewhat surprisingly, the size of the elementary charge $e^*$ plays a significant role in this charge binding
physics. 

A key question is whether the gapless boundary modes are robust to perturbations, so that these systems truly qualify as 
``fractional topological insulators.'' We have studied this question in detail and have shown that the boundary modes are 
protected for the models where the ratio $q_{\text{f}}/e^*$ is odd. That is, the edge or surface modes cannot be gapped out by 
perturbations that are time reversal symmetric and charge conserving. In fact, these modes are immune even to large 
perturbations, as long as the perturbations do not close the bulk gap or spontaneously break time reversal or charge conservation 
symmetry at the boundary.

In contrast, the gapless boundary modes are not necessarily stable when $q_{\text{f}}/e^*$ is even. In the 2D case, we 
have demonstrated this point explicitly by constructing time reversal invariant, charge conserving perturbations that open a 
gap at the edge. The situation in 3D is less clear: though our argument for proving the existence of protected 
surface modes breaks down in the models with even $q_{\text{f}}/e^*$, we have not explicitly constructed perturbations
that gap out the surface. Determining the robustness of the surface modes in these models is an interesting question 
for future research.

Our findings regarding the stability of the edge modes are intriguing for several reasons. First, we have demonstrated 
that simply having fractionally charged fermions in the right band structure does \emph{not} guarantee a system is a 
fractional topological insulator. Second, although $e^*$ is a high energy property, it evidently determines the fate 
of the low energy excitations on the edge. And third, it is interesting that the condition that guarantees stability 
the boundary modes turns out to be identical for both the two and three dimensional models considered here and those considered 
in Ref. \onlinecite{LevinStern}. 

We have focused here on a particular set of models, which does not exhaust all
the possibilities for fractional topological insulators. In two dimensions, the fractional quantum spin Hall systems
discussed by Refs. \onlinecite{BernevigZhang, LevinStern} give a whole other class of examples. Similarly, in three dimensions, there may be other families of 3D fractional topological insulators beyond the ones discussed here. Constructing microscopic 
models of the other families, and finding a complete classification of the possibilities, remain tantalizing open questions. 

\acknowledgments
ML was supported in part by an Alfred P. Sloan Research Fellowship. AS thanks the US-Israel Binational Science
Foundation, the Minerva foundation and Microsoft Station Q for financial support.

\appendix

\section{Eigenstate degeneracy of 2D lattice boson model}
In this section, we find the degeneracy of the $|q_s, b_P\>$ eigenstates of the 2D 
lattice boson model (\ref{HBose}). We consider two geometries: a
rectangular piece of square lattice with open boundary conditions (Fig. \ref{openbc}(a)), and a rectangular
piece of square lattice with periodic boundary conditions---that is, a torus (Fig. \ref{openbc}(b)). 

\subsection{Open boundary condition geometry}
\label{degopen}
The first step is to define projectors $P_{q_s}$,
$P_{b_P}$, which project onto states with $Q_s = q_s$ and $B_P = b_P$ respectively.
The degeneracy $D$ of the $q_s, b_P$ eigenspace can then be written as the
trace of the product of all the projectors:
\begin{equation}
D = Tr\left(\prod_s P_{q_s} \prod_P P_{b_P}\right)
\label{projdeg}
\end{equation}
We next write down an explicit expression for $\prod_P P_{b_P}$. To this end, we recall that the eigenvalues of 
$B_P$ are $p$th roots of unity, so the projector can be written as
\begin{equation}
P_{b_P} = \frac{1}{m} \sum_{j=0}^{m-1} \bar{b}_P^j B_P^j
\end{equation}
Expanding out the product, we have
\begin{eqnarray}
\prod_P P_{b_P} &=& \prod_P \left(\frac{1}{m} \sum_{j=0}^{m-1} \bar{b}_P^j B_P^j \right) \nonumber \\
&=& \frac{1}{m^{N_{\text{plaq}}}} \sum_{\{j_P\}} \prod_P \bar{b}_P^{j_P} \prod_P B_P^{j_P}
\label{deg2}
\end{eqnarray}
where $N_{\text{plaq}}$ is the total number of plaquettes.
Using the definition $B_P = U_{12} U_{23} U_{34} U_{41}$, we get an expression of the form
\begin{equation}
\prod_P P_{b_P} = \frac{1}{m^{N_{\text{plaq}}}} \sum_{\{j_P\}} \prod_P \bar{b}_P^{j_P} \prod_{\<ss'\>} U_{ss'}^{\Delta j_{ss'}}
\label{deg3}
\end{equation}
Here, $\Delta j_{ss'} = j_P - j_{P'}$ where $P,P'$ are the two plaquettes bordering the link $\<ss'\>$. 
If $\<ss'\>$ happens to live on the boundary of the system, then $\Delta j_{ss'} = j_P$ where $P$ is the
unique plaquette bordering this link. Substituting this into (\ref{projdeg}) gives
\begin{equation}
D = \frac{1}{m^{N_{\text{plaq}}}} \sum_{\{j_P\}} \prod_P \bar{b}_P^{j_P} \cdot Tr\left(\prod_s P_{q_s}\prod_{\<ss'\>} 
U_{ss'}^{\Delta j_{ss'}}\right)
\label{deg4}
\end{equation}

The next step is to note that the operator $U_{ss'}^r$ changes the boson number $n_{ss'}$
by $\pm r \text{ (mod m)}$. We conclude that the trace 
$Tr \left(\prod_s P_{q_s} \prod_{\<ss'\>} U_{ss'}^{\Delta j_{ss'}} \right)$ vanishes unless 
$\Delta j_{ss'} \equiv 0 \text{ (mod m)}$ everywhere, which implies $j_P = 0$ everywhere (since we are 
assuming a geometry with open boundary conditions). In this way, we see that the only
nonvanishing term in (\ref{deg4}) is the one with $j_P = 0$, so that
\begin{equation}
D = \frac{1}{m^{N_{\text{plaq}}}} \cdot Tr \left(\prod_s P_{q_s} \right)
\label{deg5}
\end{equation}
All that remains is to compute this trace, or equivalently, count the number of
states with $Q_s = q_s$. Working in the number basis, $|n_s, n_{ss'}\>$, and using the
definition (\ref{Qsop}) of $Q_s$ it is
clear that there is a one-to-one correspondence between states with $Q_s = q_s$ and 
configurations of $n_{ss'}$ satisfying 
\begin{equation}
\alpha_s \sum_{s'} n_{ss'} \equiv q_s \text{ (mod m)}
\label{constraint}
\end{equation}
We can count these configurations by comparing the number of constraints (\ref{constraint}) to the number of free parameters.
Examining (\ref{constraint}), we see that we have a constraint for every site $s$ of the lattice.
However, these constraints are not all independent, since if we add all the constraint equations together, we get
\begin{equation}
\sum_s q_s \equiv \sum_s \alpha_s \sum_{s'} n_{ss'} \equiv  0 \text{ (mod m)}
\end{equation}
which is automatically satisfied as long as $\sum_s q_s \equiv 0 \text{ (mod m)}$. (This is a special
case of the identity (\ref{Eq_Ns2})). We conclude that the
number of constraints in (\ref{constraint}) is
\begin{equation}
N_{\text{constr}} = N_{\text{site}} - 1 
\end{equation}
On the other hand, the number of free parameters is just the total number of links
\begin{equation}
N_{\text{param}} = N_{\text{link}}
\end{equation}
Combining these two calculations, we see that the total number of configurations of $n_{ss'}$ satisfying
(\ref{constraint}) is given by
\begin{equation}
N_{\text{config}} = m^{N_{\text{param}} - N_{\text{const}}} = m^{N_{\text{link}} - N_{\text{site}} + 1}
\label{nconfig}
\end{equation}
We now substitute this into our formula (\ref{deg5}) for the degeneracy, obtaining
\begin{eqnarray}
D &=& \frac{1}{m^{N_{\text{plaq}}}} N_{\text{config}} \nonumber \\
&=& m^{N_{\text{link}} - N_{\text{site}} - N_{\text{plaq}} + 1} 
\end{eqnarray}
We then note that 
\begin{equation}
N_{\text{link}} - N_{\text{site}} - N_{\text{plaq}} = -1
\label{euler}
\end{equation}
either by direct computation or by Euler's general formula $V - E + F = \chi$. We conclude 
that $D = 1$: there is a unique eigenstate $|q_s, b_P\>$ for each configuration satisfying 
$\sum_s q_s \equiv 0 \text{ (mod m)}$.

\subsection{Periodic (torus) geometry}
\label{degtorus}
The calculation in the torus geometry proceeds similarly to the open boundary condition case discussed
above. The expression (\ref{deg4}) for the degeneracy is still applicable in this case, and we can still 
deduce that the only nonvanishing terms in this sum are the ones where $\Delta j_{ss'} \equiv 0 \text{ (mod m)}$ 
everywhere. However, unlike the open boundary condition case, there are now $m$ terms satisfying 
$\Delta j_{ss'} \equiv 0 \text{ (mod m)}$---namely the terms where $j_P$ is constant. Therefore, in the torus 
geometry case, (\ref{deg4}) reduces to
\begin{eqnarray}
D &=& \frac{1}{m^{N_{\text{plaq}}}} \sum_j \prod_P \bar{b}_P^{j} \cdot Tr \left(\prod_s P_{q_s} \right) \nonumber \\
&=& \begin{cases} \frac{1}{m^{N_{\text{plaq}}-1}} \cdot Tr \left(\prod_s P_{q_s} \right) & \mbox{if }\prod_P b_P = 1 \\
0 & \mbox{otherwise}
\end{cases} 
\label{deg6}
\end{eqnarray}
Just as in the open boundary condition case, we can compute $Tr(\prod_s P_{q_s})$ by counting the number of
configurations $n_{ss'}$ satisfying (\ref{constraint}), and again this number is given by
(\ref{nconfig}). Substituting (\ref{nconfig}) into (\ref{deg6}), and specializing to the case $\prod_P b_P = 1$,
we find
\begin{equation}
D = \frac{1}{m^{N_{\text{plaq}}}} N_{config} = m^{N_{\text{link}} - N_{\text{site}} - N_{\text{plaq}}+2}
\end{equation}
At this stage, there is again a difference from the open boundary condition case. Instead
of (\ref{euler}), we have the modified relation
\begin{equation}
N_{\text{link}} - N_{\text{site}} - N_{\text{plaq}} = 0
\end{equation}
(again, this can be derived directly or via Euler's formula $V - E + F = \chi$). We conclude that there 
are $D = m^2$ degenerate eigenstates $|q_s, b_P\>$ for every configuration satisfying $\prod_P b_P = 1$ 
and $\sum_s q_s \equiv 0 \text{ (mod m)}$.
In particular, the ground state $|q_s = 0, b_P = 1\>$ is $m^2$-fold degenerate. 

\section{Eigenstate degeneracy of 3D lattice boson model}
\subsection{Open boundary condition geometry}
\label{degopen3}
We use the equation (\ref{deg4}) in an almost unchanged form, the only difference being that the quantity
$\Delta j_{ss'}$ is now defined as $j_{P_1}-j_{P_2}+j_{P_3}-j_{P_4}$, where $P_i$ are the four plaquettes meeting
at the edge $\<ss'\>$ and the labeling is clockwise as we go around the link (i.e. there is a relative minus 
sign between plaquettes which are perpendicular to each other. This is a different convention from the 2D case.). The 
only nonvanishing matrix elements are those for which:
\begin{equation}
\Delta j_{ss'} \equiv 0 \text{ (mod m)}
\label{3dconstr}
\end{equation}
The question
is thus reduced to the problem of counting possible assignments $\{j_P\}$ that satisfy those conditions.
To count the allowed configurations of $\{j_P\}$, we will construct an explicit mapping from choices of $\{ j_P \}$ satisfying 
Eq. (\ref{3dconstr}) to configurations of $Z_m$ spins living at the centers of the cubes on the lattice (i.e. $Z_m$ spins on 
the dual lattice). One can picture $j_P = 0$ as the absence of a domain wall between the cubes; more generally the value of $j_P$ 
assigns a label to the face separating any pair of cubes, which we will later identify with the change in the $Z_m$ spin between 
these cubes. The conditions (\ref{3dconstr}) ensure that domain walls can never end on an edge -- and further, that the net 
change in the spin along any closed curve is a multiple of $m$, so that an allowed configuration of $\{j_P\}$ does indeed specify 
a set of domain walls between $Z_m$ spins on adjacent cubes.

\begin{figure}[tb]
\centerline{
\includegraphics[width=0.7\columnwidth]{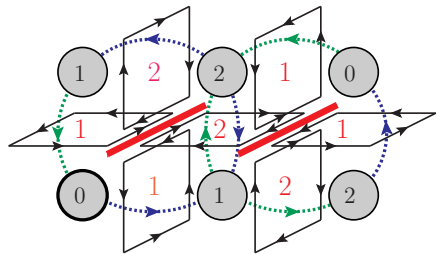}
}
\caption{An example of mapping to spin configuration for $m=3$. The numbers in red correspond to a choice of $\{j_P\}$. 
The base of the loop is the lower left corner, which for concreteness has been assigned a value 0.
The sign with which we pick up the appropriate $j_P$ (or equivalently with which the plaquette contributes to $\Delta j_{ss'}$ 
for a specific link $\<ss'\>$ enclosed by the loop) has been coded in blue for $+$ and green for $-$. The involved links have 
been colored red.}
\label{3dcountfig}
\end{figure}

The precise mapping is as follows.  Let us focus on some link $\<ss'\>$ incident
at the corner of the sample -- there are only two plaquettes $P,P'$ bordering that link and 
$ \Delta j_{ss'} = j_P - j_{P'}$, so in order for it to contribute to the trace we need $j_P = j_{P'}$ which can 
assume $m$ values. We place the corresponding value of $j_C = j_P$ in the corner cube of the lattice. We assign
a value to all other cubes sequentially using the following algorithm: start in the corner cube and construct a 
closed directed loop going through a number of other cubes. If $C,C'$ are cubes in the loop separated 
by a plaquette $P$ and the value $j_C$ is already assigned, then $j_{C'} \equiv j_C \pm j_P \text{ (mod m)}$.
 We adopt the convention that we always pick up a positive sign as we leave the corner cube and later 
it alternates as shown in figure (\ref{3dcountfig}). 
The consistency is guaranteed by the fact that as we go around a single link  $\<ss'\>$ and return to the cube $C$ 
we pick up a contribution of precisely $\pm\Delta j_{ss'} \equiv 0 \text{ (mod m)}$. More generally, if we pick 
any closed loop starting at $C$ we can decompose it into a number of small loops around individual links enclosed by the 
big loop, all of which contribute $0 \text{ (mod m)}$, which can also be seen in figure (\ref{3dcountfig}). Thus the assignment 
of spins is consistent. Once the corner cube has been assigned a value the mapping is unique. Conversely, 
given a configuration of $\{j_C\}$ in the cubes we can reconstruct the corresponding assignment of $\{j_P\}$. 
Therefore there is a bijection between the plaquette configurations and the cube configurations -- 
of which there are $m^{N_{\text{cube}}}$. We conclude that there are $m^{N_{\text{cube}}}$ terms contributing 
to the sum in equation (\ref{deg4}). 

For each plaquette configuration $\{j_P\}$, we must now evaluate the product 
\begin{equation}
\frac{1}{m^{N_{plaq} }} \prod_P \overline{b}_P^{j_P} Tr \left( \prod_s P_{q_s } \right ) \ \ \ .
\end{equation}
The evaluation of the trace is identical to the 2D case, and again gives $m^{N_{link}-N_{site}+1}$. The factor 
$\prod_P \overline{b}_P^{j_P}$ is always equal to $1$ -- as long as $\prod_{P \in C} b_P = 1$ 
for each cube $C$. (Using this relation, we can simultaneously reduce the labels on all of the faces of a given 
cube by any integer $j$ (mod $m$). This preserves the condition (\ref{3dconstr}) everywhere. Since the allowed configurations 
constitute closed domain walls, a series of these reductions can be used to reduce all of the labels to $0$, which proves that 
$\prod_P \bar{b}_P^{j_P} = \prod_P \bar{b}_P^{0} = 1$ for the allowed configurations $\{j_P\}$). Thus:
\begin{eqnarray}
D &=& \frac{m^{N_{\text{cube}}}}{m^{N_{\text{plaq}}}} \cdot Tr \left(\prod_s P_{q_s} \right) \nonumber \\
&=& m^{N_{\text{cube}}-N_{\text{plaq}}+N_{\text{link}}-N_{\text{site}}+1}
\label{3dopencount1}
\end{eqnarray}
The exponent can be evaluated by explicit counting or using the generalized Euler formula:
\begin{equation}
N_{\text{cube}}-N_{\text{plaq}}+N_{\text{link}}-N_{\text{site}}+1 = 0.
\label{3dopencount2}
\end{equation}
Thus we have shown that $D=1$, provided the constraint $\prod_{P \in C} b_P = 1$ is satisfied for all cubes.

\subsection{Periodic (torus) geometry}
\label{degtorus3}
Next, we repeat the above counting argument for a system with periodic boundary conditions in the $x$, $y$, and $z$ directions.
It is instructive to start with the open boundary case again. We may divide the configurations $\{j_P\}$ satisfying 
(\ref{3dconstr}) into bulk and boundary parts: $\{j_P\}_{P\in boundary}$ and $\{j_P\}_{P\in bulk}$. Assume we have fixed some 
bulk configuration. In the open geometry, fixing one of the boundary plaquettes (say, in the corner) to one of the $m$ possible 
values automatically fixes all other boundary plaquettes. This is because the links $\<ss'\>$ on the boundary always have only 
two incident boundary plaquettes (and possibly a bulk one, but it is already fixed). Hence, the total number of configurations 
is $m$ times the number of bulk configurations.

Let us now define the periodic geometry by identifying opposite faces of the boundary. This allows us to inherit the notion of 
bulk and boundary configurations from the open case. Note that the two cases do not differ at all in the bulk -- 
the allowed configurations satisfying (\ref{3dconstr}) are the same. However, at the boundary there are more possible 
configurations if we impose periodic boundary conditions. Let us again fix a bulk configuration. In the periodic case choosing 
a value for one of the boundary plaquettes only fixes the value of the plaquettes in the same boundary plane. The two other 
perpendicular planes still remain unfixed as each boundary link in the corner has four incident boundary plaquettes. 
Independently choosing a value for one plaquette in the remaining planes produces a factor of $m^2$. Hence the total number of 
configurations in periodic case is $m^3$ times the number of bulk configurations or $m^2$ times the total number of 
configurations in the open case. Thus we have $m^{N_{\text{cube}}+2}$ configurations. Note also, that in the periodic case there 
is an additional algebraic constraint satisfied by operators $B_P$: if we take any boundary plane $i$ (which is a 2D torus 
embedded in the 3D one) then:
\begin{equation}
\prod_{P \in plane(i)} B_P = 1. 
\label{newconstr}
\end{equation}

The rest of the calculation is unchanged:
\begin{equation}
D= m^{N_{\text{cube}}+2 -N_{\text{plaq}}+N_{\text{link}}-N_{\text{site}}+1} = m^3,
\label{3dperiodiccount1}
\end{equation}
since in the periodic case we have $N_{\text{cube}} -N_{\text{plaq}}+N_{\text{link}}-N_{\text{site}} = 0$ either by explicit 
counting or using the Euler formula. Thus there is a $D = m^3$ degeneracy, provided the constraints (\ref{bpconstr}) and 
(\ref{newconstr}) are satisfied. In particular, the ground state $|q_s=0, b_P = 1\>$ has an $m^3$ degeneracy.

\section{Relationship between the bosonic edge theory (\ref{2dedgeboson}) and the free fermion edge theory (\ref{2dedge})}
\label{edgerelation}
In this section, we show that the perturbation $U(\Lambda_0)$ (\ref{scatter}) where $\Lambda_0 = (1,0,0,0)$ 
can gap out two of the edge modes of (\ref{2dedgeboson}), leaving behind exactly the free fermion edge (\ref{2dedge}). 
We accomplish this via a change of variables,
\begin{align}
\Phi = W \tilde{\Phi} \ , \ W = \bpm 1 & 0 & 0 & 0 \\ 
			      0 & 1/m & k/m & k/m \\
			      0 & 0 & 1 & 0 \\
			      0 & 0 & 0 & 1 \epm
\end{align}
Substituting these expressions into the edge action (\ref{2dedgeboson}), we find
\begin{eqnarray}
L &=& \frac{1}{4\pi} \partial_x \tilde{\Phi}_I( \tilde{K}_{IJ} \partial_t \tilde{\Phi}_J - 
\tilde{V}_{IJ} \partial_x\tilde{\Phi}_J) \nonumber \\
&+& \frac{1}{2\pi} \tilde{t}_I\epsilon^{\mu \nu} \partial_\mu \tilde{\Phi}_I A_\nu
\end{eqnarray}
where
\begin{align}
\tilde{K}_{IJ} = \bpm 0 & 1 & 0 & 0 \\
		    1 & 0 & 0 & 0 \\
		    0 & 0 & 1 & 0 \\
		    0 & 0 & 0 & -1 \epm, \ \tilde{t}_{I} = \bpm 0 \\ 2e/m \\ (2k+m)e/m \\ (2k+m)e/m \epm
\end{align}
and $\tilde{V} = W^T V W$. In these variables, the perturbation becomes
\begin{equation}
U(\Lambda_0) = 2 U(x)\cos(\tilde{\Phi}_2 - \alpha(x))
\label{pertnewbasis}
\end{equation}
We next assume that that the interactions on the edge are tuned so that $\tilde{V} = v \delta_{IJ}$ (we can make this
assumption without any loss of generality since the velocity matrix is non-universal and can be modified by
appropriate perturbations at the edge). In this case, the edge theory can be written as a sum of two decoupled actions, 
one involving $\tilde{\Phi}_1, \tilde{\Phi}_2$ and one involving $\tilde{\Phi}_3, \tilde{\Phi}_4$:
\begin{equation}
L = L_{12} + L_{34}
\end{equation}
where
\begin{eqnarray}
L_{12} &=& \frac{1}{4\pi} \left(2\partial_x \tilde{\Phi}_1 \partial_t \tilde{\Phi}_2 
- v (\partial_x\tilde{\Phi}_1)^2 - v (\partial_x \tilde{\Phi}_2)^2 \right) \nonumber \\ 
&+& \frac{e}{m\pi} \epsilon^{\mu \nu} \partial_\mu \tilde{\Phi}_2 A_\nu
\end{eqnarray}
and
\begin{eqnarray}
L_{34} &=& \frac{1}{4\pi} \partial_x \tilde{\Phi}_3 (\partial_t \tilde{\Phi}_3 - v \partial_x \tilde{\Phi}_3) \nonumber \\
&-& \frac{1}{4\pi} \partial_x \tilde{\Phi}_4 (\partial_t \tilde{\Phi}_4 + v \partial_x \tilde{\Phi}_4) \nonumber \\
&+& \frac{(2k+m)e}{2m\pi} \epsilon^{\mu \nu} \partial_\mu (\tilde{\Phi}_3 + \tilde{\Phi}_4) A_\nu
\end{eqnarray}
It is now easy to analyze the effect of the perturbation $U(\Lambda_0)$ (\ref{pertnewbasis}): this term gaps out the non-chiral
Luttinger liquid described by $L_{12}$ by freezing the value of $\tilde{\Phi}_2$ (at least if $U(x)$ is large).
The resulting edge then has only two gapless modes, and is described by $L_{34}$. On the other hand, it is easy to check
that $L_{34}$ is nothing but the bosonized description of the free fermion edge theory (\ref{2dedge}), with 
$e^{i\tilde{\Phi}_3}, e^{-i\tilde{\Phi}_4}$ corresponding to the fermion creation operators $d_{\uparrow}^\dagger, 
d_{\downarrow}^\dagger$ respectively. \cite{WenReview,WenBook} We conclude that the perturbation $U(\Lambda_0)$
does indeed gap out two of the modes of (\ref{2dedgeboson}) leaving the free fermion edge (\ref{2dedge}).

\bibliography{3DTI}

\begin{thebibliography}{10}%
\makeatletter
\providecommand \@ifxundefined [1]{%
 \ifx #1\undefined \expandafter \@firstoftwo
 \else \expandafter \@secondoftwo
\fi
}%
\providecommand \@ifnum [1]{%
 \ifnum #1\expandafter \@firstoftwo
 \else \expandafter \@secondoftwo
\fi
}%
\providecommand \enquote [1]{``#1''}%
\providecommand \bibnamefont  [1]{#1}%
\providecommand \bibfnamefont [1]{#1}%
\providecommand \citenamefont [1]{#1}%
\providecommand\href[0]{\@sanitize\@href}%
\providecommand\@href[1]{\endgroup\@@startlink{#1}\endgroup\@@href}%
\providecommand\@@href[1]{#1\@@endlink}%
\providecommand \@sanitize [0]{\begingroup\catcode`\&12\catcode`\#12\relax}%
\@ifxundefined \pdfoutput {\@firstoftwo}{%
 \@ifnum{\z@=\pdfoutput}{\@firstoftwo}{\@secondoftwo}%
}{%
 \providecommand\@@startlink[1]{\leavevmode\special{html:<a href="#1">}}%
 \providecommand\@@endlink[0]{\special{html:</a>}}%
}{%
 \providecommand\@@startlink[1]{%
  \leavevmode
  \pdfstartlink
   attr{/Border[0 0 1 ]/H/I/C[0 1 1]}%
   user{/Subtype/Link/A<</Type/Action/S/URI/URI(#1)>>}%
  \relax
 }%
 \providecommand\@@endlink[0]{\pdfendlink}%
}%
\providecommand \url  [0]{\begingroup\@sanitize \@url }%
\providecommand \@url [1]{\endgroup\@href {#1}{\urlprefix}}%
\providecommand \urlprefix [0]{URL }%
\providecommand \Eprint[0]{\href }%
\@ifxundefined \urlstyle {%
  \providecommand \doi [1]{doi:\discretionary{}{}{}#1}%
}{%
  \providecommand \doi [0]{doi:\discretionary{}{}{}\begingroup
  \urlstyle{rm}\Url }%
}%
\providecommand \doibase [0]{http://dx.doi.org/}%
\providecommand \Doi[1]{\href{\doibase#1}}%
\providecommand \bibAnnote [3]{%
  \BibitemShut{#1}%
  \begin{quotation}\noindent
    \textsc{Key:}\ #2\\\textsc{Annotation:}\ #3%
  \end{quotation}%
}%
\providecommand \bibAnnoteFile [2]{%
  \IfFileExists{#2}{\bibAnnote {#1} {#2} {\input{#2}}}{}%
}%
\providecommand \typeout [0]{\immediate \write \m@ne }%
\providecommand \selectlanguage [0]{\@gobble}%
\providecommand \bibinfo [0]{\@secondoftwo}%
\providecommand \bibfield [0]{\@secondoftwo}%
\providecommand \translation [1]{[#1]}%
\providecommand \BibitemOpen[0]{}%
\providecommand \bibitemStop [0]{}%
\providecommand \bibitemNoStop [0]{.\EOS\space}%
\providecommand \EOS [0]{\spacefactor3000\relax}%
\providecommand \BibitemShut [1]{\csname bibitem#1\endcsname}%
\bibitem{KaneMele}%
  \BibitemOpen
  \bibfield{author}{%
  \bibinfo {author} {\bibfnamefont{C.~L.}\ \bibnamefont{Kane}}\ and\ \bibinfo
  {author} {\bibfnamefont{E.~J.}\ \bibnamefont{Mele}},\ }%
  \bibfield{journal}{%
  \Doi{10.1103/PhysRevLett.95.226801}{\bibinfo {journal} {Phys. Rev. Lett.}}\
  }%
  \textbf{\bibinfo {volume} {95}},\ \bibinfo {pages} {226801} (\bibinfo {year}
  {2005})%
  \bibAnnoteFile{NoStop}{KaneMele}%
\bibitem{KaneMele2}%
  \BibitemOpen
  \bibfield{author}{%
  \bibinfo {author} {\bibfnamefont{C.~L.}\ \bibnamefont{Kane}}\ and\ \bibinfo
  {author} {\bibfnamefont{E.~J.}\ \bibnamefont{Mele}},\ }%
  \bibfield{journal}{%
  \Doi{10.1103/PhysRevLett.95.226801}{\bibinfo {journal} {Phys. Rev. Lett.}}\
  }%
  \textbf{\bibinfo {volume} {95}},\ \bibinfo {pages} {146802} (\bibinfo {year}
  {2005})%
  \bibAnnoteFile{NoStop}{KaneMele2}%
\bibitem{BernevigZhang}%
  \BibitemOpen
  \bibfield{author}{%
  \bibinfo {author} {\bibfnamefont{B.~A.}\ \bibnamefont{Bernevig}}\ and\
  \bibinfo {author} {\bibfnamefont{S.-C.}\ \bibnamefont{Zhang}},\ }%
  \bibfield{journal}{%
  \Doi{10.1103/PhysRevLett.96.106802}{\bibinfo {journal} {Phys. Rev. Lett.}}\
  }%
  \textbf{\bibinfo {volume} {96}},\ \bibinfo {pages} {106802} (\bibinfo {year}
  {2006})%
  \bibAnnoteFile{NoStop}{BernevigZhang}%
\bibitem{HasanKaneRMP}%
  \BibitemOpen
  \bibfield{author}{%
  \bibinfo {author} {\bibfnamefont{M.~Z.}\ \bibnamefont{Hasan}}\ and\ \bibinfo
  {author} {\bibfnamefont{C.~L.}\ \bibnamefont{Kane}},\ }%
  \bibfield{journal}{%
  \bibinfo {journal} {Rev. Mod. Phys.}\ }%
  \textbf{\bibinfo {volume} {82}},\ \bibinfo {pages} {3045} (\bibinfo {year}
  {2010})%
  \bibAnnoteFile{NoStop}{HasanKaneRMP}%
\bibitem{Roy}%
  \BibitemOpen
  \bibfield{author}{%
  \bibinfo {author} {\bibfnamefont{R.}~\bibnamefont{Roy}},\ }%
  \bibfield{journal}{%
  \Doi{10.1103/PhysRevB.79.195322}{\bibinfo {journal} {Phys. Rev. B}}\ }%
  \textbf{\bibinfo {volume} {79}},\ \bibinfo {pages} {195322} (\bibinfo {year}
  {2009})%
  \bibAnnoteFile{NoStop}{Roy}%
\bibitem{FuKaneMele}%
  \BibitemOpen
  \bibfield{author}{%
  \bibinfo {author} {\bibfnamefont{L.}~\bibnamefont{Fu}}, \bibinfo {author}
  {\bibfnamefont{C.~L.}\ \bibnamefont{Kane}},\ and\ \bibinfo {author}
  {\bibfnamefont{E.~J.}\ \bibnamefont{Mele}},\ }%
  \bibfield{journal}{%
  \Doi{10.1103/PhysRevLett.98.106803}{\bibinfo {journal} {Phys. Rev. Lett.}}\
  }%
  \textbf{\bibinfo {volume} {98}},\ \bibinfo {pages} {106803} (\bibinfo {year}
  {2007})%
  \bibAnnoteFile{NoStop}{FuKaneMele}%
\bibitem{MooreBalents}%
  \BibitemOpen
  \bibfield{author}{%
  \bibinfo {author} {\bibfnamefont{J.~E.}\ \bibnamefont{Moore}}\ and\ \bibinfo
  {author} {\bibfnamefont{L.}~\bibnamefont{Balents}},\ }%
  \bibfield{journal}{%
  \Doi{10.1103/PhysRevB.75.121306}{\bibinfo {journal} {Phys. Rev. B}}\ }%
  \textbf{\bibinfo {volume} {75}},\ \bibinfo {pages} {121306} (\bibinfo {year}
  {2007})%
  \bibAnnoteFile{NoStop}{MooreBalents}%
\bibitem{FuKanepump}%
  \BibitemOpen
  \bibfield{author}{%
  \bibinfo {author} {\bibfnamefont{L.}~\bibnamefont{Fu}}\ and\ \bibinfo
  {author} {\bibfnamefont{C.~L.}\ \bibnamefont{Kane}},\ }%
  \bibfield{journal}{%
  \Doi{10.1103/PhysRevB.76.195312}{\bibinfo {journal} {Phys. Rev. B}}\ }%
  \textbf{\bibinfo {volume} {74}},\ \bibinfo {pages} {195312} (\bibinfo {year}
  {2006})%
  \bibAnnoteFile{NoStop}{FuKanepump}%
\bibitem{FuKaneHall}%
  \BibitemOpen
  \bibfield{author}{%
  \bibinfo {author} {\bibfnamefont{L.}~\bibnamefont{Fu}}\ and\ \bibinfo
  {author} {\bibfnamefont{C.~L.}\ \bibnamefont{Kane}},\ }%
  \bibfield{journal}{%
  \Doi{10.1103/PhysRevB.76.045302}{\bibinfo {journal} {Phys. Rev. B}}\ }%
  \textbf{\bibinfo {volume} {76}},\ \bibinfo {pages} {045302} (\bibinfo {year}
  {2007})%
  \bibAnnoteFile{NoStop}{FuKaneHall}%
\bibitem{LevinStern}%
  \BibitemOpen
  \bibfield{author}{%
  \bibinfo {author} {\bibfnamefont{M.}~\bibnamefont{Levin}}\ and\ \bibinfo
  {author} {\bibfnamefont{A.}~\bibnamefont{Stern}},\ }%
  \bibfield{journal}{%
  \Doi{10.1103/PhysRevLett.103.196803}{\bibinfo {journal} {Phys. Rev. Lett.}}\
  }%
  \textbf{\bibinfo {volume} {103}},\ \bibinfo {pages} {196803} (\bibinfo {year}
  {2009})%
  \bibAnnoteFile{NoStop}{LevinStern}%
\bibitem{Freedmanetal}%
  \BibitemOpen
  \bibfield{author}{%
  \bibinfo {author} {\bibfnamefont{M.}~\bibnamefont{Freedman}}, \bibinfo
  {author} {\bibfnamefont{C.}~\bibnamefont{Nayak}}, \bibinfo {author}
  {\bibfnamefont{K.}~\bibnamefont{Shtengel}}, \bibinfo {author}
  {\bibfnamefont{K.}~\bibnamefont{Walker}},\ and\ \bibinfo {author}
  {\bibfnamefont{Z.}~\bibnamefont{Wang}},\ }%
  \bibfield{journal}{%
  \Doi{10.1016/j.aop.2004.01.006}{\bibinfo {journal} {Ann. Phys.}}\ }%
  \textbf{\bibinfo {volume} {310}},\ \bibinfo {pages} {428} (\bibinfo {year}
  {2004})%
  \bibAnnoteFile{NoStop}{Freedmanetal}%
\bibitem{Swingle}%
  \BibitemOpen
  \bibfield{author}{%
  \bibinfo {author} {\bibfnamefont{B.}~\bibnamefont{Swingle}}, \bibinfo
  {author} {\bibfnamefont{M.}~\bibnamefont{Barkeshli}}, \bibinfo {author}
  {\bibfnamefont{J.}~\bibnamefont{McGreevy}},\ and\ \bibinfo {author}
  {\bibfnamefont{T.}~\bibnamefont{Senthil}},\ }%
  \bibfield{journal}{%
  \Doi{10.1103/PhysRevB.83.195139}{\bibinfo {journal} {Phys. Rev. B}}\ }%
  \textbf{\bibinfo {volume} {83}},\ \bibinfo {pages} {195139} (\bibinfo {year}
  {2011})%
  \bibAnnoteFile{NoStop}{Swingle}%
\bibitem{Maciejko}%
  \BibitemOpen
  \bibfield{author}{%
  \bibinfo {author} {\bibfnamefont{J.}~\bibnamefont{Maciejko}}, \bibinfo
  {author} {\bibfnamefont{X.-L.}\ \bibnamefont{Qi}}, \bibinfo {author}
  {\bibfnamefont{A.}~\bibnamefont{Karch}},\ and\ \bibinfo {author}
  {\bibfnamefont{S.-C.}\ \bibnamefont{Zhang}},\ }%
  \bibfield{journal}{%
  \Doi{10.1103/PhysRevLett.105.246809}{\bibinfo {journal} {Phys. Rev. Lett.}}\
  }%
  \textbf{\bibinfo {volume} {105}},\ \bibinfo {pages} {246809} (\bibinfo {year}
  {2010})%
  \bibAnnoteFile{NoStop}{Maciejko}%
\bibitem{KitaevToric}%
  \BibitemOpen
  \bibfield{author}{%
  \bibinfo {author} {\bibfnamefont{A.~Y.}\ \bibnamefont{Kitaev}},\ }%
  \bibfield{journal}{%
  \bibinfo {journal} {Annals of Physics}\ }%
  \textbf{\bibinfo {volume} {303}},\ \bibinfo {pages} {2} (\bibinfo {year}
  {2003})%
  \bibAnnoteFile{NoStop}{KitaevToric}%
\bibitem{MotrunichSenthil}%
  \BibitemOpen
  \bibfield{author}{%
  \bibinfo {author} {\bibfnamefont{O.~I.}\ \bibnamefont{Motrunich}}\ and\
  \bibinfo {author} {\bibfnamefont{T.}~\bibnamefont{Senthil}},\ }%
  \bibfield{journal}{%
  \Doi{10.1103/PhysRevB.66.277004}{\bibinfo {journal} {Phys. Rev. Lett.}}\ }%
  \textbf{\bibinfo {volume} {89}},\ \bibinfo {pages} {277004} (\bibinfo {year}
  {2002})%
  \bibAnnoteFile{NoStop}{MotrunichSenthil}%
\bibitem{SenthilMotrunich}%
  \BibitemOpen
  \bibfield{author}{%
  \bibinfo {author} {\bibfnamefont{T.}~\bibnamefont{Senthil}}\ and\ \bibinfo
  {author} {\bibfnamefont{O.}~\bibnamefont{Motrunich}},\ }%
  \bibfield{journal}{%
  \Doi{10.1103/PhysRevB.66.205104}{\bibinfo {journal} {Phys. Rev. B}}\ }%
  \textbf{\bibinfo {volume} {66}},\ \bibinfo {pages} {205104} (\bibinfo {year}
  {2002})%
  \bibAnnoteFile{NoStop}{SenthilMotrunich}%
\bibitem{ReadChakraborty}%
  \BibitemOpen
  \bibfield{author}{%
  \bibinfo {author} {\bibfnamefont{N.}~\bibnamefont{Read}}\ and\ \bibinfo
  {author} {\bibfnamefont{B.}~\bibnamefont{Chakraborty}},\ }%
  \bibfield{journal}{%
  \Doi{10.1103/PhysRevB.40.7133}{\bibinfo {journal} {Phys. Rev. B}}\ }%
  \textbf{\bibinfo {volume} {40}},\ \bibinfo {pages} {7133} (\bibinfo {year}
  {1989})%
  \bibAnnoteFile{NoStop}{ReadChakraborty}%
\bibitem{MoessnerSondhiFradkin}%
  \BibitemOpen
  \bibfield{author}{%
  \bibinfo {author} {\bibfnamefont{R.}~\bibnamefont{Moessner}}, \bibinfo
  {author} {\bibfnamefont{S.~L.}\ \bibnamefont{Sondhi}},\ and\ \bibinfo
  {author} {\bibfnamefont{E.}~\bibnamefont{Fradkin}},\ }%
  \bibfield{journal}{%
  \Doi{10.1103/PhysRevB.65.024504}{\bibinfo {journal} {Phys. Rev. B}}\ }%
  \textbf{\bibinfo {volume} {65}},\ \bibinfo {pages} {024504} (\bibinfo {year}
  {2001})%
  \bibAnnoteFile{NoStop}{MoessnerSondhiFradkin}%
\bibitem{NayakShtengel}%
  \BibitemOpen
  \bibfield{author}{%
  \bibinfo {author} {\bibfnamefont{C.}~\bibnamefont{Nayak}}\ and\ \bibinfo
  {author} {\bibfnamefont{K.}~\bibnamefont{Shtengel}},\ }%
  \bibfield{journal}{%
  \Doi{10.1103/PhysRevB.64.064422}{\bibinfo {journal} {Phys. Rev. B}}\ }%
  \textbf{\bibinfo {volume} {64}},\ \bibinfo {pages} {064422} (\bibinfo {year}
  {2001})%
  \bibAnnoteFile{NoStop}{NayakShtengel}%
\bibitem{WenReview}%
  \BibitemOpen
  \bibfield{author}{%
  \bibinfo {author} {\bibfnamefont{X.-G.}\ \bibnamefont{Wen}},\ }%
  \bibfield{journal}{%
  \bibinfo {journal} {Adv. Phys.}\ }%
  \textbf{\bibinfo {volume} {44}},\ \bibinfo {pages} {405} (\bibinfo {year}
  {1995})%
  \bibAnnoteFile{NoStop}{WenReview}%
\bibitem{WenBook}%
  \BibitemOpen
  \bibfield{author}{%
  \bibinfo {author} {\bibfnamefont{X.-G.}\ \bibnamefont{Wen}},\ }%
  \emph{\bibinfo {title} {Quantum Field Theory of Many-body Systems: From the
  Origin of Sound to an Origin of Light and Electrons}}\ (\bibinfo {publisher}
  {Oxford University Press},\ \bibinfo {year} {2007})%
  \bibAnnoteFile{NoStop}{WenBook}%
\bibitem{Einarsson}%
  \BibitemOpen
  \bibfield{author}{%
  \bibinfo {author} {\bibfnamefont{T.}~\bibnamefont{Einarsson}},\ }%
  \bibfield{journal}{%
  \Doi{10.1103/PhysRevLett.64.1995}{\bibinfo {journal} {Phys. Rev. Lett.}}\ }%
  \textbf{\bibinfo {volume} {64}},\ \bibinfo {pages} {1995} (\bibinfo {year}
  {1990})%
  \bibAnnoteFile{NoStop}{Einarsson}%
\bibitem{WittenEffect}%
  \BibitemOpen
  \bibfield{author}{%
  \bibinfo {author} {\bibfnamefont{E.}~\bibnamefont{Witten}},\ }%
  \bibfield{journal}{%
  \Doi{DOI: 10.1016/0370-2693(79)90838-4}{\bibinfo {journal} {Physics Letters
  B}}\ }%
  \textbf{\bibinfo {volume} {86}},\ \bibinfo {pages} {283 } (\bibinfo {year}
  {1979}),\ ISSN \bibinfo {issn} {0370-2693}%
  \bibAnnoteFile{NoStop}{WittenEffect}%
\bibitem{Wilczek}%
  \BibitemOpen
  \bibfield{author}{%
  \bibinfo {author} {\bibfnamefont{F.}~\bibnamefont{Wilczek}},\ }%
  \bibfield{journal}{%
  \Doi{10.1103/PhysRevLett.58.1799}{\bibinfo {journal} {Phys. Rev. Lett.}}\ }%
  \textbf{\bibinfo {volume} {58}},\ \bibinfo {pages} {1799} (\bibinfo {year}
  {1987})%
  \bibAnnoteFile{NoStop}{Wilczek}%
\bibitem{RosenbergFranz}%
  \BibitemOpen
  \bibfield{author}{%
  \bibinfo {author} {\bibfnamefont{G.}~\bibnamefont{Rosenberg}}\ and\ \bibinfo
  {author} {\bibfnamefont{M.}~\bibnamefont{Franz}},\ }%
  \bibfield{journal}{%
  \Doi{10.1103/PhysRevB.82.035105}{\bibinfo {journal} {Phys. Rev. B}}\ }%
  \textbf{\bibinfo {volume} {82}},\ \bibinfo {pages} {035105} (\bibinfo {year}
  {2010})%
  \bibAnnoteFile{NoStop}{RosenbergFranz}%
\bibitem{BernevigHughesZhang}%
  \BibitemOpen
  \bibfield{author}{%
  \bibinfo {author} {\bibfnamefont{B.~A.}\ \bibnamefont{Bernevig}}, \bibinfo
  {author} {\bibfnamefont{T.~L.}\ \bibnamefont{Hughes}},\ and\ \bibinfo
  {author} {\bibfnamefont{S.-C.}\ \bibnamefont{Zhang}},\ }%
  \bibfield{journal}{%
  \Doi{10.1126/science.1133734}{\bibinfo {journal} {Science}}\ }%
  \textbf{\bibinfo {volume} {314}},\ \bibinfo {pages} {1757} (\bibinfo {year}
  {2006})%
  \bibAnnoteFile{NoStop}{BernevigHughesZhang}%
\bibitem{Kogut}%
  \BibitemOpen
  \bibfield{author}{%
  \bibinfo {author} {\bibfnamefont{J.~B.}\ \bibnamefont{Kogut}},\ }%
  \bibfield{journal}{%
  \Doi{10.1103/RevModPhys.51.659}{\bibinfo {journal} {Rev. Mod. Phys.}}\ }%
  \textbf{\bibinfo {volume} {51}},\ \bibinfo {pages} {659} (\bibinfo {year}
  {1979})%
  \bibAnnoteFile{NoStop}{Kogut}%
\bibitem{GefenThouless}%
  \BibitemOpen
  \bibfield{author}{%
  \bibinfo {author} {\bibfnamefont{Y.}~\bibnamefont{Gefen}}\ and\ \bibinfo
  {author} {\bibfnamefont{D.~J.}\ \bibnamefont{Thouless}},\ }%
  \bibfield{journal}{%
  \Doi{10.1103/PhysRevB.47.10423}{\bibinfo {journal} {Phys. Rev. B}}\ }%
  \textbf{\bibinfo {volume} {47}},\ \bibinfo {pages} {10423} (\bibinfo {year}
  {1993})%
  \bibAnnoteFile{NoStop}{GefenThouless}%
\bibitem{LevinWenHop}%
  \BibitemOpen
  \bibfield{author}{%
  \bibinfo {author} {\bibfnamefont{M.}~\bibnamefont{Levin}}\ and\ \bibinfo
  {author} {\bibfnamefont{X.-G.}\ \bibnamefont{Wen}},\ }%
  \bibfield{journal}{%
  \Doi{10.1103/PhysRevB.67.245316}{\bibinfo {journal} {Phys. Rev. B}}\ }%
  \textbf{\bibinfo {volume} {67}},\ \bibinfo {pages} {245316} (\bibinfo {year}
  {2003})%
  \bibAnnoteFile{NoStop}{LevinWenHop}%
\bibitem{KouLevinWen}%
  \BibitemOpen
  \bibfield{author}{%
  \bibinfo {author} {\bibfnamefont{S.-P.}\ \bibnamefont{Kou}}, \bibinfo
  {author} {\bibfnamefont{M.}~\bibnamefont{Levin}},\ and\ \bibinfo {author}
  {\bibfnamefont{X.-G.}\ \bibnamefont{Wen}},\ }%
  \bibfield{journal}{%
  \Doi{10.1103/PhysRevB.78.155134}{\bibinfo {journal} {Phys. Rev. B}}\ }%
  \textbf{\bibinfo {volume} {78}},\ \bibinfo {pages} {155134} (\bibinfo {year}
  {2008})%
  \bibAnnoteFile{NoStop}{KouLevinWen}%
\bibitem{MooreBF}%
  \BibitemOpen
  \bibfield{author}{%
  \bibinfo {author} {\bibfnamefont{G.~Y.}\ \bibnamefont{Cho}}\ and\ \bibinfo
  {author} {\bibfnamefont{J.~E.}\ \bibnamefont{Moore}},\ \bibinfo {pages}
  {arXiv:1011.3485}}%
   (\bibinfo {year} {2010})%
  \bibAnnoteFile{NoStop}{MooreBF}%
\bibitem{SondhiSC}%
  \BibitemOpen
  \bibfield{author}{%
  \bibinfo {author} {\bibfnamefont{T.~H.}\ \bibnamefont{Hansson}}, \bibinfo
  {author} {\bibfnamefont{V.}~\bibnamefont{Oganesyan}},\ and\ \bibinfo {author}
  {\bibfnamefont{S.~L.}\ \bibnamefont{Sondhi}},\ }%
  \bibfield{journal}{%
  \bibinfo {journal} {Annals of Physics}\ }%
  \textbf{\bibinfo {volume} {313}},\ \bibinfo {pages} {497 } (\bibinfo {year}
  {2004})%
  \bibAnnoteFile{NoStop}{SondhiSC}%
\bibitem{WenKmatrix}%
  \BibitemOpen
  \bibfield{author}{%
  \bibinfo {author} {\bibfnamefont{X.~G.}\ \bibnamefont{Wen}}\ and\ \bibinfo
  {author} {\bibfnamefont{A.}~\bibnamefont{Zee}},\ }%
  \bibfield{journal}{%
  \Doi{10.1103/PhysRevB.46.2290}{\bibinfo {journal} {Phys. Rev. B}}\ }%
  \textbf{\bibinfo {volume} {46}},\ \bibinfo {pages} {2290} (\bibinfo {year}
  {1992})%
  \bibAnnoteFile{NoStop}{WenKmatrix}%
\bibitem{QiHughesZhang}%
  \BibitemOpen
  \bibfield{author}{%
  \bibinfo {author} {\bibfnamefont{X.-L.}\ \bibnamefont{Qi}}, \bibinfo {author}
  {\bibfnamefont{T.~L.}\ \bibnamefont{Hughes}},\ and\ \bibinfo {author}
  {\bibfnamefont{S.-C.}\ \bibnamefont{Zhang}},\ }%
  \bibfield{journal}{%
  \Doi{10.1103/PhysRevB.78.195424}{\bibinfo {journal} {Phys. Rev. B}}\ }%
  \textbf{\bibinfo {volume} {78}},\ \bibinfo {pages} {195424} (\bibinfo {year}
  {2008})%
  \bibAnnoteFile{NoStop}{QiHughesZhang}%
\bibitem{MooreEssin}%
  \BibitemOpen
  \bibfield{author}{%
  \bibinfo {author} {\bibfnamefont{A.~M.}\ \bibnamefont{Essin}}, \bibinfo
  {author} {\bibfnamefont{A.~M.}\ \bibnamefont{Turner}}, \bibinfo {author}
  {\bibfnamefont{J.~E.}\ \bibnamefont{Moore}},\ and\ \bibinfo {author}
  {\bibfnamefont{D.}~\bibnamefont{Vanderbilt}},\ }%
  \bibfield{journal}{%
  \Doi{10.1103/PhysRevB.81.205104}{\bibinfo {journal} {Phys. Rev. B}}\ }%
  \textbf{\bibinfo {volume} {81}},\ \bibinfo {pages} {205104} (\bibinfo {year}
  {2010})%
  \bibAnnoteFile{NoStop}{MooreEssin}%
\bibitem{HammaZanardiWen}%
  \BibitemOpen
  \bibfield{author}{%
  \bibinfo {author} {\bibfnamefont{A.}~\bibnamefont{Hamma}}, \bibinfo {author}
  {\bibfnamefont{P.}~\bibnamefont{Zanardi}},\ and\ \bibinfo {author}
  {\bibfnamefont{X.-G.}\ \bibnamefont{Wen}},\ }%
  \bibfield{journal}{%
  \Doi{10.1103/PhysRevB.72.035307}{\bibinfo {journal} {Phys. Rev. B}}\ }%
  \textbf{\bibinfo {volume} {72}},\ \bibinfo {pages} {035307} (\bibinfo {year}
  {2005})%
  \bibAnnoteFile{NoStop}{HammaZanardiWen}%
\bibitem{LevinSternprep}%
  \BibitemOpen
  \bibfield{author}{%
  \bibinfo {author} {\bibfnamefont{M.}~\bibnamefont{Levin}}\ and\ \bibinfo
  {author} {\bibfnamefont{A.}~\bibnamefont{Stern}},\ }%
  \bibinfo {journal} {in preparation}%
  \bibAnnoteFile{NoStop}{LevinSternprep}%
\bibitem{RingelKrausStern}%
  \BibitemOpen
\bibfield{journal}{%
    }%
  \bibfield{author}{%
  \bibinfo {author} {\bibfnamefont{Z.}~\bibnamefont{Ringel}}, \bibinfo {author}
  {\bibfnamefont{Y.~E.}\ \bibnamefont{Kraus}},\ and\ \bibinfo {author}
  {\bibfnamefont{A.}~\bibnamefont{Stern}},\ \bibinfo {pages}
  {arxiv:1105.4351}}%
   (\bibinfo {year} {2011})%
  \bibAnnoteFile{NoStop}{RingelKrausStern}%
\bibitem{Haldane}%
  \BibitemOpen
  \bibfield{author}{%
  \bibinfo {author} {\bibfnamefont{F.}~\bibnamefont{Haldane}},\ }%
  \bibfield{journal}{%
  \Doi{10.1103/PhysRevLett.74.2090}{\bibinfo {journal} {Phys. Rev. Lett.}}\ }%
  \textbf{\bibinfo {volume} {74}},\ \bibinfo {pages} {2090} (\bibinfo {year}
  {1995})%
  \bibAnnoteFile{NoStop}{Haldane}%
\end{thebibliography}%

\end{document}